\documentclass[%
 reprint,
superscriptaddress,
 amsmath,amssymb,
 aps,twocolumn,
pra,
]{revtex4-2}

\usepackage{graphicx}
\usepackage{dcolumn}
\usepackage{bm}
\usepackage[hypertexnames=false]{hyperref}
\usepackage{siunitx}
\usepackage[capitalise]{cleveref}
\usepackage{lipsum}

\usepackage{braket}

\newcommand{\ah}{\hat{a}}
\newcommand{\hc}{\mathrm{h.c.}}

\usepackage{physics} 
\usepackage{enumerate}
\usepackage{dsfont}

\newcommand{\eff}{\mathrm{eff}}
\newcommand{\emax}{\eta_{\mathrm{max}}}

\usepackage{xcolor}

\usepackage[normalem]{ulem}

\newcommand{\cc}{\mathrm{c.c.}}

\renewcommand{\Im}{\mathrm{Im}}
\renewcommand{\Re}{\mathrm{Re}}
\newcommand{\lf}{\left(\frac{\lambda}{2}\right)}
\newcommand{\E}{\mathbb{E}}

\begin{document}


\title{Non-Local Multi-Qubit Quantum Gates via a Driven Cavity} 



\author{Sven Jandura}
\affiliation{University of Strasbourg and CNRS, CESQ and ISIS (UMR 7006), aQCess, 67000 Strasbourg, France}

\author{Vineesha Srivastava}
\affiliation{University of Strasbourg and CNRS, CESQ and ISIS (UMR 7006), aQCess, 67000 Strasbourg, France}

\author{Laura Pecorari}
\affiliation{University of Strasbourg and CNRS, CESQ and ISIS (UMR 7006), aQCess, 67000 Strasbourg, France}

\author{Gavin K Brennen}
\affiliation{Center for Engineered Quantum Systems, School of Mathematical and Physical Sciences, Macquarie University, 2109 NSW, Australia}

\author{Guido Pupillo}
\affiliation{University of Strasbourg and CNRS, CESQ and ISIS (UMR 7006), aQCess, 67000 Strasbourg, France}


\date{\today}

\begin{abstract}
We present two protocols for realizing deterministic non-local multi-qubit quantum gates on qubits coupled to a common cavity mode. The protocols rely only on a classical drive of the cavity mode, while no external drive of the qubits is required. Applied to just two qubits,
both protocols provide a universal gate set for quantum computing, together with single qubit
gates. In the first protocol,  the state of the cavity follows a closed trajectory in phase space and accumulates a geometric phase depending on the state of the qubits. This geometric phase gate can be used together with global single qubit gates to generate high-fidelity GHZ states. The second protocol uses an adiabatic evolution of the combined qubit-cavity system to accumulate a dynamical phase. Repeated applications of this protocol allow for the realization of a family of phase
gates with arbitrary phases, e.g. phase-rotation gates and multi-controlled-Z gates.  For both protocols, we provide analytic solutions for the error rates, which scale as $\sim N/\sqrt{C}$ in the presence of relevant losses, with $C$ the cooperativity and $N$ the qubit number. Our protocols are applicable to a variety of systems and can be generalized by replacing the cavity by a different bosonic mode, such as a phononic mode. We provide estimates of gate fidelities and durations for atomic and molecular qubits as well as superconducting fluxonium qubits coupled to optical or microwave cavities and outline implications for quantum error correction.
\end{abstract}


\maketitle

\section{Introduction}
High-fidelity gates are essential for quantum computing, but looking towards scalable fault-tolerant computation, it is additionally highly desirable to have non-local quantum gates between two or more qubits. For example, the availability of an all-to-all connectivity can vastly reduce the circuit depth of typical quantum circuits, compared to a geometrically local connectivity \cite{holmes_impact_2020}. 
Furthermore, the ability to perform non-local multi-qubit gates would enable the usage of quantum error correction (QEC) codes with non-local stabilizers, such as LDPC codes \cite{gottesman_fault-tolerant_2014, breuckmann_quantum_2021,cohen_low-overhead_2022,bravyi2023highthreshold}, which have a significantly lower overhead than the currently leading approach of surface codes \cite{fowler_surface_2012}. 
In many physical platforms, however, neither non-local nor multi-qubit gates are natively available, but have to be costly synthesized from a sequence of local single- and two-qubit operations. 

One way to realize non-local two-qubit gates is via qubit shuttling, which has been demonstrated for trapped ions \cite{pino_demonstration_2021} and neutral atoms \cite{beugnon_two-dimensional_2007, bluvstein_quantum_2022}. Evaluating the cost of non-local operations in this case is nontrivial, as the architecture can perform certain parallel moves simultaneously, but unequal moves must be performed serially. The shuttling time overhead for atoms in a planar lattice of linear dimension $L$, relevant to performing operations in certain LDPC codes, is $O(\sqrt{L})$ \cite{xu2023constantoverhead}. 
Alternatively, non-local gates have been previously proposed or realized with neutral atoms or ions by mediating interactions between qubits via a quantized bosonic mode, using motional modes of trapped ions \cite{cirac_quantum_1995, molmer_multiparticle_1999,sackett_experimental_2000, garcia-ripoll_speed_2003} or optical cavity modes for neutral atom spin qubits \cite{pellizzari_decoherence_1995, zheng_efficient_2000, beige_quantum_2000, sorensen_measurement_2003, zheng_unconventional_2004,borregaard_heralded_2015, ramette_any--any_2021, lewalle_multi-qubit_2022}. For deterministic gates, prior art finds the fidelity error is $\mathcal{O}(C^{-1/2})$ where $C$ is the cooperativity of the cavity supporting the mode \cite{sorensen_measurement_2003}. Expending additional detector resources, heralded non-local gates are achievable with error $\mathcal{O}(C^{-1})$ but with a failure probability of $\mathcal{O}(C^{-1/2})$ \cite{borregaard_heralded_2015}. Another scheme using heralded photon transfers has an improved success probability but places stringent requirements on the level structure of the qubits so that all scattering and photon loss events are detectable \cite{ramette_any--any_2021}. In contrast, non-local entangled states can be prepared as fixed points of dissipative maps with an $\mathcal{O}(C^{-1})$  fidelity error \cite{kastoryano_dissipative_2011}, though a fixed phase relation must be maintained between the fields addressing the qubits. 
While some of the proposals above can be extended to $N$-qubit Toffoli gates \cite{borregaard_heralded_2015,lewalle_multi-qubit_2022}, for large scale digital quantum simulations and computing a unified approach that provides native implementations of larger families of multi-qubit gates would be highly desirable. 


In all proposals above, entangling quantum gates are realized by a direct drive of the qubits via a free space mode, e.g. a laser, to turn the interaction between the qubits on or off.
In this work, we explore a different approach based on simply driving the cavity mode directly with a classical field that is modulated time-dependently, without requiring an external drive of the qubits.
%
We find that this approach enables two new protocols for the implementation of large families of deterministic non-local multi-qubit quantum gates. Applied to just two qubits, both protocols provide, together with single qubit gates, a universal gate set for quantum computing, with two-qubit gate errors scaling as $\mathcal{O}(C^{-1/2})$, similar to the  protocols driving the qubits directly. Applied to more than two qubits, however, each protocol provides a family of deterministic, multi-qubit non-local gates requiring minimal control, showing a unique combination of desirable features such as versatility in gate design, speed, and robustness.


The first protocol (A) operates in the limit of a strong drive on the cavity. It implements a family of geometric phase gates $U_A = \exp(i\theta \hat{n}^2)$, where $\hat{n}$ is the number operator of qubits in state $\ket{1}$, by displacing the state of the cavity in a closed loop in phase space. Any angle $\theta$ can be achieved by choosing an appropriate drive $\eta(t)$. 
A particularly important application of protocol A is the generation of multi-qubit GHZ states \cite{hein_multi-party_2004, leibfried_toward_2004}, a task for which viable protocols for qubits coupled via a cavity are rare and require a direct drive on the qubits \cite{zhao_creation_2021}. 
One distinguishing feature of protocol A is its speed: In many previous proposals, the cavity is far detuned from the qubit frequency to avoid a large number of photons in the cavity and thus a large error through photon losses. This comes at the cost of a long gate duration of the order $\Delta/g^2$, where $\Delta$ is the detuning of the cavity and $g$ is the coupling between the qubits and the cavity. In protocol A, the cavity is also far detuned, but a driving strength which is of the order of $\Delta$ and adapted to the photon loss rate allows for gate durations of order $g^{-1}$. An additional advantage of protocol A is its robustness: Similar to the M{\o}lmer-S{\o}rensen gate for trapped ions \cite{molmer_multiparticle_1999}, $U_A$ is independent of the initial state of the cavity mode, which is of particular importance if the cavity mode is in the microwave regime and may exhibit significant thermal population. Furthermore, protocol A is inherently robust against pulse imperfections in the drive of the cavity, since only the area enclosed and not the exact trajectory in phase space determines $U_A$. 

The second protocol (B) operates in the limit of a weak drive and thus in the opposite limit of protocol A. It makes use of an adiabatic evolution of the joint cavity-qubit system to implement a family of phase gates $U_B = \exp[ic_1/(c_2-\hat{n})]$, where $c_1$ and $c_2$ are parameters depending on the intensity, duration and detuning of the applied drive. The distinguishing feature of protocol B is its versatility: Since $U_B$ depends \emph{nonlinearly} on $c_2$, the repeated application of $U_B$ with different values of $c_1$ and $c_2$ can be used to synthesise \emph{arbitrary} phase gates $\exp(i\varphi(\hat{n}))$.
This can for example be used to implement phase-rotation gates $\exp(i\alpha \sigma_z^{(1)} \otimes...\otimes \sigma_z^{(N)})$, which appear in many variational quantum algorithms for fermionic systems \cite{barkoutsos_quantum_2018, grimsley_adaptive_2019}. It can also be used to implement multi-controlled Z gates, enabling generalized Toffoli gates which are frequently used as primitives in QEC to perform majority voting circuits for syndrome extraction and for measurement free QEC \cite{paz-silva_fault_2010,crow_improved_2016,ercan_measurement-free_2018}. Note that synthesizing multi-controlled Z gates using only single- and two-qubit gates either requires circuits of large depths or additional ancilla qubits \cite{groen_efficient_2022}, both of which can be avoided using protocol B.

\begin{figure}
    \centering
    \includegraphics[width=\linewidth]{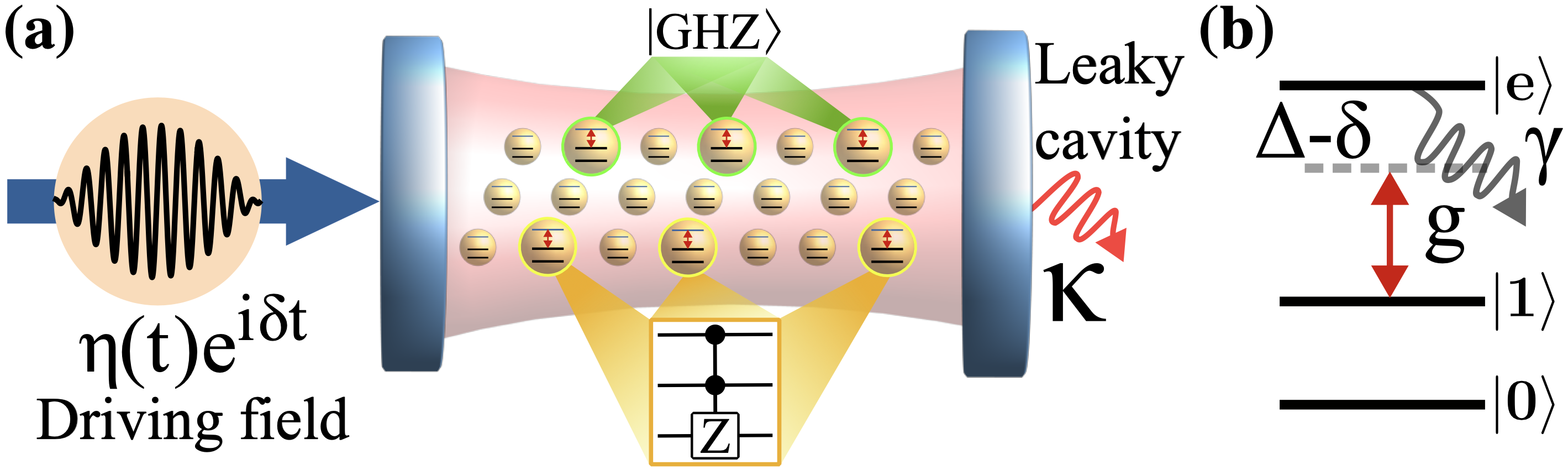}
    \caption{(a) A register of qubits is coupled to a common cavity with decay rate $\kappa$. By simply driving the cavity with a single classical field $\eta(t)$ detuned by $\delta$ from the resonance frequency of the cavity, a non-local entangled state like $\ket{{\rm GHZ}}$ is generated, or, with a sequence of drives, non-local gates like a C$_{2}$Z are implemented. (b) Level scheme for each qubit consisting of the computational basis states $\ket{0}$ and $\ket{1}$ (with infinite lifetime), and an ancillary excited state $\ket{e}$ (with lifetime $1/\gamma$). 
    The $\ket{1} \leftrightarrow \ket{e}$ transition is coupled to the cavity with coupling strength $g$ and detuned from the cavity resonance by $\Delta - \delta$.}
    \label{fig:setup}
\end{figure}


There are several main implications of this work. While there are proposals for $N$-qubit Toffoli gates on qubits coupled via a cavity \cite{borregaard_heralded_2015, lewalle_multi-qubit_2022}, our protocols give the first native implementation for a large family of other multi-qubit gates. In particular, protocol A introduces for the first time a way to implement geometric phase gates for more than two qubits on these systems, while Protocol B even allows for the implementation of native \emph{arbitrary} phase gates without decomposing them into single- and two-qubit gates. This significantly enhances the prospect of realizing non-local stabilizers and quantum error correction schemes such as LDPC codes with reduced qubit overheads compared to current leading schemes, in particular if our protocols are parallelized in architectures that exploit multiple modes (e.g. frequency, polarization, spatial modes for overlapping cavities) as necessary for parallel operations to support QEC. For near term applications, protocol A enhances the toolbox for the generation of large high-fidelity entangled states such as GHZ states, while the arbitrary phase gates implementable by protocol B are of significant interest for quantum simulation. All of these tasks can for the first time be accomplished without the need of an external drive of the qubits. Additionally, both protocols applied to just two qubits form, together with single qubit gates, a universal gate set for quantum computation. These protocols may in principle also be  applied to other leading qubit platforms for quantum computing that exploit delocalized boson modes, such as trapped ions coupled via a motional mode.




The paper is structured as follows: In Sec~\ref{sec:setup} we introduce the Hamiltonian considered in this work. In Secs.~\ref{sec:protocol_A} and~\ref{sec:protocol B} we present protocols A and B, respectively, and derive their fidelities in the presence of the relevant losses. In Sec.~\ref{sec:fidelity_estimates_in_realistic_system} we estimate the achievable infidelities for our protocols for four different systems: atoms coupled via an optical cavity, Rydberg atoms coupled via a microwave cavity, polar molecules coupled via a microwave cavity, and superconducting fluxonium qubits coupled via a microwave cavity. Finally, in Sec.~\ref{sec:application_to_quantum_error_correction} we comment on the usage of both protocols in quantum error correction schemes.

\section{Setup}
\label{sec:setup}

The setup we have in mind is rather general and is shown in Fig. \ref{fig:setup}. It consists of $N$ three-level systems with computational basis states $\ket{0}$ and $\ket{1}$ and an excited state $\ket{e}$, with transition frequencies $\omega_0$  for the $\ket{1} \leftrightarrow \ket{0}$ and $\omega_e$  for the $\ket{1}\leftrightarrow\ket{e}$ transition. A cavity mode with annihilation (creation) operators $a(a^{\dagger})$ and frequency $\omega_c$ couples the states $\ket{1}$ and $\ket{e}$ with coupling strength $g$. We assume that all qubits couple with the identical coupling strength $g$ to the cavity mode. The effects of inhomogeneities in the coupling strength are discussed in Appendix~\ref{app:protA_effects_of_inhomogeneities} and~\ref{app:protB_effects_of_inhomogeneities}.

The cavity mode is driven by a (complex) classical field of strength $\eta(t)$ according to $H_{\rm drive}=2|\eta(t)| \sin(\omega_L t-\arg(\eta(t)))(\ah^{\dagger}+\ah)$.
This classical field is detuned from the cavity and the $\ket{1}\leftrightarrow\ket{e}$ transition by $\delta=\omega_c-\omega_L$ and $\Delta=\omega_e-\omega_L$, respectively.
The Hamiltonian in the rotating wave approximation and in the rotating frame defined by 
$\hat U_r(t)=\exp\left[it(\omega_L (\ah^\dag \ah +\hat{n}_e) +\sum_j  \omega_0 \ket{0_j}\bra{0_j})\right]$ reads ($\hbar = 1$)
\begin{equation}
    H(t) = \delta \ah^\dag \ah + (\Delta-i\gamma/2)\hat{n}_e + [(g\hat S^- +i\eta(t))\ah^\dag + \hc],
    \label{eq:Hamiltonian}
\end{equation}
with $\hat{n}_e = \sum_j \ket{e_j}\bra{e_j}$ the population of state $\ket{e}$ with lifetime $1/\gamma$, and $\hat S^+ = \sum_j \ket{e_j}\bra{1_j}$, $\hat S^- = (\hat S^+)^\dag$ collective raising and lowering operators, respectively.  

The system evolves under the Lindblad master equation $\dot{\rho} = -iH\rho +i\rho H^\dag +L\rho L^\dag - \{L^\dag L, \rho\}/2$ with the jump operator $L=\sqrt{\kappa}\ah$ and $1/\kappa$  the lifetime of excitations in the cavity mode. The decay of $\ket{e}$ is treated as population leakage, described by a non-hermitian term in $H$.  Thus, expressions for the  fidelity of the gate protocols derived below will be exact if none of the decay channels of $\ket{e}$ can repopulate $\ket{0}$ or $\ket{1}$, and otherwise will provide a lower bound. 
For both protocols, a time-dependent pulse $\eta(t)$ of duration $T$ and with $\eta(0)=\eta(T)=0$ is applied, while $g$, $\delta$ and $\Delta$ are kept constant in time.

 To make our gates address only a subset of qubits in a register, we can map the $\ket{1}$ state for qubits that should be spectators to an ancillary state $\ket{a}$ that does not couple to the bosonic mode either due to being far off resonant or by virtue of selection rules which forbid direct coupling to $\ket{e}$. Alternatively, a spatially addressable off-resonant laser beam can be used to apply an ac-Stark shift to the qubits, shifting the $\ket{1}\leftrightarrow\ket{e}$ transition far enough out of resonance with the cavity to neglect the coupling \cite{labuhn_single-atom_2014, wang_single-qubit_2016}. 

\section{Protocol A}
\label{sec:protocol_A}
In this section we discuss the first of our two protocols, Protocol A, which for $N=2$ forms a universal gate set for quantum computation -- together with single qubit gates --, while for arbitrary $N$ it can be used to generate GHZ states. 

Protocol A operates in the limit of a large detuning $\Delta$ between the cavity and the $\ket{1} \leftrightarrow \ket{e}$ transition and of a cavity driving strength $\eta$ of the same order (i.e., $\Delta, \eta \rightarrow \infty$ and  $\Delta = \mathcal{O}(\eta))$. We take $\delta$ to be of order $\mathcal{O}(g)$ and choose the pulse duration $T$ to be of the order of $\mathcal{O}(g^{-1})$, such that it does not diverge in the limit $\Delta, \eta \rightarrow \infty$. In the following, we start by deriving an effective Hamiltonian valid in the limits given above, by first applying a time-dependent basis transformation on the cavity in Sec.~\ref{subsec:prot_A_basis_transformation_on_the_cavity}, followed by a time-dependent basis transformation on the qubits in Sec.~\ref{subsec:protocol_A_elimination_of_e} to eliminate the state $\ket{e}$. The resulting effective Hamiltonian is similar to that of a M{\o}lmer-Sorensen-Gate for trapped ions \cite{molmer_multiparticle_1999} and is then used in Sec.~\ref{subsec:protocol_A_implementation_of_a_quantum_gate} to derive a family of geometric gates $U_A = \exp(i\theta \hat{n}^2)$. The fidelity of the gate for aribtrary $N$ as a function of $\gamma$ and $\kappa$ is calculated analytically in Sec.~\ref{subsec:protocol_A_performance_in_the_presence_of_losses}.  Section~\ref{subsec:protocol_A_numerical_results} verifies the analytical results against numerical simulations of the full Lindblad dynamics, finding excellent agreement.

\subsection{Basis Transformation on the Cavity}
\label{subsec:prot_A_basis_transformation_on_the_cavity}
To motivate the first  basis transformation, acting on the cavity, we note that due to the limit $\eta \rightarrow \infty$ the cavity typically contains many ($\sim |\eta|^2/\delta^2$) photons. However, due to the simultaneous limit $\Delta \rightarrow \infty$ the number of photons only weakly depends on $n$, the number of qubits in state $\ket{1}$. It is thus useful to switch into a time-dependent frame of the cavity which reduces the number of photons. For this, we choose a frame which is given by the evolution that the cavity would undergo if it were not coupled to the qubits. This corresponds to the case $n=0$, where all qubits are in state $\ket{0}$.


Such a frame is given by the simple displacement $D(\alpha) = \exp(\alpha \ah^\dag - \alpha^* \ah)$, where $\alpha(t)$ is the solution of
\begin{equation}
\dot{\alpha} = -\eta-(i\delta+\kappa/2)\alpha, \qquad \alpha(t=0)=0.
\label{eq:alpha_eta_relation}
\end{equation}
If $n=0$, a cavity starting in the empty state $\ket{0}$ will be in the coherent state $\ket{\psi_{\mathrm{cav}}(t)} = \ket{-\alpha(t)}$ at time $t$, so that $D(\alpha)\ket{\psi_{\mathrm{cav}}(t)} = \ket{0}$. Note that the cavity remains in a pure state at all times, even if it undergoes decay.



Given the evolution of $\rho$ for $n=0$, we now treat the evolution of the joint cavity-qubit system for a general $n$. For this, we now proceed with the basis transformation $\tilde{\rho}(t) = D(\alpha(t))\rho(t) D(\alpha(t))^\dag$. For general $n$, the evolution of $\tilde{\rho}$ is then given by (see Appendix~\ref{app:protA_first_basis_trafo_cavity})
\begin{equation}
    \tilde{H} = \delta \ah^\dag \ah + (\Delta-i\gamma/2)\hat{n}_e + g[(\ah^\dag - \alpha^*)\hat S^- +\hc]
    \label{eq:H_tilde}
\end{equation}
and $\tilde{L} = L = \sqrt{\kappa}\ah$. Hence, the drive of the cavity mode is converted into an effective drive of the qubits with strength $-ig\alpha$.  Because the decay in the original frame is compensated by a $\kappa$-dependent choice of $\alpha$,  in this new frame there are no excitations in the cavity mode and no decay events if $n=0$ -- even if in the original frame there may be many excitations and decay events. 

\subsection{Basis Transformation on the Qubits and Derivation of the Effective Hamiltonian}
\label{subsec:protocol_A_elimination_of_e}
In order to derive an effective Hamiltonian on the computational states $\ket{0}, \ket{1}$, and the cavity and to eliminate the state $\ket{e}$, we now use the limit $\Delta\rightarrow\infty$. For this, we consider $\tilde{H}^{(0)} = \Delta\hat{n}_e - (g\alpha^* S^- + \hc)$, which is the part of $\tilde{H}$ which scales with $\Delta$. (Recall that as $\Delta \rightarrow \infty$ we also consider the limit $\eta \rightarrow \infty$, and thus $|\alpha| \rightarrow \infty$). We perform a time-dependent basis transformation on the qubits so that the new basis states are the instantaneous eigenvectors of $\tilde{H}^{(0)}$. Such a basis transformation is given by (see Appendix~\ref{app:protA_second_basis_trafo_qubits})
\begin{equation}
    \hat U = \exp\left[\frac{\lambda}{2}\left(-e^{i\mu}\hat S^++e^{-i\mu}\hat S^-\right)\right]
\end{equation}
with  $\cos\lambda = \Delta/\sqrt{4g^2|\alpha|^2+\Delta^2}$ and $\mu = \arg(\alpha)$. In this new basis, the Hamiltonian is given by $\bar{H} = \hat U\tilde{H}\hat U^\dag + i\dot{\hat U}\hat U^\dag$. The inertial term is of the form $i\dot{\hat U}\hat U^\dag = \mathcal{O}(1)S^+ + \hc$, where $\mathcal{O}(1)$ denotes a term which does not diverge as $\Delta \rightarrow \infty$. Crucially, since the gap between the eigenstates of $H^{(0)}$ diverges as $\Delta \rightarrow \infty$ and we consider a pulse duration $T$ independent of $\Delta$, the inertial term can be neglected as $\Delta \rightarrow \infty$, leaving $\bar{H} = \hat U\tilde{H}\hat U^\dag$.

A direct calculation (see Appendix~\ref{app:protA_second_basis_trafo_qubits}) now shows
\begin{eqnarray}
 \bar{H} &=& \delta \ah^\dag \ah + \left(\varepsilon_1-i\frac{\gamma_1}{2}\right) \hat{n} +  \left(\varepsilon_e -i\frac{\gamma_e}{2}\right) \hat{n}_e \label{eq:H_bar}\\ \nonumber
    &+& (\zeta \ah^\dag + \zeta^* \ah)(\hat n-\hat n_e) 
\end{eqnarray}
where 
\begin{eqnarray}\label{eq:parameters1}
    \varepsilon_{e/1} &=& (\Delta\pm \sqrt{\Delta^2+4g^2|\alpha|^2})/2 + \mathcal{O}(1) \\\label{eq:parameters2}
    \zeta &=& \frac{g^2\alpha}{\sqrt{4g^2|\alpha|^2 + \Delta^2}} \label{eq:zeta} \\ 
     \gamma_{e/1} &=& \frac{\gamma}{2}\left(1\pm\sqrt{1-4|\zeta|^2/g^2}\right)\label{eq:parameters3}
\end{eqnarray}
where the expressions above are evaluated with the $+$ sign for $\varepsilon_e$ and $\gamma_e$ and the $-$ sign for $\varepsilon_1$ and $\gamma_1$.  We note that Eqs.~\eqref{eq:parameters1}--\eqref{eq:parameters3} are time-dependent. In Sec.~\ref{subsec:protocol_A_implementation_of_a_quantum_gate} below we use the time-dependency of $\zeta$ to implement the desired quantum gate. 

If we assume that none of the qubits start in state $\ket{e}$, the terms in Eq.~\eqref{eq:H_bar} proportional to $\hat{n}_e$ can be neglected. Furthermore, the $\varepsilon_1\hat{n}$ term just corresponds to a frequency shift of the qubits, which can be compensated for either by single qubit $z$-rotation at the end of the gate, or by a change of reference frame. We are thus left with the effective Hamiltonian
\begin{equation}
    H_\eff = \delta \ah^\dag \ah +  \left(-i\frac{\gamma_1}{2} + \zeta \ah^\dag + \zeta^* \ah\right)\hat{n}.
    \label{eq:H_eff}
\end{equation}
This effective Hamiltonian simply describes a driven cavity, where the driving strength $\zeta\hat{n}$ depends on the number $n$ of qubits in state $\ket{1}$. It is thus analogous to the Hamiltonian for a M{\o}lmer-S{\o}rensen gate \cite{molmer_multiparticle_1999}.

The finite lifetime $\gamma$ of the state $\ket{e}$ leads to an effective error rate $\gamma_1\hat{n}$. Note that since the basis transformation in this section only affected the space of the qubits,  the Lindblad operator $L_\eff = L = \sqrt{\kappa}\ah$ is unchanged. We discuss the influence of these error sources in Sec.~\ref{subsec:protocol_A_performance_in_the_presence_of_losses} below.

\subsection{Implementation of a Quantum Gate}
\label{subsec:protocol_A_implementation_of_a_quantum_gate}
In this section we use the effective Hamiltonian~\eqref{eq:H_eff} to derive a shape of $\zeta(t)$ which implements a quantum gate $U_A = \exp(i\theta \hat{n}^2)$ on the qubits only and leaves the system in a state with no entanglement between the photons and the qubits. We first consider the loss free case $\gamma=\kappa=0$, while the infidelity for finite values of $\gamma$ and $\kappa$ is calculated in the next section. 

Below in Sec.~\ref{subsub:derivation_geometric} we show that if $\zeta(t)$ is chosen of the form $\zeta(t) = -\delta f(t) + i\dot{f}(t)$, the effective Hamiltonian~\eqref{eq:H_eff} implements the quantum gate $U_A = \exp(i\theta \hat{n}^2)$ with
\begin{equation}
    \theta=\delta \int_0^T \mathrm{d}t f(t)^2.
    \label{eq:protocol_A_theta_final}
\end{equation}
We note that here $f$ can be any \emph{real} function satisfying $f(0)=f(T)=0$, $\dot{f}(0)=\dot{f}(T)=0$, and $\delta^2 f(t)^2 + \dot{f}(t)^2 < g^2/4$ for all $t$. These constraints follow from the two points below: 

(i) To find a pulse $\eta(t)$ in the original Hamiltonian~\eqref{eq:Hamiltonian} which leads to the desired $\zeta(t)$ in the effective Hamiltonian~\eqref{eq:H_eff}, Eqs.~\eqref{eq:zeta} and~\eqref{eq:alpha_eta_relation} have to be inverted to first find $\alpha(t)$ and then $\eta(t)$. Equation~\eqref{eq:zeta} is only invertible if $|\zeta(t)| < g/2$, which imposes the constraint $\delta^2 f(t)^2 + \dot{f}(t)^2 < g^2/4$ on the choice of $f$, while Eq.~\eqref{eq:alpha_eta_relation} can be solved for $\eta(t)$ for any differentiable $\alpha(t)$. 

(ii) However, we require $\alpha(0)=\alpha(T)=0$, so that the new frame introduced in Sec.~\ref{subsec:protocol_A_elimination_of_e} coincides with the lab frame at $t=0$ and $t=T$. This is guaranteed by $f(0)=f(T)=0$ and $\dot{f}(0)=\dot{f}(T)=0$.


\subsubsection{Derivation of the geometric phase for $\gamma=\kappa=0$}\label{subsub:derivation_geometric}
Now we show that the choice $\zeta(t) = -\delta f(t) + i\dot{f}(t)$ indeed leads to the implementation of $U_A$ with the phase $\theta$ given by Eq.~\eqref{eq:protocol_A_theta_final}. While this derivation is analogous to that of a M{\o}lmer-Sorensen gate \cite{molmer_multiparticle_1999}, we rederive it here in a way which allows for the easy addition of the effects of finite $\gamma$ and $\kappa$ in the next section.

We first assume that the qubits start in a computational basis state $\ket{q}$ ($q \in \{0,1\}^N$) with exactly $n = \sum_{j=1}^N q_j$ qubits in state $\ket{1}$ (i.e. $\hat{n}\ket{q} = n \ket{q}$). Additionally, we assume that the cavity starts in a coherent state $\ket{\beta(0)}$. Since any initial state of the joint cavity-qubit system can be written as a superposition of states of the form $\ket{\psi(0)} = \ket{\beta(0)} \otimes \ket{q}$, those states suffice to uniquely determine the dynamics of the system under $H_\eff$ for any initial state.

We now make the Ansatz $\ket{\psi(t)} = e^{i\varphi_n(t)}\ket{\beta_n(t)}\otimes\ket{q}$, which indeed satisfies the Schrödinger equation for $H_\eff$ if
\begin{eqnarray}
    \dot{\beta}_n &=& -i\delta\beta_n -in\zeta \label{eq:beta_no_decay}\\ 
    \dot{\varphi}_n &=& -n\Re(\zeta^*\beta_n) \label{eq:theta_no_decay}
\end{eqnarray}
Making the choice $\zeta(t) = -\delta f(t) + i\dot{f}(t)$, the solution to Eq.~\eqref{eq:beta_no_decay} is given by
\begin{equation}
    \beta_n(t) = \beta(0)e^{-i\delta t} + n f(t)
\end{equation}
Plugging this into Eq.~\eqref{eq:theta_no_decay} yields
\begin{eqnarray}
    \varphi_n(T) &=& -n \Re\left[\int_0^T \mathrm{d}t (-\delta f(t) -i\dot{f}(t))\left(\beta(0)e^{-i\delta t} + n f(t)\right)  \right] \nonumber \\
    &=& -n \Re\left[\int_0^T \mathrm{d}t \left(-n\delta f(t)^2 + \dot{h}(t) \right)\right],
\end{eqnarray}
where 
\begin{equation}
    h(t) = -i\left(\beta(0)f(t)e^{-i\delta t} + \frac{1}{2}nf(t)^2\right).
\end{equation}
Using that $h(0)=h(T) = 0$, we obtain $\varphi_n(T) = n^2\theta$.

Thus, the final state at time $t=T$ is $\ket{\psi(T)} = e^{i\theta n^2} \ket{\beta(0)e^{-i\delta T}}\otimes \ket{q} =  \ket{\beta(0)e^{-i\delta T}} \otimes (U_A\ket{q})$. Since the final state of the cavity is independent of $n$, there is no entanglement between the qubits and the cavity at time $T$. Furthermore, since $U_A$ is \emph{independent} of $\beta(0)$, and any arbitrary initial state of the cavity can be written as a superposition of different coherent states $\ket{\beta(0)}$, the implemented unitary is in fact independent of the initial state of the cavity.

\subsubsection{Generation of GHZ states}
The unitary $U_A$ can be used together with global single qubit gates to generate the GHZ state \cite{hein_multi-party_2004, leibfried_toward_2004}
\begin{equation}
    \ket{\mathrm{GHZ}} = (\ket{0...0}+\ket{1...1})/\sqrt{2} 
\end{equation}
on $N$ qubits as follows: Start by preparing the system in $\ket{+}^{\otimes N}$, where $\ket{+}=(\ket{0}+\ket{1})/\sqrt{2}$. Then apply $U_A$ for $\theta = \pi/2$, followed by the single qubit gate $U_{sq} = U_3U_2U_1$ on each qubit, where $U_1 = \exp(i\pi\sigma_z/4)$, $U_2 = (\sigma_x+\sigma_z)/\sqrt{2}$ and $U_3 = \exp(i\pi(N+1)\sigma_z/(4N))$. For convenience, we restate the proof of this known result in Appendix~\ref{app:protA_generation_of_GHZ_states}. 


\subsection{Performance in the presence of losses}
\label{subsec:protocol_A_performance_in_the_presence_of_losses}
In this section we calculate the gate fidelity for the implementation of $U_A$ in the presence of losses. In contrast to the previous section we restrict ourselves to an initial state $\ket{0}$ of the cavity. We start by solving the Lindblad equation with the effective Hamiltonian $H_\eff$ and the jump operator $L$. This allows us to find the quantum channel $\mathcal{E}$ acting on the qubits which is obtained if the cavity is traced out after the gate. Given $\mathcal{E}$, we then find an expression for the infidelity. In the limit $\gamma, \kappa \rightarrow 0$ and $T\rightarrow\infty$, the infidelity is found analytically to be
\begin{equation}
    1-F = \left(\frac{\kappa}{4(1+2^{-N})\delta} + \frac{\gamma\delta}{2g^2}\right)N\theta.
    \label{eq:protA_infidelity_final}
\end{equation}
To our knowledge this is the first analytical solution of $1-F$ for Hamiltonians of the type of Eq.~\eqref{eq:H_eff} in the presence of the relevant losses.

\subsubsection{Solution of the Lindblad Equation}
To solve the Lindblad equation for $H_\eff$ of Eq.~\eqref{eq:H_eff} and $L$, we proceed analogously to Sec.~\ref{subsec:protocol_A_implementation_of_a_quantum_gate} by first providing an Ansatz for the density matrix of the joint cavity-qubit system and then verifying that this Ansatz provides the correct solution of the time-dependent Lindblad equation.

To determine $\mathcal{E}$, it is sufficient to consider initial operator of the form $\rho(0) = \ket{0}\bra{0} \otimes \ket{q}\bra{q'}$ of the joint cavity-qubit system, where $\ket{q}$ and $\ket{q'}$ ($q,q' \in \{0,1\}^N$) are computational basis states with exactly $n= \sum_j q_j$ and $m=\sum_j q_j'$ qubits in state $\ket{1}$.

We now make the Ansatz
\begin{equation}
    \rho(t) = e^{i\varphi_{nm}} \ket{\beta_n}\bra{\beta_m} \otimes \ket{q   }\bra{q'}/\braket{\beta_n}{\beta_m}. 
\end{equation}

In Appendix~\ref{app:protA_analytic_solution_of_time_evolution} we show that this Ansatz solves the Lindblad equation if
\begin{eqnarray}
    \dot{\beta}_n &=& -(i\delta + \kappa/2)\beta_n -in\zeta \label{eq:beta_decay} \\
    \dot{\varphi}_{nm} &=& (m-n)(\zeta\beta_m+\zeta^*\beta_n) +i(m+n)\gamma_1/2 \label{eq:theta_decay}
\end{eqnarray}
The quantum operation on the Hilbert space of the qubits is given by 
\begin{equation}
    \mathcal{E}(\ket{q}\bra{q'}) = \tr_{\mathrm{cav}}(\rho(T)) = e^{i\varphi_{nm}(T)}\ket{q}\bra{q'}.
    \label{eq:prot_A_quantum_Channel}
\end{equation}
This latter expression for $\mathcal{E}$ is used in the next subsection to determine the fidelity $F$.

\subsubsection{Analytic calculation of the fidelity for $\gamma$, $\kappa\neq 0$}

With Eq.~\eqref{eq:prot_A_quantum_Channel}, the averaged gate fidelity can be computed as
\begin{eqnarray}
    \label{eq:protocol_A_fidelity}
    F &=& \int \mathrm{d}\psi \bra{\psi}e^{-i\theta \hat{n}^2 } \mathcal{E}(\ket{\psi}\bra{\psi})e^{i\theta\hat{n}^2 }\ket{\psi}\\ \nonumber &=&\frac{\sum_{n=0}^N \binom{N}{n} e^{i\varphi_{nn}} 
    + \sum_{n,m=0}^N \binom{N}{n} \binom{N}{m}e^{i\varphi_{nm}-i(n^2-m^2)\theta} }{2^N(2^N+1)},
\end{eqnarray}
where the integral is taken over the whole computational subspace, and the second expression follows from~\cite{pedersen_fidelity_2007}.

Equation \eqref{eq:protocol_A_fidelity} now allows us to calculate the gate fidelity for arbitrary values of $\delta$, $T$, $\gamma$ and $\kappa$ by inserting the solutions of Eqs.~\eqref{eq:beta_decay} and~\eqref{eq:theta_decay}, given by
\begin{equation}
    \beta_n(t) = -in \int_0^t \mathrm{d}t' \zeta(t') e^{-(i\delta+\kappa/2)(t-t')} 
\end{equation}
and
\begin{eqnarray}
    \varphi_{nm}(T) = \int_0^T  &\Big[& (m-n)(\zeta(t) \beta_m(t)^* + \zeta(t)^*\beta_n(t)) \nonumber \\ 
     &+& i(m+n)\gamma_-(t)/2\Big] \mathrm{d}t
     \label{eq:protA_phinm},
\end{eqnarray}
respectively.
In the limit $\gamma,\kappa \rightarrow 0$ and $T\rightarrow\infty$ this can be evaluated to Eq.~\eqref{eq:protA_infidelity_final}. From Eq.~\eqref{eq:protA_infidelity_final} we observe that $\delta$ can be used to trade between the infidelity arising from the decay of photons in the cavity (proportional to $\kappa$) and decay of the ancillary state $\ket{e}$ (proportional to $\gamma$).
The infidelity is minimized for $\delta = \sqrt{\kappa/[2(1+2^{-N})\gamma]}g$, which gives 
\begin{equation}
    1-F = N\theta/\sqrt{2(1+2^{-N})C},
    \label{eq:prot_A_infidelity_optimal_delta}
\end{equation}
where $C = g^2/(\gamma\kappa)$ denotes the cooperativity. 

\subsection{CZ gate: Numerical Results}
\label{subsec:protocol_A_numerical_results}
\begin{figure}
    \centering
    \includegraphics[width = \linewidth]{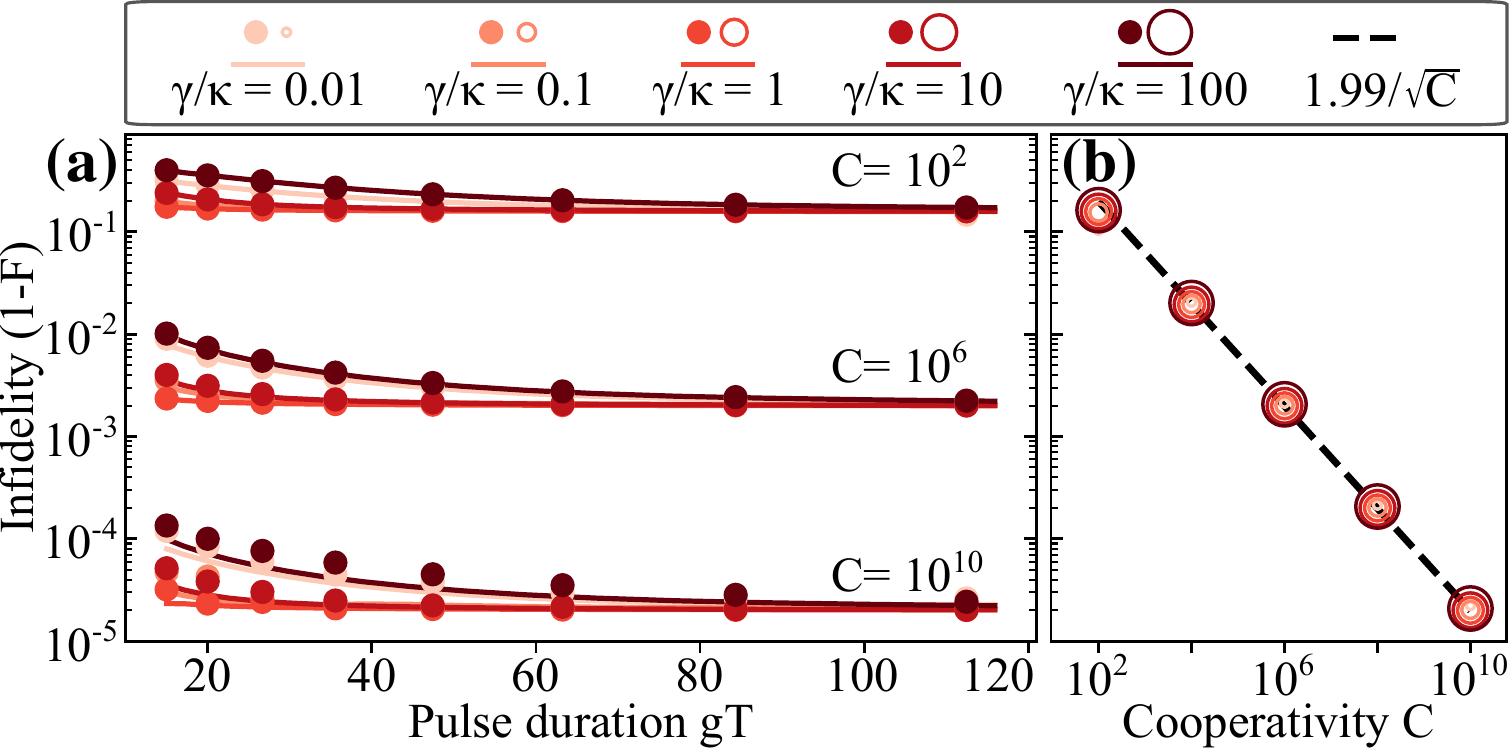}
     \caption{Protocol A: (a) Infidelity of a CZ gate vs pulse duration $T$ for different values of $C$ and $\gamma/\kappa$. Solid lines show the infidelity (analytic result) in the $\Delta \rightarrow \infty$ limit, circles show the infidelity (numerical calculation) at a finite value of $\Delta$, chosen such that $\max_t |\eta(t)|=30g$. For each $T$, $C$ and $\gamma/\kappa$, $\delta$ is optimized to obtain the minimal 1-$F$. (b) Numerical (circles) and analytical (dashed line) value of the infidelity vs $C$ in the $\Delta, T\rightarrow \infty$ limit for different values of $\gamma/\kappa$.}
    \label{fig:ProtocolA_main_plot}
\end{figure}
In the following, we confirm our analysis above and find the infidelity of the $U_A$ gate away from the limit $\Delta,\eta \rightarrow\infty$ via a numerical simulation of the full Lindblad equation for the specific case of the CZ gate ($N=2$). The latter is implemented, up to single qubit gates, for $\theta=\pi/2$. To achieve this, we choose $f(t) = \sqrt{4\pi/(3\delta T)}\sin^2(\pi t/T)$, which satisfies the requirement $\delta \int_0^T f(t)^2\mathrm{d}t = \pi/2$ (see Eq.~\eqref{eq:protocol_A_theta_final}). We numerically verify that there is a $\delta$ with $|\zeta(t)| < g/2$ for all  $t$ as long as $Tg \geq 8.3$.

For the chosen $f$, the infidelity  $1-F$ is shown in Fig.~\ref{fig:ProtocolA_main_plot}(a) as a function of the pulse duration $T$ for several values of the cooperativity $C$ and ratios $\gamma/\kappa$. The solid lines show the infidelity in the limit $\Delta\rightarrow\infty$ calculated analytically using Eqs.~\eqref{eq:protocol_A_fidelity}-\eqref{eq:protA_phinm}. The choice of $\delta$ has been optimized to achieve the best fidelity at each value of the pulse duration $T$. As $T\rightarrow\infty$ the infidelity approaches its asymptotic value,  which is as predicted by Eq.~\eqref{eq:protA_infidelity_final} independent of $\gamma/\kappa$ and only depends on the cooperativity $C$. For shorter pulse durations, there is a slight dependency on $\gamma/\kappa$, with the best infidelity always being achieved at $\gamma/\kappa\sim 1$. Note that the asymptotic value of the infidelity as $T\rightarrow\infty$ is often already closely approached for durations $T \sim 20g^{-1}$, underlining the fast speed of protocol A. 

The dots in Fig.~\ref{fig:ProtocolA_main_plot}(a) show the infidelity which is achieved at a finite value of $\Delta$, chosen such that $\max_t \eta(t)=30g$. These values were found through a numerical integration of the Lindblad equation given by $H$ [Eq.~\eqref{eq:Hamiltonian}] and the jump operator $L$. Only small deviations between the numerical and the analytical results can be observed, showing that a maximum driving strength of $30g$ is sufficient to implement protocol A with high fidelity.

Finally, Fig.~\ref{fig:ProtocolA_main_plot}(b) compares the asymptotic value of the infidelity from Fig.~\ref{fig:ProtocolA_main_plot}(a) with its analytical prediction $1-F = 1.99/\sqrt{C}$ from Eq.~\eqref{eq:prot_A_infidelity_optimal_delta}. A good agreement is observed for all values of $C$ and $\gamma/\kappa$.

 This concludes our discussion of protocol A. We have demonstrated that by driving the strongly detuned cavity with a strong drive $\eta$, the unitary $U_A = \exp(i\theta \hat{n}^2)$ can be implemented through the proper choice of $\eta(t)$. We derived the infidelity of protocol A and showed that it agrees with numerical simulations. Finally, we demonstrated how protocol A can be used together with single qubit gates to generate a GHZ state on $N$ qubits.
 

\section{Protocol B}
\label{sec:protocol B}
In contrast to protocol A, protocol B is an adiabatic protocol that operates in the limit $\eta \rightarrow 0$, with detunings $\Delta, \delta = \mathcal{O}(g)$, and a pulse duration $T = \mathcal{O}(\eta^{-2})$. In Sec.~\ref{subsec:protocol_B_loss_free_case} we discuss protocol B in the absence of losses, followed by the calculation of the infidelity for finite values of $\gamma$ and $\kappa$ in Sec.~\ref{subsec:protocol_B_performance_with_losses}. We confirm our analysis through a numerical simulation in Sec.~\ref{subsec:prot_B_CZ_numerical_results}. In Sec.~\ref{subsec:protocol_B_implementation_of_arbitrary_phase_gates} we discuss how several repetitions of protocol B can be used to implement \emph{arbitrary} phase gates. 

\subsection{Loss free case}
\label{subsec:protocol_B_loss_free_case}
We start by assuming $\gamma=\kappa=0$. We consider an initial state $\ket{\psi(0)} = \ket{0}\otimes\ket{q}$, with the cavity starting in state $\ket{0}$ and the qubits in a computational basis state $\ket{q}$ ($q \in \{0,1\}^N)$, with exactly $n=\sum_j q_j$ qubits in state $\ket{1}$. Note that $\ket{\psi(0)}$ is an eigenstate of the Hamiltonian $H$ [Eq.~\eqref{eq:Hamiltonian}] for $\eta=0$. If now $\eta$ is varied slowly enough, the system will stay in an eigenstate of $H$ and accumulate a dynamical phase. Since at the final time we have again $\eta(T)=0$, we obtain $\ket{\psi(T)} = e^{i\varphi_n}\ket{0}\otimes\ket{q}$, where the dynamical phase is given by 
\begin{equation}
    \varphi_n = -\int_0^T \bra{\psi_n(t)}H(t)\ket{\psi_n(t)} \mathrm{d}t. 
\end{equation}
Using second order perturbation theory, one obtains (see Appendix~\ref{app:protB_eigenenergies_in_perturbation_theory})
\begin{equation}
    \varphi_n = -\frac{I}{\delta -  n g^2/\Delta},
    \label{eq:protB_phin}
\end{equation}
where $I = \int_0^T |\eta(t)|^2\mathrm{d}t$ is the pulse energy. Thus, the pulse implements a unitary $U_B = \exp[-iI/(\delta - \hat n g^2/\Delta)]$. 

\subsection{Performance in the presence of losses}
\label{subsec:protocol_B_performance_with_losses}
For $\gamma, \kappa \neq 0$ the quantum operation on the space of the qubits can be approximated by (see Appendix~\ref{app:protB_effect_of_losses})
\begin{equation}
    \mathcal{E}(\ket{q}\bra{q'}) = c_{nm} e^{i(\varphi_n-\varphi_m)} \ket{q}\bra{q'}.
\end{equation}
Again, $\ket{q}$ and $\ket{q'}$ are computational basis states of the qubits with exactly $n=\sum_j q_j$ and $m=\sum_j q_j'$ qubits in state $\ket{1}$, respectively.
The coefficients $c_{nm}$ are given by
\begin{equation}
    c_{nm} = 1- [(\gamma_n+\gamma_m+(s_n-s_m)^2], 
\end{equation}
with
\begin{eqnarray}
    \gamma_n &=& \frac{\gamma n g^2}{(\Delta \delta -ng^2)^2}I = - \frac{\gamma}{\Delta} \frac{ng^2}{\Delta\delta -ng^2}\varphi_n \label{eq:gamma_n} \\
    s_n &=& \frac{\sqrt{\kappa} \Delta}{\Delta \delta -ng^2}\sqrt{I} =\pm \frac{\sqrt{\kappa\Delta}}{\sqrt{|\Delta\delta -ng^2|}}\sqrt{|\varphi_n|} \label{eq:s_n}.
\end{eqnarray}
where in the last equality the sign is $+$ if $\Delta/(\Delta\delta -ng^2)>0$ and $-$ otherwise.

The fidelity can be calculated analogously to Eq.~\eqref{eq:protocol_A_fidelity} as
\begin{equation}
    F = \frac{\sum_{n=0}^N \binom{N}{n} c_{nn} 
    + \sum_{n,m=0}^N \binom{N}{n} \binom{N}{m}c_{nm} }{2^N(2^N+1)}
    \label{eq:fidelity_protocol_B}
\end{equation}

To implement a CZ gate ($N=2$), up to single qubit gates, $I$ has to be chosen such that $|\varphi_2-2\varphi_1+\varphi_0| = \pi$. Given this choice, the values of $\delta$ and $\Delta$ that maximize $F$ can be found numerically as $\delta = 0.529\sqrt{\kappa/\gamma}g$, $\Delta = -2.09\sqrt{\gamma/\kappa}g$, which gives $1-F = 1.79/\sqrt{C}$.

The scaling of the optimal $\delta$ and $\Delta$ with $\gamma$ and $\kappa$ can be explained as follows: Inserting the second expressions from Eq.~\eqref{eq:gamma_n} and~\eqref{eq:s_n} into Eq.~\eqref{eq:fidelity_protocol_B} shows that for any given phases $\varphi_0,...,\varphi_N$, the infidelity is of the form $1-F = \gamma h_1(\delta\Delta)/|\Delta| + \kappa h_2(\delta\Delta)|\Delta|$, where $h_1$ and $h_2$ are positive functions independent of $\gamma$ and $\kappa$ which only depend on $\delta$ and $\Delta$ through their product $\delta\Delta$.  At a fixed value of $\delta\Delta$, the optimal choice of $\Delta$ is thus $|\Delta| = \sqrt{\gamma/\kappa}\sqrt{h_1(\delta\Delta)/h_2(\delta\Delta)}$, and the infidelity is $1-F = \sqrt{2\gamma\kappa h_1(\delta\Delta)h_2(\delta\Delta)}$. Since $h_1$ and $h_2$ are independent of $\gamma$ and $\kappa$, the optimal value of the product $\delta\Delta$ is also independent of $\gamma$ and $\kappa$. Since $\Delta \propto \sqrt{\gamma/\kappa}$ it follows $\delta \propto \sqrt{\kappa/\gamma}$.

\subsection{CZ gate: Numerical Results}
\label{subsec:prot_B_CZ_numerical_results}
\begin{figure}
    \centering
    \includegraphics[width=\linewidth]{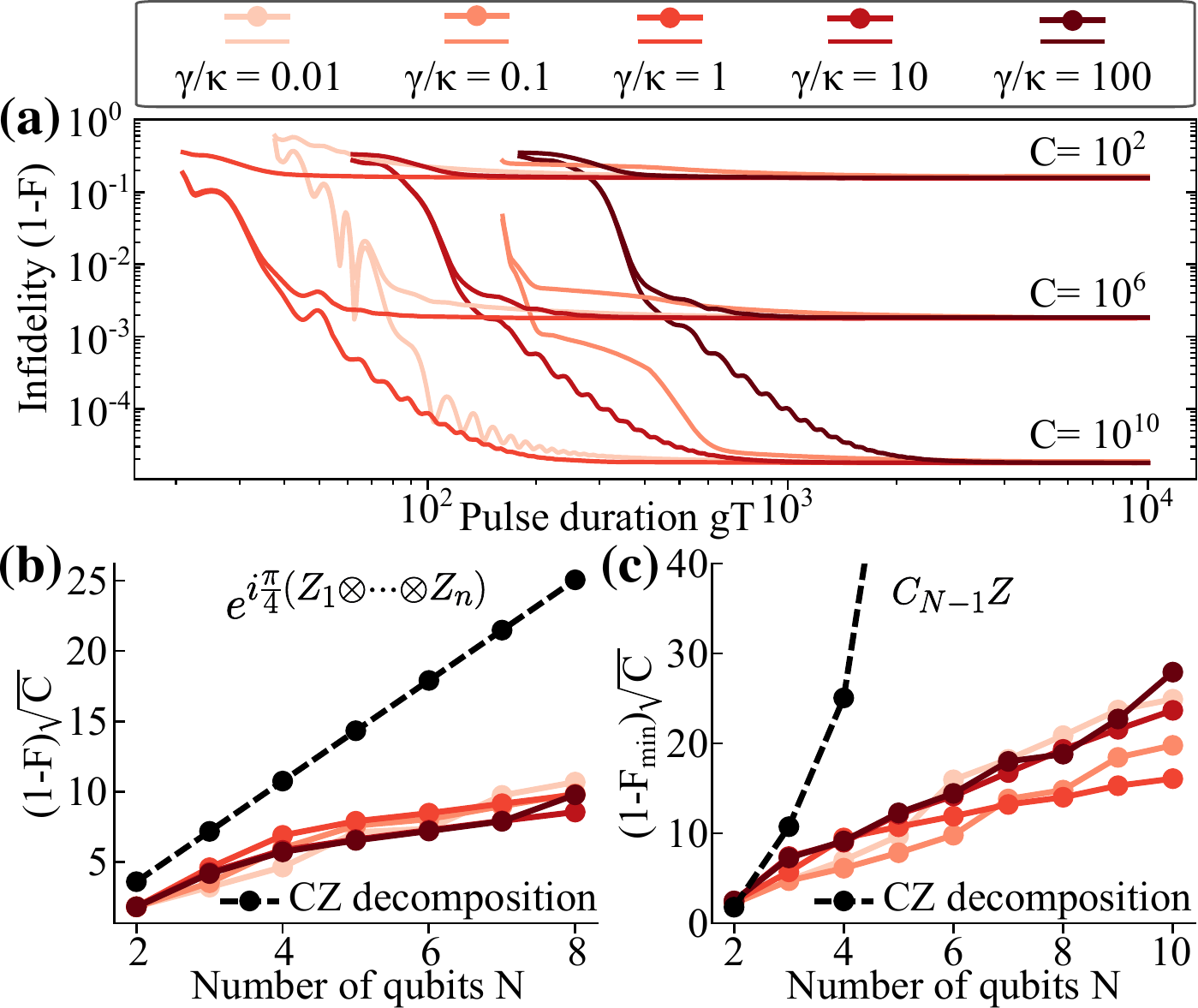}
    \caption{Protocol B: (a) Infidelity (numerical calculation) for a CZ gate as a function of pulse duration $T$ for different values of $C$ and $\gamma/\kappa$. (b) Infidelity of a phase-rotation gate with $\alpha = \pi/4$ in the $T\rightarrow \infty$ limit as a function of $N$. Also shown is the infidelity of the same gate implemented using a decomposition into CZ and single-qubit gates using the circuit from Ref.~\cite{barkoutsos_quantum_2018}. (c) Infidelity of a C$_{N-1}$Z gate vs $N$ when implemented using protocol B and by decomposition into CZ and single qubit gates using an ancilla-free 
    Gray-code \cite{groen_efficient_2022}.}
    \label{fig:ProtocolB_main_plot}
\end{figure}

To confirm our formula for the infidelity and to determine the infididelity for finite pulse durations $T$, we numerically solve the Lindblad equations for different pulse durations $T$ and different values of $\gamma$ and $\kappa$. 
To achieve adiabaticity, $\eta(t)$ is chosen as a flat-top pulse,  rising to its maximium value $\emax$ with a  $\sin^2$-shaped flank of duration $T_0 \leq T/2$, staying at $\emax$ for a duration $T-2T_0$, and then falling back to 0 in a $\sin^2$-shaped flank. $T_0$ and $\emax$ are numerically chosen to satisfy $|\varphi_2-2\varphi_1+\varphi_0| = \pi$ with the minimal possible slope $\max_t|\dot{\eta}(t)|$.

Figure~\ref{fig:ProtocolB_main_plot}(a) shows the infidelity as a function of $T$ for a CZ gate using protocol B for different values of $C$ and $\gamma/\kappa$. 
We find that $1-F$ approaches its asymptotic value  $1.79/\sqrt{C}$ for pulse durations $10^2g^{-1} \lesssim T \lesssim 10^{3}g^{-1}$, while for smaller $T$ it increases due to diabatic errors. The different behavior for different $\gamma/\kappa$ ratios arises due to a nontrivial behavior of $|\varphi_2-2\varphi_1+\varphi_0|$ away from the perturbative approximation made above. 

\subsection{Implementation of arbitrary phase gates}
\label{subsec:protocol_B_implementation_of_arbitrary_phase_gates}

\subsubsection{Derivation}
In the following we show how $N-1$ repetitions of protocol B can be used to implement an \emph{arbitrary} symmetric phase gate $\exp(i\varphi(\hat{n}))$ for \emph{any} function $\varphi(\hat{n})$, up to single qubit gates and a global phase. 

To see this, let us consider applying protocol B $N-1$ times, with different detunings $\delta_1,...,\delta_{N-1}$ and $\Delta_1,...,\Delta_{N-1}$ and different pulse energies $I_1,...,I_{N-1}$ in each pulse. We require the that $\Delta_k-\delta_k$ is independent of $k$, so that the different pulses can be implemented by only changing the amplitude, duration and detuning of the external drive of the cavity, while the detuning between the cavity frequency and the $\ket{1}\leftrightarrow\ket{e}$ transition stays constant. Each of these pulses now implements a phase gate $\exp(i\varphi_k(\hat{n}))$ with $\varphi_k(\hat{n})$ given through Eq.~\eqref{eq:protB_phin}. Taking all pulses together and adding a global phase $\theta_g$ and a single qubit phase $\theta_s$, the implemented phase gate is
\begin{equation}
    \varphi(\hat{n}) = \theta_g + \hat{n}\theta_s-\sum_{k=1}^{N-1} \frac{I_k}{\delta_k-\hat{n}g^2\Delta_k}
    \label{eq:protoB_linear_dependency}
\end{equation}

Observe that the $\varphi(\hat{n})$ depend \emph{linearly} on the $N+1$ variables $\theta_g$, $\theta_s$ and $I_1,...,I_{N-1}$. Thus, since there are $N+1$ possible values of $n$ (from 0 to $N$), Eq.~\eqref{eq:protoB_linear_dependency} has a unique solution of the $\theta_g$, $\theta_s$ and $I_1,...,I_{N-1}$ as a function of $\varphi(\cdot)$ and the $\delta_k$ and $\Delta_k$. Hence there are pulse energies $I_1,...,I_{N-1}$ to implement $\exp(i\varphi(\hat{n}))$ up to single qubit gates and a global phase. Note that such $I_1,...,I_{N-1}$ can be found for \emph{any} choice of the $\delta_k$ and $\Delta_k$. In Appendix~\ref{app:protB_using_protB_for_arbitrary_phase_gates} we give a method based on linear programming to find the  $\delta_k$ and $\Delta_k$ which minimize the gate infidelity.


\subsubsection{Arbitrary Multi-Qubit Phase Gates: Numerical Results}
We exemplify the procedure described above for two classes of multi-qubit gates: Phase-rotation gates $\exp(-i \alpha Z_1 \otimes ... \otimes Z_n)$  -- corresponding to  phases $\varphi_n = -\alpha (-1)^{n}$ -- and $N$-qubit multi-controlled-Z gates (C$_{N-1}$Z gates), i.e. phase gates with $\varphi_N=\pi$ and $\varphi_n=0$ for $n < N$. 
The infidelity for both multi-qubit gates as a function  of $N$ is shown in Fig.~\ref{fig:ProtocolB_main_plot}(b,c) for different values of $\gamma/\kappa$. Here, we take $\delta_k-\Delta_k = 2.09 g/\sqrt{\kappa/\gamma}+0.529 g\sqrt{\kappa/\gamma}$ (the optimal choice for $N=2$), and choose the $\delta_1,...,\delta_{N-1}$ to maximize the fidelity (See Appendix~\ref{app:protB_using_protB_for_arbitrary_phase_gates}). Note that for C$_{N-1}$Z gates we  consider the minimal fidelity $F_{\min} = \min_{\ket{\psi}} \bra{\psi}C_{N-1}Z \mathcal{E}(\ket{\psi}\bra{\psi})C_{N-1}Z\ket{\psi}$ instead of the average gate fidelity for a fair comparison between different $N$. An approximately linear scaling of the infidelity with $N$ is observed for both gates in Fig.~\ref{fig:ProtocolB_main_plot}(b,c). Our protocol outperforms implementations using decompositions into individual  CZ and (perfect) single qubit gates in both cases for any $N>2$.


This concludes our discussion of protocol B. We showed that by driving the cavity with a weak and slowly changing pulse $\eta(t)$, a multi-qubit quantum gate can be implemented by adiabatic evolution. Like for protocol A, the infidelity in the limit $T\rightarrow\infty$ only depends on the cooperativity and not on the ratio $\gamma/\kappa$. We also showed how $N-1$ repetitions of protocol B with different pulse parameters can be used to implement \emph{any} symmetric phase gate.

\section{Fidelity estimates in realistic systems}
\label{sec:fidelity_estimates_in_realistic_system}
In this section we provide estimates for the achievable gate fidelity and pulse duration for protocols A and B for three different physical systems. We discuss atoms coupled to an optical cavity in Sec.~\ref{subsec:fid_estimates_neutral_atoms}, Rydberg atoms coupled to a microwave cavity in Sec.~\ref{subsec:fid_estimates_rydberg_atoms}, polar molecules coupled to a microwave cavity in Sec.~\ref{subsec:fid_estimates_polar_molecules}, and superconducting fluxonium qubits coupled to a microwave cavity in Sec.~\ref{susection:fluxoniumqubits}.
\subsection{Neutral atoms coupled to an optical cavity}
\label{subsec:fid_estimates_neutral_atoms}
 As a first example, we consider neutral $^{87}$Rb atoms trapped in optical tweezers and coupled to a fiber Fabry-Perot cavity as pioneered in Refs. \cite{hunger_fiber_2010-1, uphoff_frequency_2015, barontini_deterministic_2015}. As qubit states, we choose the electronic groundstates $\ket{0} = \ket{5 ~^2S_{1/2}\,F=1\, m_F=0}$ and $\ket{1} = \ket{5 ~^2S_{1/2}\,F=2\, m_F=0}$, while the ancillary state $\ket{e}$ is the electronically excited state $\ket{e} = \ket{5 ~^2P_{3/2}\,F=3\, m_F=0}$. The linewidth of the $\ket{1}\leftrightarrow\ket{e}$ transition ($\lambda = 780$ nm) is $\gamma = 2\pi \times 6$ MHz (FWHM). 
 
 For the cavity we assume a finesse $\mathcal{F}\approx 2\times 10^5$, a waist radius $w_r \approx 2\,\mu$m and a length $L \approx 40\, \mu$m \cite{hunger_fiber_2010-1,uphoff_frequency_2015,barontini_deterministic_2015} resulting in a cooperativity of $C = 3\lambda^2\mathcal{F}/(2\pi^3w_r^2) \approx 1500$ with a coupling strength of $g = \sqrt{3\lambda^2c\gamma/(2\pi^2w_r^2L)} \approx 2\pi \times 400$ MHz and $\kappa = \pi c/L\mathcal{F} \approx 2\pi \times 20$ MHz (FWHM), so that $\gamma/\kappa \approx 0.3$. These values for the cavity parameters are within experimental reach 
  \cite{hunger_fiber_2010-1}.
  
With the numbers above, for protocol A a CZ gate on two atoms can be achieved with a fidelity of $1-F \approx 5.1\%$ in the limit $T, \Delta \rightarrow \infty$. 
Finite values for $\Delta$ can be chosen as long as $\Delta < \omega_0$, with the latter the energy separation between the states $\ket{0}$ and $\ket{1}$ (which is about 6.8 GHz for the states given above). For example, for a detuning $\Delta =1$~GHz, the infidelity only slightly increases to $1-F = 6.4\%$ for a choice of a finite (fast) pulse duration $T = 80 $~ns. Other choices of $\Delta$ and $T$ are possible [see plot Fig. \ref{fig:ProtocolA_main_plot}(a)]. 

For protocol B, the infidelity in the limit $T \rightarrow \infty$ is given by $1-F = 4.6\%$. Similar infidelities can be achieved for finite pulse durations, e.g.  $1-F=5.0\%$ for $T = 120$~ns. These gate speeds compare favorably to those for current fast neutral atom gates \cite{jaksch_fast_2000,morgado_quantum_2021}, however the fidelity is limited by the cooperativity $C$. We note while these infidelities are comparatively large, they are sufficient for specific tasks, e.g. for linking error corrected qubits \cite{ramette_fault-tolerant_2023}. 

\subsection{Rydberg atoms coupled to a microwave cavity}
\label{subsec:fid_estimates_rydberg_atoms}
Higher fidelities at the expense of longer gate durations can be achieved by taking both $\ket{1}$ and $\ket{e}$ to be Rydberg states e.g. $\ket{1}=\ket{90 ~^2P_{3/2}}$ and $\ket{e}=\ket{90 ~^2S_{1/2}}$ in Cs with lifetime 2 ms and 820 $\mu$s, respectively , while $\ket{0}$ is chosen as a long-lived state in the ground manifold of the atoms. In this case, the $\ket{1}\leftrightarrow\ket{e}$ transition has the frequency  $\omega_e \approx 2\pi \times 5$ GHz, and is thus in the microwave regime.

The states $\ket{1}$ and $\ket{r}$ may be coupled via a superconducting microwave resonator with reasonable coupling strength $g\approx 2\pi \times 4$ MHz \cite{pritchard_hybrid_2014}. Quality factors $Q> 3\times 10^8$ have been reported for microwave stripline resonators 
\cite{lei_high_2020}, yielding $\kappa = \omega_e/Q \approx 2\pi \times 17$ Hz, and thus a cooperativity $C = 5\times 10^9$, with $\gamma/\kappa \simeq 12$. 

We include the decay of the state $\ket{1}$ in our analysis, which is important as the latter is now a Rydberg state with a lifetime comparable to that of $\ket{e}$. Therefore, the minimal infidelity is not achieved anymore as $\Delta, T\rightarrow\infty$, but at finite values of $T$. For protocol A, as an example we choose $\Delta = 2\pi \times 400$ MHz, which is much smaller than the spacing of $\ket{1}$ and $\ket{e}$ to adjacent Rydberg states (approximately 5 GHz). The minimal infidelity of $1-F = 2.3\times 10^{-4}$ is then achieved at $T=800 $ns. Furthermore, protocol A could be used to generate a GHZ state on 40 qubits with an infidelity below $10^{-2}$, in a duration of $T= 800$ ns (the same duration as for a CZ gate). 

Since protocol B is much slower than protocol A, due to its adiabatic nature, it is also affected more strongly by the decay of $\ket{e}$. However, the minimal achievable infidelity of $1-F = 2.1 \times 10^{-3}$ at $T= 6.0\, \mu$s still significantly outperforms the infidelities from the previous section (Sec.\ref{subsec:fid_estimates_neutral_atoms}). 


\subsection{Polar Molecules coupled to a microwave cavity}
\label{subsec:fid_estimates_polar_molecules}
As a final platform we consider ultracold polar molecules coupled to a microwave resonator \cite{andre_coherent_2006, rabl_hybrid_2006, sawant_ultracold_2020}. Following Ref.~\cite{andre_coherent_2006}, we assume Ca$^{79}$Br molecules \cite{childs_hyperfine_1981} trapped electrostatically in the vicinity of a  superconducting microwave stripline resonator. 
We chose the computational basis states $\ket{0} = \ket{N=1, m_N=0,F=1, m_F=0}$, $\ket{1} = \ket{N=1, m_N=0,F=2, m_F=1}$ to be different hyperfine levels of the first excited rotational manifold of the molecule, where $N$ is the rotational quantum number.   The ancillary state $\ket{e}$ is then chosen in the second excited rotational level, $\ket{e} = \ket{N=2, m_N=0,F=2, m_F=1}$. These states are chosen to ensure that $\ket{0}, \ket{1}$ and $\ket{e}$ are simultaneously trappable \cite{andre_coherent_2006}.

For this choice of states the frequency of the $\ket{1}\leftrightarrow\ket{e}$ transition is given by $\omega_e = 2\pi \times 11$ GHz and is thus in the microwave regime. Coupling strengths up to $g=2\pi \times 400$ kHz can be achieved with realistic experimental parameters \cite{andre_coherent_2006}. 
Assuming $Q = 3\times 10^8$ (as in Sec.~\ref{subsec:fid_estimates_rydberg_atoms}) yields $\kappa = \omega_e/Q \approx 2\pi \times 37$ Hz, while the decay from $\ket{e}$ is $\gamma\lesssim 10^{-2}$Hz \cite{buhmann_surface-induced_2008} and can be neglected. If we assume $\gamma=0$, arbitrarily low fidelities can be reached if we allow for arbitrarily long gate times. 
At finite pulse duration, protocol A can achieve an infidelity of $1-F = 1.0\times 10^{-5}$ already at a pulse duration of $T = 80\,\mu$s, at $\Delta = 2\pi \times 1.2 \, \mathrm{MHz} \gg g$. Again, other choices of $T$ and $\Delta$ are possible, see Fig.~\ref{fig:ProtocolA_main_plot}(a).

At the same pulse duration, the infidelity of protocol B is given by  $1-F = 8.7 \times 10^{-5}$. This is almost one order of magnitude worse than protocol A, but nevertheless still sufficient for most quantum information processing tasks. 



\subsection{Superconducting Fluxonium qubits}\label{susection:fluxoniumqubits}




Our protocols can also be applied to superconducting qubits coupled via a driven microwave resonator. 
For our purposes, we consider fluxonium qubits \cite{Manucharyan_2009} 
which have a level structure compatible with our protocols. Considering the experimental parameters from \cite{Abdelhafez_2020} at an external flux of $\Phi_\text{ext}=0.49\Phi_0$ (near, but not at the so-called sweet spot of $0.5\Phi_0$), we obtain the ground state $\ket{0}$ and the long lived state $\ket{1}$ ($T_1$ in the millisecond regime \cite{PhysRevLett.130.267001}) separated by $\sim 2\pi \times 100$MHz, while the next higher excited state $\ket{e}$ is separated by $2\pi \times 3.5$GHz from $\ket{0}$ \cite{Groszkowski_2021, chitta2022computeraided}.

The dominant error in this regime is a finite dephasing time $T_2^*\approx 20 \mu$s of the $\ket{0}\leftrightarrow\ket{1}$ transition \cite{PhysRevX.9.041041}, which is in fact not included in the error analysis in Secs.~\ref{sec:protocol_A} and~\ref{sec:protocol B}. We estimate the effect of this errors as an additional $T/T_2^*$ contribution to the infidelity.

We assume a coupling strength of $g=2\pi \times 10 \mathrm{MHz}$ (much smaller than the spacing between $\ket{0}$ and $\ket{1}$) to selectively couple the $\ket{0}\leftrightarrow\ket{e}$ transition to the microwave resonator. We choose this transition instead of the $\ket{1}\leftrightarrow\ket{e}$ transition and flip the roles of $\ket{0}$ and $\ket{1}$ since the coupling of the $\ket{1}\leftrightarrow\ket{e}$ transition to the cavity is weaker than the coupling of $\ket{0}\leftrightarrow\ket{e}$, further suppressing the unwanted coupling of $\ket{1}\leftrightarrow\ket{e}$  to the cavity. Further, we take $1/\gamma = 5\,\mu$s (including decay \emph{and} dephasing of $\ket{e}$) and $Q=3\times 10^8$. For protocol A, these values allow for an infidelity of 4.0\% for a CZ gate (dominated by the contribution from $T_2^*$ of 3.2\%) at $\Delta = 2\pi\times30$ MHz, $\delta=2\pi \times 13$ MHz and $T=640$ ns. The $T_2^*$ contribution could be reduced by using a larger coupling strength $g$, which would however lead to an increased unwanted coupling of the $\ket{1}\leftrightarrow\ket{e}$ transition to the cavity, the effect of which is beyond the scope of this work. For protocol B, a slightly lower infidelity of 3.7\% can be achieved at the same pulse duration.

We note that while these fidelity estimates are lower than that for traditional nearest neighbor two qubit gates on fluxonium qubits \cite{Moskalenko2022}, the non-local nature of our gate allows to couple distant qubits, which is only possible with several CZ gates -- implying a reduced fidelity in systems with only nearest neighbor coupling.
We also note that dynamical decoupling schemes have been proposed to extend $T_2^*$ of a fluxonium qubit beyond $100\mu s$ \cite{dynamical_dec}, which would reduce the infidelities of protocol A and B to 1.5\% and 1.1\%, respectively.

\section{Applications to Quantum Error Correction}
\label{sec:application_to_quantum_error_correction}


The methods described here together with fast and reliable near neighbour gates may be integrated into an architecture to support fault tolerant quantum computation. For an application of non-local entangled states prepared using Protocol A, consider a setup where each register qubit has one or more neighbouring ancillary qubits that are addressable. A useful primitive gate is the non-local measurement of a Pauli operator with support on a set of distantly separated register qubits 
$\{q_k\}$. This could be used for stabilizer measurements or a non-destructive Pauli measurement gate in an LDPC code for example. Such measurements could be performed \a`a la Shor \cite{PhysRevLett.77.3260} using a $\ket{{\rm GHZ}}$ on the set of ancilla $\{a_k\}$ neighbouring the $\{q_k\}$ and prepared via Protocol A, so that the ancilla controlled gates targeting the register qubits would be spatially near neighbour. The usual $\ket{{\rm GHZ}}$ verification steps before the controlled operations would themselves be non-local, but they can be obviated using the Aliferis-DiVincenzo method \cite{divincenzo_effective_2007}. In that procedure, errors in the $\ket{{\rm GHZ}}$ preparation (encoding) can be accounted for by unpreparing (decoding) the state, which is achievable non-locally since our method is unitary, and measuring the ancilla to infer errors which can be accounted for by adapting the Pauli frame of the computation. 

The method above is particularly advantageous when measurements are slow and can be made fault tolerant by performing a second level repetition code on the ancilla. That is achievable via near neighbour controlled operations between the $\{a_k\}$ and a second set of neighbouring ancilla 
$\{b_k\}$ after the controlled operations acting on the 
$\{q_k\}$. A syndrome measurement compatible with a bit flip error in both ancilla sub-blocks implies a fault in the non-local encoding that propagated to the data register and can be accounted for in subsequent gates. Otherwise the error likely occured during the non-local decoding and the process can be repeated.
Note an alternative method for non-local stabilizer measurments is to use flag qubits \cite{Chao_2018}. This uses fewer ancilla (as few as $2$) but would require faster resets and $k$ non-local CZ gates to measure a weight $k$ non-local stabilizer. Using protocol A, for $C\gg 1$ the fidelity for preparing a $\ket{{\rm GHZ}_k}$ state vs. a circuit of $k$ CZ gates are comparable but the time to prepare the former for a fixed fidelity is shorter, essentially independent of $k$, potentially favoring the former approach in this context.

An application of Protocol B is to perform non-local C$_{N-1}$Z gates, which are locally equivalent to multi-controlled Toffoli gates, for majority voting circuits. These are frequently used e.g. in measurement free quantum error correction \cite{paz-silva_fault_2010,crow_improved_2016,ercan_measurement-free_2018}. Even though for $N>2$ the gate is not Clifford and our implementation is not fault tolerant, the gate {\it can} be used for fault tolerant quantum error correction when it involves controls that are ancilla that carry error syndrome data that is classical \cite{paz-silva_fault_2010}.


\section{Conclusion}
\label{sec:conclusion}
We have presented two new protocols for implementing a large family of non-local multi-qubit quantum gates on qubits coupled to a common cavity mode. These protocols are implemented by applying a classical drive to the cavity mode, while no external drive on the qubits is required. Applied to just two qubits, both protocols form, together with single qubit gates, a universal gate set for quantum computing. Applied on more than two qubits, protocol A can be used to generate a GHZ state, while protocol B can be used to implement arbitrary phase gates, such as phase-rotation or multi-controlled Z gates. We evaluated the fidelity of both protocols in the presence of a finite lifetime of the ancillary state $\ket{e}$ of the qubits and of the photons in the cavity, finding that the infidelity scales as $\mathcal{O}(C^{-1/2})$. For Rydberg atoms or polar molecules coupled via a microwave cavity, we expect that our protocols can achieve infidelities below $10^{-3}$ with realistic parameters, while for neutral atoms coupled via optical cavities infidelities of the order of a few percent can be reached. 

Our protocols allow for the first time the realization of a large family of deterministic non-local multi-qubit quantum gates with applications in digital quantum simulations, metrology, cryptography and error correction, by controlling the system via only a simple classical drive of the cavity. 
In a quantum computing architecture, our protocol could be applied in several manners, either as the only entangling gate of the architecture, or in conjunction with other, local, entangling protocols. For example, in an array of Rydberg atoms, entangling operations between nearby atoms could be performed using the Rydberg blockade mechanism, while entangling atoms further apart could be done with our protocols. It is also possible to use our protocols only for certain error correction tasks, while other entangling operations are done by local gates.  Finally, our protocols could also be extended to overlapping cavities \cite{ramette_any--any_2021} to connect even more atoms. 

In this work we modeled each qubit as a three level system and the cavity as a single bosonic mode. We expect that our protocols are generalizable to more complicated models with e.g. several excited states, a nonzero coupling from $\ket{0}$ to $\ket{e}$, or several bosonic modes (e.g. light modes of different polarizations) supported in the cavity. For example, the derivations of both protocols can be extended in a straightforward manner to include the coupling of $\ket{1}$ to a second exited state $\ket{e'}$. Such an additional coupling would only
effect the parameter $\theta$ in protocol A and the dependency of $\varphi(\hat{n})$ on $\hat{n}$ in protocol B. Finding the optimal gate parameters and the achievable fidelities for more general models of the qubit and the cavity will be subject to future work. 

We expect that our protocols may significantly benefit from optimization of the time-dependent pulse-shape $\eta(t)$.  In particular, while the infidelity for the various gates in the limit $T\rightarrow\infty$ is independent of the exact choice of $\eta$, we expect that the infidelity at finite $T$ could be improved by applying quantum optimal control techniques to optimize the pulse-shape of $\eta(t)$ \cite{glaser_training_2015, wilhelm_introduction_2020}, making our protocols both higher-fidelity and faster.


\begin{acknowledgments}
We gratefully acknowledge discussions with Shannon Whitlock. This research has received funding from the European Union’s Horizon 2020 research
and innovation programme under the Marie Sk{\l}odowska-Curie project 955479 (MOQS) and project 847471(QUSTEC), the Horizon Europe programme HORIZON-CL4-2021-DIGITAL-EMERGING-01-30 via the project 101070144 (EuRyQa) and from the French National
Research Agency under the Investments of the Future
Program projects ANR-21-ESRE-0032 (aQCess) and ANR-22-CE47-0013-02 (CLIMAQS). G.K.B. acknowledges
support from the Australian Research Council Centre
of Excellence for Engineered Quantum Systems (Grant
No. CE 170100009).
\end{acknowledgments}

\appendix

\section{Supporting calculations for protocol A}
\label{app:protA}
\subsection{First basis transformation on the cavity}
\label{app:protA_first_basis_trafo_cavity}
Here we discuss the first time-dependent basis transformation on the subsystem of the cavity. For a function $\alpha(t)$ consider the displacement operator $D(\alpha(t)) = \exp(\alpha(t) \ah^\dag - \alpha^*(t) \ah)$. Recall that it satisfies $D(\alpha)\ah D(\alpha)^\dag = \ah - \alpha$ and $D(\alpha)\ah ^\dag D(\alpha)^\dag = \ah^\dag - \alpha^*$, and furthermore
\begin{equation}
\begin{array}{lll}
    \frac{\mathrm{d}}{\mathrm{d}t}D(\alpha) &=& [\dot{\alpha}\ah^\dag - \dot{\alpha}^* \ah + i\Im(\dot{\alpha}^*\alpha)]D(\alpha) \\
    &=& D(\alpha)[\dot{\alpha} \ah^\dag - \dot{\alpha}^*\ah -i\Im(\dot{\alpha}^*\alpha)].
    \end{array}
\end{equation}
Now we define $\tilde{\rho} = D(\alpha)\rho D(\alpha)^\dag$. It satisfies
\begin{equation}
     \begin{array}{lll}
     \dot{\tilde{\rho}} &=& -i(H'\tilde{\rho} - \rho (H')^\dag) + L'\tilde{\rho} (L')^\dag -\frac{1}{2}\{(L')^\dag L', \tilde{\rho}\}\\
     &+& \left(\frac{\mathrm{d}}{\mathrm{d}t}D(\alpha)\right)D(\alpha)^\dag \tilde{\rho}
     + \tilde{\rho}D(\alpha)\left(\frac{\mathrm{d}}{\mathrm{d}t}D(\alpha)^\dag\right)
     \end{array}
\label{eq:si_adiabatic_lindblad_tilde}
\end{equation}
where $H' = D(\alpha)H D(\alpha)^\dag$ and $L' = D(\alpha)L D(\alpha)^\dag$.

We calculate
\begin{equation}
\begin{array}{lll}
    H' &=& \delta \ah^\dag \ah + (\Delta -i \gamma/2)\hat n_e + g(\ah^\dag \hat S^- + \ah \hat S^+) \\
    &+& (i\eta - \delta\alpha)\ah^\dag 
    - (i\eta^* + \delta\alpha^*) \ah + -g\alpha^* \hat S^-\\
    &-&g \alpha \hat S^+ + \delta |\alpha|^2 + i(\eta\alpha^*-\eta^*\alpha)
    \end{array}
\end{equation}

\begin{equation}
\begin{array}{lll}
    &L'\tilde{\rho}(L')^\dag& - \frac{1}{2}\{(L')^\dag L, \tilde{\rho}\} = L\tilde{\rho} L^\dag -\frac{1}{2}\{L^\dag L, \tilde{\rho}\}+\\
    && \frac{\kappa}{2}(-\alpha^* \ah\rho - \alpha \rho \ah^\dag + \alpha^* \rho \ah + \alpha \ah^\dag \rho)
    \end{array}
\end{equation}
and
\begin{equation}
    \left(\frac{\mathrm{d}}{\mathrm{d}t}D(\alpha)\right)D(\alpha)^\dag \tilde{\rho} + \tilde{\rho}D(\alpha)\left(\frac{\mathrm{d}}{\mathrm{d}t}D(\alpha)^\dag\right) = [\dot{\alpha}\ah^\dag - \dot{\alpha}^*\ah, \tilde{\rho}]
\end{equation}
Plugging this into Eq.~\eqref{eq:si_adiabatic_lindblad_tilde} gives
\begin{equation}
    \dot{\tilde{\rho}} = -i\tilde{H}\tilde{\rho} +i\tilde{\rho}\tilde{H}^\dag+ L\rho L^\dag -\frac{1}{2}\{L^\dag L, \rho\}
\end{equation}
with
\begin{eqnarray}
    \tilde{H} &=& \delta \ah^\dag \ah + (\Delta -i \gamma/2)\hat n_e + g(\ah^\dag \hat S^- + \ah \hat S^+)
    \\ \nonumber &-& g\alpha^* \hat S^- -g \alpha \hat S^+ +  \left[(i\eta - (\delta - i\kappa/2)\alpha + i\dot{\alpha})\ah^\dag + \hc \right]
\end{eqnarray}

Now we take $\alpha(t)$ such that
\begin{equation}
    \dot{\alpha} = -\eta  -(i\delta +\kappa/2)\alpha
\end{equation}
which is satisfied by
\begin{equation}
    \alpha(t) = \int_0^t \mathrm{d}t' \eta(t') e^{-(i\delta+\kappa/2)(t-t')}.
\end{equation}
With this choice, $\tilde{H}$ becomes simply
\begin{equation}
     \tilde{H} = \delta \ah^\dag \ah + (\Delta -i \gamma/2)\hat n_e + g(\ah^\dag \hat S^- + \ah \hat S^+)  -g\alpha^* \hat S^- -g \alpha \hat S^+
\end{equation}

\subsection{Second basis transformation on the qubits}
\label{app:protA_second_basis_trafo_qubits}
We perform the time dependent basis transformation $\bar{H} = U\tilde{H}U^\dag + i\dot{U}U^\dag$ for 
\begin{equation}
    U = \exp\left[\frac{\lambda}{2}\left(-e^{i\mu}\hat S^+ +e^{-i\mu}\hat S^-\right)\right]
\end{equation}
and
\begin{equation}
    \tilde{H} = \delta \ah^\dag \ah + (\Delta-i\gamma/2)\hat{n}_e + g\ah^\dag \hat S^- + g\ah\hat S^+ -g\alpha \hat S^+ -g\alpha^*\hat S^-.
\end{equation}
We calculate
\begin{equation}
\begin{array}{lll}
U &=& \Big[\ket{0}\bra{0} + \cos\lf\left(\ket{1}\bra{1}+\ket{e}\bra{e}\right)\\
&+& \sin\lf\left(-e^{i\mu}\ket{e}\bra{1}+e^{-i\mu}\ket{1}\bra{e}\right) \Big]^{\otimes N}
\end{array}
\end{equation}
which gives
\begin{eqnarray}
\nonumber U\hat S^+U^\dag &=& \left(U\hat S^-U^\dag\right)^\dag = \cos^2\lf \hat S^+ - e^{-2i\mu}\sin^2\lf \hat S^-  \\ &+& e^{-i\mu}\sin\lf \cos\lf (\hat{n}-\hat{n}_e),
\end{eqnarray}
and 
\begin{eqnarray}
    U\hat{n}_e U^\dag &=& \cos^2\lf \hat{n}_e + \sin^2\lf \hat{n} \\ \nonumber &+& \sin\lf \cos\lf \left(e^{-i\mu}\hat S^-+e^{i\mu}\hat S^+\right)
\end{eqnarray}
Now first consider $H^{(0)} = \Delta \hat{n}_e -g\alpha \hat S^+ -g\alpha^*\hat S^-$, the part of $\tilde{H}$ that scales with $\Delta$. We choose $\lambda$ and $\mu$ so that $U H^{(0)} U^\dag$ is diagonal. We find
\begin{eqnarray}
\nonumber
&U &H^{(0)} U^\dag  = \left[\Delta \cos^2\lf +\frac{g}{2}\left(\alpha e^{-i\mu}+\cc\right)\sin\left(\lambda\right)\right]\hat{n}_e \\
&+& \left[\Delta \sin^2\lf -\frac{g}{2}\left(\alpha e^{-i\mu}+\cc\right)\sin\left(\lambda\right)\right]\hat{n} \\ \nonumber
&+& \left[\frac{\Delta}{2} e^{i\mu}\sin\left(\lambda\right) + e^{2i\mu}\sin^2\lf g\alpha^* -g\alpha\cos^2\lf \right]\hat S^+ \\ \nonumber
&+& \left[\frac{\Delta}{2} e^{-i\mu}\left(\lambda\right) + e^{-2i\mu}\sin^2\lf g\alpha -g\alpha^*\cos^2\lf \right]\hat S^- 
\end{eqnarray}
The coefficients of $\hat S^+$ and $\hat S^-$ vanish for $\mu = \arg(\alpha)$ and $\lambda$ such that $\Delta\sin\lf\cos\lf = g\alpha(\cos^2\lf - \sin^2\lf)$, which is satisfied for $\cos\lambda = \Delta/\sqrt{\Delta^2 + 4g^2|\alpha|^2}$. We denote by $\varepsilon_-$ and $\varepsilon_+$ the coefficients of $\hat{n}$ and $\hat{n}_e$,respectively, and find
\begin{eqnarray}
\varepsilon_- &=& \Delta \sin^2\lf -2g|\alpha|\sin\lf\cos\lf \\
\nonumber &=& \frac{1}{2}\left(\Delta - \sqrt{\Delta^2+4g^2|\alpha|^2}\right) \\
\varepsilon_+ &=& \Delta \cos^2\lf +2g|\alpha|\sin\lf\cos\lf \\ \nonumber &=& \frac{1}{2}\left(\Delta + \sqrt{\Delta^2+4g^2|\alpha|^2}\right)
\end{eqnarray}
Now we consider $\tilde{H}-H^{(0)} = \delta \ah^\dag \ah -i\frac{\gamma}{2}\hat{n}_e + g \ah \hat S^+ + g\ah^\dag \hat S^-$ and calculate $U(\tilde{H}-H^{(0)})U^\dag$ term by term (the notation $\mathcal{O}(1)$ refers to the limit $\Delta\rightarrow\infty$). 
\begin{equation}
U \ah^\dag \ah U^\dag = \ah^\dag \ah
\end{equation}
\begin{eqnarray}
 \nonumber &U&\hat{n}_e U^\dag = \frac{1-\cos\lambda}{2}\hat{n} + \frac{1+\cos\lambda}{2}\hat{n}_e + \mathcal{O}(1)\hat S^+ + \mathcal{O}(1)\hat S^-  \\
&=& \frac{\hat{n} + \hat{n}_e }{2} -\frac{\Delta \left(\hat{n} - \hat{n}_e\right) }{2\sqrt{\Delta^2+4g^2|\alpha|^2}}+ \mathcal{O}(1)\hat S^+ + \mathcal{O}(1)\hat S^-
\end{eqnarray}
\begin{equation}
U \hat S^+ U^\dag = \frac{\alpha^*}{\sqrt{\Delta^2+4g^2|\alpha|^2}}(\hat{n}-\hat{n}_e) + \mathcal{O}(1)\hat S^+ + \mathcal{O}(1)\hat S^-
\end{equation}
so that in total we find
\begin{eqnarray}
U\tilde{H}U^\dag &=& \delta \ah^\dag \ah + (\varepsilon_1-i\gamma_1/2)\hat{n}+(\varepsilon_e-i\gamma_e/2)\hat{n}_e \\
\nonumber &+& (\zeta \ah^\dag + \zeta^* \ah)(\hat{n}-\hat{n}_e) + \mathcal{O}(1)\hat S^+ + \mathcal{O}(1)\hat S^-
\end{eqnarray}
where
\begin{equation}
\gamma_\pm = \frac{\gamma}{2}\left(1\pm \frac{\Delta}{\sqrt{\Delta^2+4g^2|\alpha|^2}}\right),
\end{equation}

\begin{eqnarray}
    \zeta = \frac{g^2\alpha}{\sqrt{\Delta^2+4g^2|\alpha|^2}}.
\end{eqnarray}
Now using the fact that $i\dot{U}U^\dag$ is $\mathcal{O}(1)$ and acts on the qubits only (i.e. contains no $a$ or $a^\dag$ terms) we obtain the expression (5) from the main text for $\bar{H}$.

\subsection{Analytic solution of time evolution under $H_\eff$}
\label{app:protA_analytic_solution_of_time_evolution}
In this section we find the analytic solution of the Lindbladt equation $\dot{\rho} = -iH\rho+i\rho H^\dag +L\rho L^\dag -\frac{1}{2}\{L^\dag L, \rho\}$ under $H = \delta \ah^\dag \ah + (-i\gamma_1(t)/2 + \zeta(t)\ah^\dag + \zeta(t)^* \ah)\hat{n}$ and $L = \sqrt{\kappa}\ah$ for an arbitrary drive $\zeta(t)$ and time dependent decay rate $\gamma_1(t)$. For this, we assume an initial state $\rho(0) = \ket{\beta_n(0)}\bra{\beta_m(0)}\otimes\ket{q}\bra{q'}$, where $\beta_n$ and $\beta_m$ are coherent states and $\ket{q}$($\ket{q'}$) are computational basis states with $n$($m$) qubits in state $\ket{1}$. Note that initial states of this form are a basis of space of all possible initial density matrices, so solving  the Lindbladt equation for the initial state $\rho(0)$ suffices to solve it for an arbitrary initial state.

In the following we show that the solution to the Lindblad equation is given by 
\begin{equation}
\rho(t) = e^{i\varphi_{nm}(t)}\frac{\ket{\beta_n(t)}\bra{\beta_m(t)} \otimes \ket{q}\bra{q'}}{\braket{\beta_m(t)}{\beta_n(t)}}
\label{eq:si_rho_exact}
\end{equation}
where $\dot{\beta}_n = -(i\delta+\kappa/2)\beta_n - in\zeta$, i.e.
\begin{equation}
\beta_n(t) = \beta_n(0)e^{-(i\delta+\kappa/2)t} -in\int_0^t \mathrm{d}t' \zeta(t')e^{-(i\delta+\kappa/2)(t-t')}
\label{eq:si_beta_n}
\end{equation}
and
\begin{eqnarray}
\varphi_{nm}(t) &=& \int_0^t \mathrm{d}t'  \left[ (m-n)(\zeta(t') \beta_m(t')^* \right.  \\ \nonumber  &+& \left.  \zeta(t')^*\beta_n(t'))  + i(m+n)\gamma_1(t')/2 \right].
\label{eq:si_theta_nm}
\end{eqnarray}
Tracing out the cavity then gives the reduced density matrix $\rho_\eff = e^{i\varphi_{nm}(t)}\ket{q}\bra{q'}$ discussed in the main text.

To show Eq.~\eqref{eq:si_rho_exact}, we make the Ansatz $\rho = \rho_{nm} \otimes \ket{q}\bra{q'}$ with $\rho_{nm}(t) = c_{nm}(t) \ket{\beta_n(t)}\bra{\beta_m(t)}$. The Lindblad equation gives 
\begin{equation}
\dot{\rho}_{nm} = -iH_n\rho_{nm} +i\rho H_m^\dag  + L\rho_{nm} L^\dag -\frac{1}{2}\{L^\dag L, \rho_{nm}\},
\label{eq:si_lindblad_equation_Heff}
\end{equation}
 with $H_n = \delta \ah^\dag \ah + (-i\gamma_1/2 + \zeta \ah^\dag + \zeta^* \ah)n$.

We start by calculating the left side of Eq.~\eqref{eq:si_lindblad_equation_Heff}. It is a property of coherent states that
\begin{equation}
    \frac{\mathrm{d}}{\mathrm{d}t}\ket{\beta_n(t)} = \dot{\beta}_n \ah^\dag \ket{\beta_n}-\frac{1}{2}\frac{\mathrm{d}|\beta_n|^2}{\mathrm{d}t}\ket{\beta_n}
\end{equation}
so that
\begin{eqnarray}
    \dot{\rho}_{nm} &=& c_{nm}\dot{\beta}_n \ah^\dag \ket{\beta_n}\bra{\beta_m} + c_{nm}\dot{\beta}_m^* \ket{\beta_n}\bra{\beta_m}\ah \\
    \nonumber &+& \left(\dot{c}_{nm}-\frac{c_{nm}}{2}\frac{\mathrm{d}(|\beta_n|^2+|\beta_m|^2)}{\mathrm{d}t}\right)\ket{\beta_n}\bra{\beta_m}
    \label{eq:error_cavity_lhs}
\end{eqnarray}
Now we evaluate the right side of Eq.~\eqref{eq:si_lindblad_equation_Heff}: 
\begin{eqnarray}
    H_n\rho_{nm}/c_{nm} &=& \left[\left(\delta\beta_n + n\zeta \right)\ah^\dag \right. \\ \nonumber &+& \left. n\zeta^*\beta_n  - in \gamma_1/2 \right]  \ket{\beta_n}\bra{\beta_m} \label{eq:si_decay_error_Lnm1} \\
    \rho_{nm}H_m^\dag/c_{nm} &=& \left[ \left(\delta\beta_m^* + m\zeta^* \right)\ah \right. \\ \nonumber &+& \left. m\zeta \beta_m^* + im\gamma_1/2 \right] \ket{\beta_n}\bra{\beta_m} \label{eq:si_decay_error_Lnm2}\\
    L\rho_{nm} L^\dag/c_{nm} &=& \kappa \beta_n\beta_m^*\ket{\beta_n}\bra{\beta_m} \label{eq:si_decay_error_Lnm3} \\
    \left\{L^\dag L, \rho_{nm} \right\}/c_{nm}  &=& \left[ \kappa\beta_n \ah^\dag + \kappa\beta_m^* \ah \right]\ket{\beta_n}\bra{\beta_m}\
    \label{eq:si_decay_error_Lnm4}
\end{eqnarray}
Together, Eq.~\eqref{eq:si_decay_error_Lnm1}-\eqref{eq:si_decay_error_Lnm4} give
\begin{eqnarray}
   \nonumber &&-iH_n\rho_{nm} +i\rho_{nm} H_m^\dag  + L\rho_{nm} L^\dag -\frac{1}{2}\{L_{nm}^\dag L, \rho_{nm}\}\\ \nonumber &=& c_{nm}(-i\delta\beta_n -in\zeta-\kappa\beta_n/2)\ah^\dag\ket{\beta_n}\bra{\beta_m} \\
    &+& c_{nm}(i\delta\beta_m^* +im\zeta^*-\kappa\beta_m^*/2)\ket{\beta_n}\bra{\beta_m} \ah \nonumber \\
    &+& c_{nm}(-in\zeta^*\beta_n + im\zeta\beta_m^*+\kappa \beta_n\beta_m^* \nonumber\\ &-&(n+m)\gamma_1/2)\ket{\beta_n}\bra{\beta_m}
    \label{eq:error_cavity_rhs}
\end{eqnarray}
Equating Eq.~\eqref{eq:error_cavity_lhs} and Eq.~\eqref{eq:error_cavity_rhs} gives Eq.~\eqref{eq:si_beta_n}, as well as
\begin{eqnarray}
\dot{c}_{nm}/c_{nm} &=& \frac{1}{2}\frac{\mathrm{d}(|\beta_n|^2+|\beta_m|^2)}{\mathrm{d}t} \\ \nonumber &-&in\zeta^*\beta_n + im\zeta\beta_m^*+\kappa \beta_n\beta_m^*-(n+m)\gamma_1/2
\end{eqnarray}
Now we take $c_{nm} = e^{i\varphi_{nm}}/\braket{\beta_m|\beta_n}$. Using $\braket{\beta_m|\beta_n} = \exp\left(-\frac{1}{2}(|\beta_n|^2+|\beta_m|^2)+\beta_m^*\beta_n\right)$ we obtain
\begin{eqnarray}
i\dot{\varphi}_{nm} &=& \frac{\dot{c}_{nm}}{c_{nm}} + \frac{\mathrm{d}}{\mathrm{d}t}\left(-\frac{1}{2}(|\beta_n|^2+|\beta_m|^2)+\beta_m^*\beta_n\right)\label{eq:si_theta_deriv}\\ \nonumber
&=& -in\zeta^*\beta_n + im\zeta\beta_m^*+\kappa \beta_n\beta_m^* \\ \nonumber &-&(n+m)\gamma_1/2 + \dot{\beta}_m^*\beta_n + \beta_m^*\dot{\beta_n}\\ \nonumber
&=& i(m-n)\zeta^*\beta_n + i(m-n)\zeta\beta_m^*-(n+m)\gamma_1/2
\end{eqnarray}
where in the last equality we inserted $\dot{\beta}_n = -(i\delta+\kappa/2)\beta_n - in\zeta$. Integrating Eq.~\eqref{eq:si_theta_deriv} gives Eq.~\eqref{eq:si_theta_nm}.

\subsection{Calculation of the Fidelity in the limit $T\rightarrow\infty$}
\label{app:protA_calculation_of_fidelity_in_limit}
In the following we show that in the limit $T\rightarrow\infty$ and to first order in $\gamma$ and $\kappa$ the infidelity of protocol A is given by
\begin{equation}
 1-F = \left(\frac{\kappa}{4(1+2^{-N})\delta} + \frac{\gamma\delta}{2g^2}\right)N\theta.
 \label{eq:si_protocol_A_infid}
\end{equation}
In the limit $T\rightarrow\infty$ the solution to $\dot{\beta}_n = -(i\delta+\kappa/2)\beta_n - in\zeta$ can be obtained by an adiabatic approximation. For this, we insert $\dot{\beta}_n = 0$ and obtain 
\begin{equation}
\beta_n = \frac{-in\zeta}{i\delta+\kappa/2} \approx -\frac{n\zeta}{\delta} \left(1+i\frac{\kappa}{2\delta}\right).
\end{equation}
With Eq. (7) and (10) (main text) we obtain
\begin{equation}
    \varphi_{nm} = (n^2-m^2)\theta +(m-n)^2\frac{i\kappa}{2\delta}\theta + i(m+n)\int_0^T \mathrm{d}t \gamma_1(t)/2
\end{equation}
where $\theta = \frac{1}{\delta}\int_0^T \mathrm{d}t |\zeta(t)|^2$.
Since in the limit $T\rightarrow\infty$ we have $\zeta \rightarrow 0$ we approximate
\begin{equation}
    \gamma_1 = \frac{\gamma}{2}\left(1-\sqrt{1-4|\zeta|^2/g^2}\right) \approx \frac{\gamma|\zeta|^2}{g^2}
\end{equation}
so that
\begin{eqnarray}
    \frac{\varphi_{nm}}{\theta} = n^2-m^2 + (m-n)^2\frac{i\kappa}{2\delta} + (m+n)\frac{i\gamma\delta}{2g^2}
\end{eqnarray}
Inserting this into Eq. (22) (main text) and using that
\begin{equation}
    \sum_{n,m=0}^N \binom{N}{n}\binom{N}{m} (m-n)^2 = 4^N\frac{N}{2}
\end{equation}
and
\begin{equation}
    \sum_{n=0}^N \binom{N}{n} (n+n) = 2^NN,
\end{equation}
\begin{equation}
    \sum_{n,m=0}^N \binom{N}{n}\binom{N}{m} (n+m) = 4^NN,
\end{equation} 
we obtain Eq.~\eqref{eq:si_protocol_A_infid}.

\subsection{Effects of coupling inhomogeneities on the fidelity}
\label{app:protA_effects_of_inhomogeneities}
In this section we calculate the effect different couplings $g_1,...,g_N$ of each qubit to the cavity on the gate fidelity. We assume that the $g_1,...,g_N$ are independent and identically distributed random variables and have the quadratic mean $\bar{g} = \sqrt{\E[g_j^2]}$. Furthermore assume that the drive $\eta(t)$, and thus $\alpha(t)$, is chosen as given in the main text, with the homogeneous coupling $g$ replaced by $\bar{g}$.

To be able to obtain analytical solutions we restrict ourselves to the case $T \rightarrow \infty$, but expect a similar scaling for finite $T$. 

Following the same steps as in the main text, an effective Hamiltonian can be found as
\begin{equation}
    H_\eff = \delta a^\dag a + \sum_{q\in \{0,1\}^N} (\zeta_q  a^\dag + \zeta_q^* a)\ket{q}\bra{q}
\end{equation}
where 
\begin{equation}
    \zeta_q = \sum_{j=1}^N q_j \frac{g_j^2\alpha}{\sqrt{4g_j^2|\alpha|^2 + \Delta^2}} \approx \frac{\alpha}{\Delta}\sum_{j=1}^N q_j g_j^2
\end{equation}
where the last approximation holds in the $T \rightarrow \infty$ limit, where $|\alpha| \ll \Delta$.

Starting in the initial state $\ket{\psi(0)} = \ket{0}\otimes \ket{q}$ for a computational basis state $q \in \{0,1\}^N$, the state at the final time $T$ is given by $\ket{\psi(T)} = e^{i\varphi_q(T)}\ket{\beta_q}\otimes \ket{q}$, where $\dot{\beta}_q = -i\delta\beta_q -i\zeta_q$ and $\dot{\varphi}_q = -\Re(\zeta_q^*\beta_q)$. In the limit $T \rightarrow \infty$ we obtain $\beta_q(t)=-\zeta_q(t)/\delta$ and
\begin{eqnarray}
    \varphi_q(T) &=& \left(\sum_j q_j g_j^2\right)^2 \int_0^T \mathrm{d}t \frac{|\alpha(t)|^2}{\Delta \delta} \\\nonumber &=& \left(\sum_j q_j g_j^2\right)^2\frac{\theta}{\bar{g}^4} \\ \nonumber &\approx& n\theta^2 + \frac{2n\theta}{\bar{g}^2}\sum_j q_j (g_j^2-g^2)
\end{eqnarray}
where $n = \sum_j q_j$ is the number of qubits in state $\ket{1}$. 
Crucially, $\beta_q(T)=0$, so that the action of the gate can still be described by a unitary operation, given by $U = \sum_q e^{i\varphi_q(T)} \ket{q}\bra{q}$. In the following, we will evaluate the averaged fidelity for the difference $U_A^\dag U$ between the gate $U_A$ which we aim to implement, and the gate $U$ which is actually implemented.

The averaged fidelity can be evaluated as \cite{pedersen_fidelity_2007}
\begin{eqnarray}
    1-F &=& \frac{1}{2^N(2^N+1)}\left(2^N + \left|\sum_{q\in\{0,1\}^N} e^{i(\varphi_q(T)-n\theta^2)}\right|^2\right)\nonumber\\ \nonumber &\approx& 1 + \frac{1}{2^N(2^N+1)}\Bigg[\left(\sum_q (\varphi_q -n^2\theta)\right)^2\\  &-&2^N\sum_q (\varphi_q-n^2\theta)^2\Bigg]
\end{eqnarray}
We obtain the upper bounds
\begin{eqnarray}
    1-F &\leq& \frac{1}{2^N}\sum_q (\varphi_q-n^2\theta)^2\\ \nonumber &=& \frac{4\theta^2}{\bar{g}^4}\frac{1}{2^N}\sum_{q \in \{0,1\}^N}\left(n \sum_j q_j(g_j^2-\bar{g}^2)\right)^2
    \label{eq:protocol_A_inhomos_infidelity_bound}
\end{eqnarray}

The expected value of the infidelity can be upper bounded, using the independence of the $g_j$, as 
\begin{eqnarray}
    \E[1-F] &\leq& \frac{4\theta^2}{\bar{g}^4}\sum_q n^2\sum_j q_j^2 \E[(g_j^2-\bar{g}_j^2)^2]\\
    &=& \frac{4\theta^2}{\bar{g}^4}\mathrm{Var}[g_1^2]\frac{1}{2^N}\sum_{n=0}^N \binom{N}{n}n^3 \\
    &=& N^2(N+3)\frac{\theta^2}{2\bar{g}^4}\mathrm{Var}[g_1^2].
\end{eqnarray}
Note that since we assume that the $g_j$ are independent and indetically distributed, the  $\mathrm{Var}[g_1^2]$ can be replace by $\mathrm{Var}[g_j^2]$ for any $j$.

\subsection{Generation of GHZ states}
\label{app:protA_generation_of_GHZ_states}

In the following, we show that a GHZ state can be generated by applying $U_A$ with $\theta = \pi/2$ to $\ket{+}^{\otimes N}$, followed by the three single qubit gates $U_1 = \exp(i\pi\sigma_z/4)$, $U_2 = (\sigma_x+\sigma_z)/\sqrt{2}$ and $U_3 = \exp(i\pi(N+1)\sigma_z/(4N))$. For this, we proceed in two steps. First, we show that  $\ket{\psi} = U_1^{\otimes N}U_A\ket{+}^{\otimes N}$ is the graph state of the complete graph with $N$ vertices. Then we employ the known result that this graph state is equivalent to a GHZ state up to global single qubit gates \cite{hein_multi-party_2004, leibfried_toward_2004} to explicitly find $U_2$ and $U_3$ to convert the graph state to the GHZ state.

To see that $\ket{\psi}$ is this graph state, we calculate that (up to an irrelevant global phase)
\begin{eqnarray}
      U_1^{\otimes N}U_A &=& \exp(-i\pi\hat{n}/2)\exp(i\pi\hat{n}^2/2) \\ \nonumber
      &=& \exp\left(i\pi\sum_{j<k}\hat n^{(j)}\hat n^{(k)}\right) 
\end{eqnarray}
 where $\hat n^{(j)}$ is the single qubit operator $\ket{1}\bra{1}$ on qubit $j$, so that $\hat n = \sum_{j=1}^N \hat n^{(j)}$. Thus, $U_1^{\otimes N}U_A$ applies a CZ gate simultaneously on each qubit pair. Hence, the state $\ket{\psi}$ is indeed the graph state of the complete graph with $N$ vertices

To convert $\ket{\psi}$ into a GHZ state, we use that $\ket{\psi}$ is stabilized by the $N$ independent stabilizers 
 \begin{eqnarray}
     S_1 &=& \sigma_x \otimes \sigma_z \otimes ... \otimes \sigma_z \\
     S_2 &=& \sigma_z \otimes \sigma_x \otimes \sigma_z \otimes...\otimes \sigma_z \\ \nonumber
     &\vdots & \\
     S_N &=& \sigma_z \otimes ... \otimes \sigma_z \otimes \sigma_x
 \end{eqnarray}
 i.e. it satisfies $S_k\ket{\psi} = \ket{\psi}$ for each $k$. Thus, $U_2^{\otimes N}\ket{\psi}$ is stabilized by $U_2^{\otimes N}S_kU_2^{\otimes N}$ (note that $U_2^\dag = U_2)$. A direct calculation shows that also the state
 \begin{equation}
     \ket{\mathrm{GHZ}_\alpha} = (\ket{0...0}+e^{i\alpha}\ket{1...1})/\sqrt{2}
 \end{equation}
 with $\alpha = \pi(N+1)/2$ is stabilized by the $U_2^{\otimes N}S_kU_2^{\otimes N}$. Hence, we conclude $U_2^{\otimes N}\ket{\psi} = \ket{\mathrm{GHZ}_\alpha}$, up to an irrelevant global phase. A GHZ state can now be generated by applying $U_3^{\otimes N}$ to $U_2^{\otimes N}\ket{\psi}$ to convert $\ket{\mathrm{GHZ}_\alpha}$ to the GHZ state.
 
Another way to see this is to write $U_A=e^{i \frac{\pi}{2}\hat{n}^2}=e^{i \frac{\pi}{2} (N\hat{S}^z+(\hat{S}^z)^2)}$. Since $U_A$ is permutation symmetric, if we start in the symmetric state $\ket{+}^{\otimes N}$ then we stay in the maximally symmetric subspace of the spins with total spin $S=N/2$ spanned by the Dicke states $\{\ket{M}\}$ for $M=-N/2,\ldots, N/2$. The action of $U_A$ on this space is $U_A\ket{M}=e^{i\frac{\pi}{2} M(N+M)}\ket{M}$. For $N=2\ell$ (even), then up to a global phase the action on this space is $U=e^{(-1)^{\ell+1} i\frac{\pi}{4} (P^+-P^-)}$, where $P^{\pm}$ is the projector onto states with even(odd) Hamming weight. Writing the parity operator $\sigma_z^{\otimes N}=P^+-P^-$ we have $U_A=e^{(-1)^{\ell+1} i\frac{\pi}{4} \sigma_z^{\otimes N}}$. We see that applying the uniform Hadamard gate  afterward prepares the GHZ state up to a local phase gate 
\begin{equation}
\begin{array}{lll}
  H^{\otimes N}U_A\ket{+}^{\otimes N}
  &=& H^{\otimes N}U_A H^{\otimes N} \ket{0}^{\otimes N}\\
  &=&e^{(-1)^{\ell+1} i\frac{\pi}{4} \sigma_x^{\otimes N}}\ket{0}^{\otimes N}\\
  &=&(\ket{0\ldots 0}-i^{N+1}\ket{1\ldots 1})/\sqrt{2}. 
  \end{array}
\end{equation}
A similar argument follows for $N$ odd.

\section{Supporting calculations for protocol B}
\label{app:protB}
\subsection{Eigenenergies of $H$ in perturbation theory}
\label{app:protB_eigenenergies_in_perturbation_theory}
In this section we calculate perturbations of the eigenenergies of $H$ in the limit $\eta\rightarrow 0$. To find the eigenenergy for a computational basis state $\ket{q}$ with $n$ qubits in state $\ket{1}$ it is sufficient to consider the three states $\ket{0,q}$, $\ket{1,q}$ and $\ket{\chi}=\hat S^+\ket{0,q}/\sqrt{n}$, where the first entry in a ket vector denotes the number of excitations in the cavity mode, and the second entry denotes the state of the qubits. Projected onto these three states, $H$ is given by
\begin{eqnarray}
    H &=& \underbrace{\delta \ket{1,q}\bra{1,q} + \Delta\ket{\chi}\bra{\chi} + g\sqrt{N}(\ket{1,q}\bra{\chi} + \ket{\chi}\bra{1,q})}_{H_0} \nonumber\\  &+& \underbrace{i\eta \ket{1,q}\bra{0,q} -i\eta^*\ket{0,q}\bra{1,q}}_{V}
\end{eqnarray}
Denote by $\ket{p_\pm}$ the eigenstates of $H_0$ and by $E_\pm$ their corresponding energies. The second order perturbation of the eigenenergy of $\ket{0,q}$ is
\begin{eqnarray}
        \varepsilon_n &=& -\sum_j \frac{|\bra{0,q}V\ket{p_j}|^2}{E_j}\\ \nonumber
        &=& -\eta^2 \bra{1,q}H_0^{-1}\ket{1,q}\\ \nonumber
        &=& -\frac{|\eta|^2\Delta}{\Delta\delta - ng^2}
\end{eqnarray}

The corresponding eigenstate is
\begin{equation}
    \ket{\psi_q(t)} = \ket{0,q} -i\frac{\eta(t)\left(\Delta \ket{1,q} - g\sqrt{n}\ket{\chi}\right)}{\Delta\delta - ng^2}
\end{equation}

\subsection{Effect of losses}
\label{app:protB_effect_of_losses}
In this section we calculate process $\mathcal{E}$ of protocol B to first order in $\gamma$ and $\kappa$ in the adiabatic limit. We assume an initial state $\rho(0)= \ket{0,q}\bra{0,q'}$. Let $U(t)$ be unitary evolution in the absence of noise, and let $\tilde{\rho}(t) = U(t)^\dag \rho(t) U(t)$. Then
\begin{equation}
\begin{array}{lll}
    \dot{\tilde{\rho}} &=& -\frac{\gamma}{2}U^\dag \hat{n}_eU \tilde{\rho} -\frac{\gamma}{2} \tilde{\rho}  U^\dag \hat{n}_eU + \kappa U^\dag \ah U\tilde{\rho} U^\dag \ah^\dag U\\
    &-& \frac{\kappa}{2} U^\dag \ah^\dag \ah U \tilde{\rho} - \frac{\kappa}{2} \tilde{\rho} U^\dag \ah^\dag \ah U 
    \end{array}
\end{equation}
To first order in $\gamma$ and $\kappa$ we thus find using the adiabatic approximation $U(t)\ket{0,q} = e^{-i\varphi_n(t)}\ket{\psi_q(t)}$ with $\varphi_n(t)=\int_0^t \mathrm{d}t' \varepsilon_n(t')$ that
\begin{eqnarray}
    \tilde{\rho}(T) &=& \ket{0,q}\bra{0,q'}  \\ \nonumber
    &+& \int_0^T \mathrm{d}t\Big[ -\frac{1}{2}e^{-i\varphi_n(t)}U^\dag(t) (\gamma \hat{n}_e + \kappa \ah^\dag \ah)\ket{\psi_q(t)}\bra{0,q'}\\ \nonumber &-&\frac{1}{2}e^{i\varphi_m(t)} \ket{0,q}\bra{\psi_{q'}(t)} (\gamma \hat{n}_e + \kappa \ah^\dag \ah)U(t)  \\ \nonumber &+& \kappa e^{-i(\varphi_n(t)-\varphi_m(t))}U^\dag(t) \ah \ket{\psi_q(t)}\bra{\psi_{q'}(t)}\ah^\dag U(t) \Big]
\end{eqnarray}
We obtain
\begin{eqnarray}
    c_{nm}&:=&e^{i(\varphi_n(t)-\varphi_m(t))}\bra{q}\mathcal{E}(\ket{q}\bra{q'})\ket{q'}\\ \nonumber
    &=& \sum_{k=0}^\infty \bra{k,q}\tilde{\rho}(T)\ket{k,q'}
\end{eqnarray}
Up to second order in $\eta$ only terms with $k=0$ contribute, so we obtain
\begin{equation}
\begin{array}{lll}
 c_{nm}&=& 1 + \int_0^T \mathrm{d}t-[\frac{1}{2}\bra{\psi_q(t)}(\gamma \hat{n}_e + \kappa \ah^\dag \ah)\ket{\psi_q(t)} \\
 &-&  \frac{1}{2}\bra{\psi_q'(t)}(\gamma \hat{n}_e + \kappa \ah^\dag \ah)\ket{\psi_q'(t)}\\ 
 &+& \kappa \bra{\psi_q(t)}\ah\ket{\psi_q(t)}\bra{\psi_q'(t)}\ah^\dag\ket{\psi_q'(t)}\Big].
 \end{array}
\end{equation}

Using that $\bra{\psi_q}\hat{n}_e\ket{\psi_q} = |\eta|^2g^2n/(\Delta\delta-ng^2)^2$, $\bra{\psi_q}\ah^\dag \ah \ket{\psi_q} = |\eta|^2\Delta^2/(\Delta\delta-ng^2)^2$ and $\bra{\psi_q(t)}\ah\ket{\psi_q(t)} = -i\eta\Delta/(\Delta\delta-ng^2)$ we find
\begin{equation}
    c_{nm} = 1 - \frac{\gamma_n+\gamma_m}{2} - \frac{s_n^2+s_m^2-2s_ns_m}{2}
\end{equation}
with
\begin{equation}
     \gamma_n = \frac{\gamma n g^2}{(\Delta \delta -ng^2)^2}I \qquad
    s_n = \frac{\sqrt{\kappa} \Delta}{\Delta \delta -ng^2}\sqrt{I}.
\end{equation}

\subsection{Effects of coupling inhomogeneities on the fidelity}
\label{app:protB_effects_of_inhomogeneities}
Analogously to Sec.~\ref{app:protB_effects_of_inhomogeneities} we now calculate the effect of inhomogeneities in the coupling strength in protocol B.  We assume again that the $g_1,...,g_N$ are independent and identically distributed random variables and have the quadratic mean $\bar{g} = \sqrt{\E[g_j^2]}$.

Repeating the derivation in the main text with couplings $g_1,...,g_N$ which are different for each qubit gives a phase
\begin{equation}
\begin{array}{lll}
    \varphi_q &=& -\frac{I}{\delta - \frac{1}{\Delta}\sum_{j=1}^N q_jg_j^2}\\
    &\approx& -\frac{I}{\delta - n\bar{g}^2/\Delta} - \frac{I\Delta}{(\delta\Delta-n\bar{g}^2)^2}\sum_j q_j (g_j^2-\bar{g}^2)
    \end{array}
\end{equation}
which is accumulated when starting with the qubits in state $\ket{q}$ (for $q \in \{0,1\}^N$. Here, as in Sec.~\ref{app:protB_effects_of_inhomogeneities}, we use $n = \sum_j q_j$. Analogously to Eq.~\eqref{eq:protocol_A_inhomos_infidelity_bound} we obtain
\begin{equation}
    1-F \leq \frac{1}{2^N}\sum_{q\in\{0,1\}^N}\left[\frac{I\Delta}{(\delta\Delta-n\bar{g}^2)^2}\sum_j q_j (g_j^2-\bar{g}^2)\right]^2
\end{equation}
so that
\begin{equation}
    \E[1-F] \leq \mathrm{Var}[g_1^2] \frac{1}{2^N}\sum_{n=0}^N \binom{N}{n} n \left[\frac{I\Delta}{(\delta\Delta-n\bar{g}^2)^2}\right]^2
\end{equation}

\subsection{Using protocol B for arbitrary phase gates}
\label{app:protB_using_protB_for_arbitrary_phase_gates}
In this section we discuss how protocol B can be used to implement  phase gates $\exp(i\varphi(\hat{n}))$ for arbitrary $\boldsymbol{\varphi} = (\varphi(0),...,\varphi(N))$ (Here and in the following a bold font is used to indicate vector quantities). We aim to do this by applying $K$ pulses with detunings $\delta_1,...,\delta_K$ and $\Delta_1,...,\Delta_K$, as well as driving fields $\eta_1(t),...,\eta_K(t)$ and corresponding pulse energies $\boldsymbol{I} = (I_1,...,I_K)$ with $I_k = \int_0^{T_k}|\eta_k(t)|^2\mathrm{dt}$. With this, we implement a phase gate with $\boldsymbol{\varphi} = A\boldsymbol{I}$, where $A$ is a $(N+1)\times K$ matrix with
\begin{equation}
    A_{nk} = -\frac{1}{\delta_k-n g^2/\Delta_k}.
\end{equation}
By adding the infidelities of the individual pulses we obtain an average gate infidelity $\boldsymbol{b}\cdot\boldsymbol{I}$, where
\begin{equation}
    b_k = \frac{1}{2^N(2^{N+1})} \left(\sum_{n=0}^N \binom{N}{n} \epsilon_k^{(n,n)} +\sum_{n,m=0}^N \binom{N}{n}\binom{N}{m} \epsilon_k^{(n,m)}   \right)
    \label{eq:si_bk_linear_program}
\end{equation}
with
\begin{equation}
\begin{array}{lll}
    \epsilon_k^{(n,m)} &=& \gamma g^2 \left(\frac{n}{(\Delta_k\delta_k-ng^2)^2}+ \frac{m}{(\Delta_k\delta_k-mg^2)^2}\right)\\
    &+& \kappa\Delta^2 \left(\frac{1}{\Delta_k\delta_k-ng^2}- \frac{1}{\Delta_k\delta_k-ng^2}\right)^2.
    \end{array}
\end{equation}
Thus, for a fixed set of detunings $\delta_1,...,\delta_K$ and $\Delta_1,...,\Delta_K$, finding the optimal values of $\boldsymbol{I}$ becomes a linear programming problem:
\begin{align*}
    \text{Find}\quad & \boldsymbol{I}\\
    \text{ that minimizes} \quad& \boldsymbol{b}\cdot \boldsymbol{I}\\
    \text{ subject to }&\quad A\boldsymbol{I} = \boldsymbol{\varphi}\\
    \text{ and } &\quad\boldsymbol{I} \geq 0
\end{align*}
The solution to this linear program can be readily found using the simplex method, which is implemented in various software packages.

Since the solution of the given linear program is always on an extremal point of the simplex given by $\quad A\boldsymbol{I} = \boldsymbol{\varphi}$ and $\boldsymbol{I} \geq 0$, there are exactly $N+1$ indices $k$ such that $I_k \neq 0$. To find the optimal pulse detunings one can thus take the following approach: First take $K \gg N$ and take the $\delta_1,...,\delta_K$ to form a uniformly spaced grid. Take $\Delta_k-\delta_k$ some constant independent of $k$ to ensure that the different pulses can be implemented by only changing the pulse frequency, not the frequency of the cavity or the $\ket{1}\leftrightarrow\ket{e}$ transition. Now the linear program given above is solved, giving $N+1$ indices $k_1$,..., $k_{N+1}$ at which $I_k \neq 0$. To implement the phase gate given by $\boldsymbol{\varphi}$, $N+1$ pulses with detunings $\delta_{k_1}$,...,$\delta_{k_{N+1}}$ and $\Delta_{k_1}$,...,$\Delta_{k_{N+1}}$ as well as pulse energies $I_{k_1},...,I_{k_{N+1}}$ have to be applied.

A reduction to $N-1$ instead of $N+1$ required pulses is obtained if one only aims to implement the phase gate $\exp(i\varphi(\hat{n}))$ up to a global phase and single qubit gates. Formally, this means replacing the constraint $A\boldsymbol{I} = \boldsymbol{\varphi}$ by
\begin{equation}
    \forall n \geq 2 \qquad (A\boldsymbol{I})_n - n(A\boldsymbol{I})_1+(n-1)(A\boldsymbol{I})_0 = \varphi_n
\end{equation}
Since this condition is still linear in $\boldsymbol{I}$, the optimal $\boldsymbol{I}$ can be found as before through a linear program.

With the procedure outlined above, $\boldsymbol{I}$ is chosen to maximize the average gate fidelity. For implementing a C$_{N-1}$Z gate, our goal is instead to maximize the minimal fidelity. This is not possible with our linear programming approach in a straight forward manner, so we resort to a heuristic approach. For this,  we replace the $b_k$ (Eq.~\eqref{eq:si_bk_linear_program}) by
\begin{equation}
    b_k = \frac{1}{(N+1)^2}\sum_{n,m=0}^N \epsilon_k^{(n,n)}
\end{equation}
and solve the corresponding linear program. Compared to Eq.~\eqref{eq:si_bk_linear_program} this approach has the advantage that it weights the performance of the gate for all $n$ and $m$ equally, while Eq.~\eqref{eq:si_bk_linear_program} weights terms with $n, m\sim N/2$ higher than terms with extreme $n$ and $m$. The resulting $\boldsymbol{I}$ are then used to evaluate the minimal fidelity.

\bibliography{library}

\begin{thebibliography}{61}%
\makeatletter
\providecommand \@ifxundefined [1]{%
 \@ifx{#1\undefined}
}%
\providecommand \@ifnum [1]{%
 \ifnum #1\expandafter \@firstoftwo
 \else \expandafter \@secondoftwo
 \fi
}%
\providecommand \@ifx [1]{%
 \ifx #1\expandafter \@firstoftwo
 \else \expandafter \@secondoftwo
 \fi
}%
\providecommand \natexlab [1]{#1}%
\providecommand \enquote  [1]{``#1''}%
\providecommand \bibnamefont  [1]{#1}%
\providecommand \bibfnamefont [1]{#1}%
\providecommand \citenamefont [1]{#1}%
\providecommand \href@noop [0]{\@secondoftwo}%
\providecommand \href [0]{\begingroup \@sanitize@url \@href}%
\providecommand \@href[1]{\@@startlink{#1}\@@href}%
\providecommand \@@href[1]{\endgroup#1\@@endlink}%
\providecommand \@sanitize@url [0]{\catcode `\\12\catcode `\$12\catcode
  `\&12\catcode `\#12\catcode `\^12\catcode `\_12\catcode `\%12\relax}%
\providecommand \@@startlink[1]{}%
\providecommand \@@endlink[0]{}%
\providecommand \url  [0]{\begingroup\@sanitize@url \@url }%
\providecommand \@url [1]{\endgroup\@href {#1}{\urlprefix }}%
\providecommand \urlprefix  [0]{URL }%
\providecommand \Eprint [0]{\href }%
\providecommand \doibase [0]{https://doi.org/}%
\providecommand \selectlanguage [0]{\@gobble}%
\providecommand \bibinfo  [0]{\@secondoftwo}%
\providecommand \bibfield  [0]{\@secondoftwo}%
\providecommand \translation [1]{[#1]}%
\providecommand \BibitemOpen [0]{}%
\providecommand \bibitemStop [0]{}%
\providecommand \bibitemNoStop [0]{.\EOS\space}%
\providecommand \EOS [0]{\spacefactor3000\relax}%
\providecommand \BibitemShut  [1]{\csname bibitem#1\endcsname}%
\let\auto@bib@innerbib\@empty
\bibitem [{\citenamefont {Holmes}\ \emph {et~al.}(2020)\citenamefont {Holmes},
  \citenamefont {Johri}, \citenamefont {Guerreschi}, \citenamefont {Clarke},\
  and\ \citenamefont {Matsuura}}]{holmes_impact_2020}%
  \BibitemOpen
  \bibfield  {author} {\bibinfo {author} {\bibfnamefont {A.}~\bibnamefont
  {Holmes}}, \bibinfo {author} {\bibfnamefont {S.}~\bibnamefont {Johri}},
  \bibinfo {author} {\bibfnamefont {G.~G.}\ \bibnamefont {Guerreschi}},
  \bibinfo {author} {\bibfnamefont {J.~S.}\ \bibnamefont {Clarke}},\ and\
  \bibinfo {author} {\bibfnamefont {A.~Y.}\ \bibnamefont {Matsuura}},\
  }\bibfield  {title} {\bibinfo {title} {Impact of qubit connectivity on
  quantum algorithm performance},\ }\href
  {https://doi.org/10.1088/2058-9565/ab73e0} {\bibfield  {journal} {\bibinfo
  {journal} {Quantum Science and Technology}\ }\textbf {\bibinfo {volume}
  {5}},\ \bibinfo {pages} {025009} (\bibinfo {year} {2020})}\BibitemShut
  {NoStop}%
\bibitem [{\citenamefont {Gottesman}(2014)}]{gottesman_fault-tolerant_2014}%
  \BibitemOpen
  \bibfield  {author} {\bibinfo {author} {\bibfnamefont {D.}~\bibnamefont
  {Gottesman}},\ }\bibfield  {title} {\bibinfo {title} {Fault-{{Tolerant}}
  quantum computation with constant overhead},\ }\href
  {https://doi.org/10.26421/QIC14.15-16-5} {\bibfield  {journal} {\bibinfo
  {journal} {Quantum Information and Computation}\ }\textbf {\bibinfo {volume}
  {14}},\ \bibinfo {pages} {1339} (\bibinfo {year} {2014})}\BibitemShut
  {NoStop}%
\bibitem [{\citenamefont {Breuckmann}\ and\ \citenamefont
  {Eberhardt}(2021)}]{breuckmann_quantum_2021}%
  \BibitemOpen
  \bibfield  {author} {\bibinfo {author} {\bibfnamefont {N.~P.}\ \bibnamefont
  {Breuckmann}}\ and\ \bibinfo {author} {\bibfnamefont {J.~N.}\ \bibnamefont
  {Eberhardt}},\ }\bibfield  {title} {\bibinfo {title} {Quantum {{Low-Density
  Parity-Check Codes}}},\ }\href {https://doi.org/10.1103/PRXQuantum.2.040101}
  {\bibfield  {journal} {\bibinfo  {journal} {PRX Quantum}\ }\textbf {\bibinfo
  {volume} {2}},\ \bibinfo {pages} {040101} (\bibinfo {year}
  {2021})}\BibitemShut {NoStop}%
\bibitem [{\citenamefont {Cohen}\ \emph {et~al.}(2022)\citenamefont {Cohen},
  \citenamefont {Kim}, \citenamefont {Bartlett},\ and\ \citenamefont
  {Brown}}]{cohen_low-overhead_2022}%
  \BibitemOpen
  \bibfield  {author} {\bibinfo {author} {\bibfnamefont {L.~Z.}\ \bibnamefont
  {Cohen}}, \bibinfo {author} {\bibfnamefont {I.~H.}\ \bibnamefont {Kim}},
  \bibinfo {author} {\bibfnamefont {S.~D.}\ \bibnamefont {Bartlett}},\ and\
  \bibinfo {author} {\bibfnamefont {B.~J.}\ \bibnamefont {Brown}},\ }\bibfield
  {title} {\bibinfo {title} {Low-overhead fault-tolerant quantum computing
  using long-range connectivity},\ }\href
  {https://doi.org/10.1126/sciadv.abn1717} {\bibfield  {journal} {\bibinfo
  {journal} {Science Advances}\ }\textbf {\bibinfo {volume} {8}},\ \bibinfo
  {pages} {eabn1717} (\bibinfo {year} {2022})}\BibitemShut {NoStop}%
\bibitem [{\citenamefont {Bravyi}\ \emph {et~al.}(2023)\citenamefont {Bravyi},
  \citenamefont {Cross}, \citenamefont {Gambetta}, \citenamefont {Maslov},
  \citenamefont {Rall},\ and\ \citenamefont {Yoder}}]{bravyi2023highthreshold}%
  \BibitemOpen
  \bibfield  {author} {\bibinfo {author} {\bibfnamefont {S.}~\bibnamefont
  {Bravyi}}, \bibinfo {author} {\bibfnamefont {A.~W.}\ \bibnamefont {Cross}},
  \bibinfo {author} {\bibfnamefont {J.~M.}\ \bibnamefont {Gambetta}}, \bibinfo
  {author} {\bibfnamefont {D.}~\bibnamefont {Maslov}}, \bibinfo {author}
  {\bibfnamefont {P.}~\bibnamefont {Rall}},\ and\ \bibinfo {author}
  {\bibfnamefont {T.~J.}\ \bibnamefont {Yoder}},\ }\href@noop {} {\bibinfo
  {title} {High-threshold and low-overhead fault-tolerant quantum memory}}
  (\bibinfo {year} {2023}),\ \Eprint {https://arxiv.org/abs/2308.07915}
  {arXiv:2308.07915 [quant-ph]} \BibitemShut {NoStop}%
\bibitem [{\citenamefont {Fowler}\ \emph {et~al.}(2012)\citenamefont {Fowler},
  \citenamefont {Mariantoni}, \citenamefont {Martinis},\ and\ \citenamefont
  {Cleland}}]{fowler_surface_2012}%
  \BibitemOpen
  \bibfield  {author} {\bibinfo {author} {\bibfnamefont {A.~G.}\ \bibnamefont
  {Fowler}}, \bibinfo {author} {\bibfnamefont {M.}~\bibnamefont {Mariantoni}},
  \bibinfo {author} {\bibfnamefont {J.~M.}\ \bibnamefont {Martinis}},\ and\
  \bibinfo {author} {\bibfnamefont {A.~N.}\ \bibnamefont {Cleland}},\
  }\bibfield  {title} {\bibinfo {title} {Surface codes: {{Towards}} practical
  large-scale quantum computation},\ }\href
  {https://doi.org/10.1103/PhysRevA.86.032324} {\bibfield  {journal} {\bibinfo
  {journal} {Physical Review A}\ }\textbf {\bibinfo {volume} {86}},\ \bibinfo
  {pages} {032324} (\bibinfo {year} {2012})}\BibitemShut {NoStop}%
\bibitem [{\citenamefont {Pino}\ \emph {et~al.}(2021)\citenamefont {Pino},
  \citenamefont {Dreiling}, \citenamefont {Figgatt}, \citenamefont {Gaebler},
  \citenamefont {Moses}, \citenamefont {Allman}, \citenamefont {Baldwin},
  \citenamefont {{Foss-Feig}}, \citenamefont {Hayes}, \citenamefont {Mayer},
  \citenamefont {{Ryan-Anderson}},\ and\ \citenamefont
  {Neyenhuis}}]{pino_demonstration_2021}%
  \BibitemOpen
  \bibfield  {author} {\bibinfo {author} {\bibfnamefont {J.~M.}\ \bibnamefont
  {Pino}}, \bibinfo {author} {\bibfnamefont {J.~M.}\ \bibnamefont {Dreiling}},
  \bibinfo {author} {\bibfnamefont {C.}~\bibnamefont {Figgatt}}, \bibinfo
  {author} {\bibfnamefont {J.~P.}\ \bibnamefont {Gaebler}}, \bibinfo {author}
  {\bibfnamefont {S.~A.}\ \bibnamefont {Moses}}, \bibinfo {author}
  {\bibfnamefont {M.~S.}\ \bibnamefont {Allman}}, \bibinfo {author}
  {\bibfnamefont {C.~H.}\ \bibnamefont {Baldwin}}, \bibinfo {author}
  {\bibfnamefont {M.}~\bibnamefont {{Foss-Feig}}}, \bibinfo {author}
  {\bibfnamefont {D.}~\bibnamefont {Hayes}}, \bibinfo {author} {\bibfnamefont
  {K.}~\bibnamefont {Mayer}}, \bibinfo {author} {\bibfnamefont
  {C.}~\bibnamefont {{Ryan-Anderson}}},\ and\ \bibinfo {author} {\bibfnamefont
  {B.}~\bibnamefont {Neyenhuis}},\ }\bibfield  {title} {\bibinfo {title}
  {Demonstration of the trapped-ion quantum {{CCD}} computer architecture},\
  }\href {https://doi.org/10.1038/s41586-021-03318-4} {\bibfield  {journal}
  {\bibinfo  {journal} {Nature}\ }\textbf {\bibinfo {volume} {592}},\ \bibinfo
  {pages} {209} (\bibinfo {year} {2021})}\BibitemShut {NoStop}%
\bibitem [{\citenamefont {Beugnon}\ \emph {et~al.}(2007)\citenamefont
  {Beugnon}, \citenamefont {Tuchendler}, \citenamefont {Marion}, \citenamefont
  {Ga{\"e}tan}, \citenamefont {Miroshnychenko}, \citenamefont {Sortais},
  \citenamefont {Lance}, \citenamefont {Jones}, \citenamefont {Messin},
  \citenamefont {Browaeys},\ and\ \citenamefont
  {Grangier}}]{beugnon_two-dimensional_2007}%
  \BibitemOpen
  \bibfield  {author} {\bibinfo {author} {\bibfnamefont {J.}~\bibnamefont
  {Beugnon}}, \bibinfo {author} {\bibfnamefont {C.}~\bibnamefont {Tuchendler}},
  \bibinfo {author} {\bibfnamefont {H.}~\bibnamefont {Marion}}, \bibinfo
  {author} {\bibfnamefont {A.}~\bibnamefont {Ga{\"e}tan}}, \bibinfo {author}
  {\bibfnamefont {Y.}~\bibnamefont {Miroshnychenko}}, \bibinfo {author}
  {\bibfnamefont {Y.~R.~P.}\ \bibnamefont {Sortais}}, \bibinfo {author}
  {\bibfnamefont {A.~M.}\ \bibnamefont {Lance}}, \bibinfo {author}
  {\bibfnamefont {M.~P.~A.}\ \bibnamefont {Jones}}, \bibinfo {author}
  {\bibfnamefont {G.}~\bibnamefont {Messin}}, \bibinfo {author} {\bibfnamefont
  {A.}~\bibnamefont {Browaeys}},\ and\ \bibinfo {author} {\bibfnamefont
  {P.}~\bibnamefont {Grangier}},\ }\bibfield  {title} {\bibinfo {title}
  {Two-dimensional transport and transfer of a single atomic qubit in optical
  tweezers},\ }\href {https://doi.org/10.1038/nphys698} {\bibfield  {journal}
  {\bibinfo  {journal} {Nature Physics}\ }\textbf {\bibinfo {volume} {3}},\
  \bibinfo {pages} {696} (\bibinfo {year} {2007})}\BibitemShut {NoStop}%
\bibitem [{\citenamefont {Bluvstein}\ \emph {et~al.}(2022)\citenamefont
  {Bluvstein}, \citenamefont {Levine}, \citenamefont {Semeghini}, \citenamefont
  {Wang}, \citenamefont {Ebadi}, \citenamefont {Kalinowski}, \citenamefont
  {Keesling}, \citenamefont {Maskara}, \citenamefont {Pichler}, \citenamefont
  {Greiner}, \citenamefont {Vuleti{\'c}},\ and\ \citenamefont
  {Lukin}}]{bluvstein_quantum_2022}%
  \BibitemOpen
  \bibfield  {author} {\bibinfo {author} {\bibfnamefont {D.}~\bibnamefont
  {Bluvstein}}, \bibinfo {author} {\bibfnamefont {H.}~\bibnamefont {Levine}},
  \bibinfo {author} {\bibfnamefont {G.}~\bibnamefont {Semeghini}}, \bibinfo
  {author} {\bibfnamefont {T.~T.}\ \bibnamefont {Wang}}, \bibinfo {author}
  {\bibfnamefont {S.}~\bibnamefont {Ebadi}}, \bibinfo {author} {\bibfnamefont
  {M.}~\bibnamefont {Kalinowski}}, \bibinfo {author} {\bibfnamefont
  {A.}~\bibnamefont {Keesling}}, \bibinfo {author} {\bibfnamefont
  {N.}~\bibnamefont {Maskara}}, \bibinfo {author} {\bibfnamefont
  {H.}~\bibnamefont {Pichler}}, \bibinfo {author} {\bibfnamefont
  {M.}~\bibnamefont {Greiner}}, \bibinfo {author} {\bibfnamefont
  {V.}~\bibnamefont {Vuleti{\'c}}},\ and\ \bibinfo {author} {\bibfnamefont
  {M.~D.}\ \bibnamefont {Lukin}},\ }\bibfield  {title} {\bibinfo {title} {A
  quantum processor based on coherent transport of entangled atom arrays},\
  }\href {https://doi.org/10.1038/s41586-022-04592-6} {\bibfield  {journal}
  {\bibinfo  {journal} {Nature}\ }\textbf {\bibinfo {volume} {604}},\ \bibinfo
  {pages} {451} (\bibinfo {year} {2022})}\BibitemShut {NoStop}%
\bibitem [{\citenamefont {Xu}\ \emph {et~al.}(2023)\citenamefont {Xu},
  \citenamefont {Ataides}, \citenamefont {Pattison}, \citenamefont
  {Raveendran}, \citenamefont {Bluvstein}, \citenamefont {Wurtz}, \citenamefont
  {Vasic}, \citenamefont {Lukin}, \citenamefont {Jiang},\ and\ \citenamefont
  {Zhou}}]{xu2023constantoverhead}%
  \BibitemOpen
  \bibfield  {author} {\bibinfo {author} {\bibfnamefont {Q.}~\bibnamefont
  {Xu}}, \bibinfo {author} {\bibfnamefont {J.~P.~B.}\ \bibnamefont {Ataides}},
  \bibinfo {author} {\bibfnamefont {C.~A.}\ \bibnamefont {Pattison}}, \bibinfo
  {author} {\bibfnamefont {N.}~\bibnamefont {Raveendran}}, \bibinfo {author}
  {\bibfnamefont {D.}~\bibnamefont {Bluvstein}}, \bibinfo {author}
  {\bibfnamefont {J.}~\bibnamefont {Wurtz}}, \bibinfo {author} {\bibfnamefont
  {B.}~\bibnamefont {Vasic}}, \bibinfo {author} {\bibfnamefont {M.~D.}\
  \bibnamefont {Lukin}}, \bibinfo {author} {\bibfnamefont {L.}~\bibnamefont
  {Jiang}},\ and\ \bibinfo {author} {\bibfnamefont {H.}~\bibnamefont {Zhou}},\
  }\href@noop {} {\bibinfo {title} {Constant-overhead fault-tolerant quantum
  computation with reconfigurable atom arrays}} (\bibinfo {year} {2023}),\
  \Eprint {https://arxiv.org/abs/2308.08648} {arXiv:2308.08648 [quant-ph]}
  \BibitemShut {NoStop}%
\bibitem [{\citenamefont {Cirac}\ and\ \citenamefont
  {Zoller}(1995)}]{cirac_quantum_1995}%
  \BibitemOpen
  \bibfield  {author} {\bibinfo {author} {\bibfnamefont {J.~I.}\ \bibnamefont
  {Cirac}}\ and\ \bibinfo {author} {\bibfnamefont {P.}~\bibnamefont {Zoller}},\
  }\bibfield  {title} {\bibinfo {title} {Quantum {{Computations}} with {{Cold
  Trapped Ions}}},\ }\href {https://doi.org/10.1103/PhysRevLett.74.4091}
  {\bibfield  {journal} {\bibinfo  {journal} {Physical Review Letters}\
  }\textbf {\bibinfo {volume} {74}},\ \bibinfo {pages} {4091} (\bibinfo {year}
  {1995})}\BibitemShut {NoStop}%
\bibitem [{\citenamefont {M{\o}lmer}\ and\ \citenamefont
  {S{\o}rensen}(1999)}]{molmer_multiparticle_1999}%
  \BibitemOpen
  \bibfield  {author} {\bibinfo {author} {\bibfnamefont {K.}~\bibnamefont
  {M{\o}lmer}}\ and\ \bibinfo {author} {\bibfnamefont {A.}~\bibnamefont
  {S{\o}rensen}},\ }\bibfield  {title} {\bibinfo {title} {Multiparticle
  {{Entanglement}} of {{Hot Trapped Ions}}},\ }\href
  {https://doi.org/10.1103/PhysRevLett.82.1835} {\bibfield  {journal} {\bibinfo
   {journal} {Physical Review Letters}\ }\textbf {\bibinfo {volume} {82}},\
  \bibinfo {pages} {1835} (\bibinfo {year} {1999})}\BibitemShut {NoStop}%
\bibitem [{\citenamefont {Sackett}\ \emph {et~al.}(2000)\citenamefont
  {Sackett}, \citenamefont {Kielpinski}, \citenamefont {King}, \citenamefont
  {Langer}, \citenamefont {Meyer}, \citenamefont {Myatt}, \citenamefont {Rowe},
  \citenamefont {Turchette}, \citenamefont {Itano}, \citenamefont {Wineland},\
  and\ \citenamefont {Monroe}}]{sackett_experimental_2000}%
  \BibitemOpen
  \bibfield  {author} {\bibinfo {author} {\bibfnamefont {C.~A.}\ \bibnamefont
  {Sackett}}, \bibinfo {author} {\bibfnamefont {D.}~\bibnamefont {Kielpinski}},
  \bibinfo {author} {\bibfnamefont {B.~E.}\ \bibnamefont {King}}, \bibinfo
  {author} {\bibfnamefont {C.}~\bibnamefont {Langer}}, \bibinfo {author}
  {\bibfnamefont {V.}~\bibnamefont {Meyer}}, \bibinfo {author} {\bibfnamefont
  {C.~J.}\ \bibnamefont {Myatt}}, \bibinfo {author} {\bibfnamefont
  {M.}~\bibnamefont {Rowe}}, \bibinfo {author} {\bibfnamefont {Q.~A.}\
  \bibnamefont {Turchette}}, \bibinfo {author} {\bibfnamefont {W.~M.}\
  \bibnamefont {Itano}}, \bibinfo {author} {\bibfnamefont {D.~J.}\ \bibnamefont
  {Wineland}},\ and\ \bibinfo {author} {\bibfnamefont {C.}~\bibnamefont
  {Monroe}},\ }\bibfield  {title} {\bibinfo {title} {Experimental entanglement
  of four particles},\ }\href {https://doi.org/10.1038/35005011} {\bibfield
  {journal} {\bibinfo  {journal} {Nature}\ }\textbf {\bibinfo {volume} {404}},\
  \bibinfo {pages} {256} (\bibinfo {year} {2000})}\BibitemShut {NoStop}%
\bibitem [{\citenamefont {{Garc{\'i}a-Ripoll}}\ \emph
  {et~al.}(2003)\citenamefont {{Garc{\'i}a-Ripoll}}, \citenamefont {Zoller},\
  and\ \citenamefont {Cirac}}]{garcia-ripoll_speed_2003}%
  \BibitemOpen
  \bibfield  {author} {\bibinfo {author} {\bibfnamefont {J.~J.}\ \bibnamefont
  {{Garc{\'i}a-Ripoll}}}, \bibinfo {author} {\bibfnamefont {P.}~\bibnamefont
  {Zoller}},\ and\ \bibinfo {author} {\bibfnamefont {J.~I.}\ \bibnamefont
  {Cirac}},\ }\bibfield  {title} {\bibinfo {title} {Speed {{Optimized Two-Qubit
  Gates}} with {{Laser Coherent Control Techniques}} for {{Ion Trap Quantum
  Computing}}},\ }\href {https://doi.org/10.1103/PhysRevLett.91.157901}
  {\bibfield  {journal} {\bibinfo  {journal} {Physical Review Letters}\
  }\textbf {\bibinfo {volume} {91}},\ \bibinfo {pages} {157901} (\bibinfo
  {year} {2003})}\BibitemShut {NoStop}%
\bibitem [{\citenamefont {Pellizzari}\ \emph {et~al.}(1995)\citenamefont
  {Pellizzari}, \citenamefont {Gardiner}, \citenamefont {Cirac},\ and\
  \citenamefont {Zoller}}]{pellizzari_decoherence_1995}%
  \BibitemOpen
  \bibfield  {author} {\bibinfo {author} {\bibfnamefont {T.}~\bibnamefont
  {Pellizzari}}, \bibinfo {author} {\bibfnamefont {S.~A.}\ \bibnamefont
  {Gardiner}}, \bibinfo {author} {\bibfnamefont {J.~I.}\ \bibnamefont
  {Cirac}},\ and\ \bibinfo {author} {\bibfnamefont {P.}~\bibnamefont
  {Zoller}},\ }\bibfield  {title} {\bibinfo {title} {Decoherence, {{Continuous
  Observation}}, and {{Quantum Computing}}: {{A Cavity QED Model}}},\ }\href
  {https://doi.org/10.1103/PhysRevLett.75.3788} {\bibfield  {journal} {\bibinfo
   {journal} {Physical Review Letters}\ }\textbf {\bibinfo {volume} {75}},\
  \bibinfo {pages} {3788} (\bibinfo {year} {1995})}\BibitemShut {NoStop}%
\bibitem [{\citenamefont {Zheng}\ and\ \citenamefont
  {Guo}(2000)}]{zheng_efficient_2000}%
  \BibitemOpen
  \bibfield  {author} {\bibinfo {author} {\bibfnamefont {S.-B.}\ \bibnamefont
  {Zheng}}\ and\ \bibinfo {author} {\bibfnamefont {G.-C.}\ \bibnamefont
  {Guo}},\ }\bibfield  {title} {\bibinfo {title} {Efficient {{Scheme}} for
  {{Two-Atom Entanglement}} and {{Quantum Information Processing}} in {{Cavity
  QED}}},\ }\href {https://doi.org/10.1103/PhysRevLett.85.2392} {\bibfield
  {journal} {\bibinfo  {journal} {Physical Review Letters}\ }\textbf {\bibinfo
  {volume} {85}},\ \bibinfo {pages} {2392} (\bibinfo {year}
  {2000})}\BibitemShut {NoStop}%
\bibitem [{\citenamefont {Beige}\ \emph {et~al.}(2000)\citenamefont {Beige},
  \citenamefont {Braun}, \citenamefont {Tregenna},\ and\ \citenamefont
  {Knight}}]{beige_quantum_2000}%
  \BibitemOpen
  \bibfield  {author} {\bibinfo {author} {\bibfnamefont {A.}~\bibnamefont
  {Beige}}, \bibinfo {author} {\bibfnamefont {D.}~\bibnamefont {Braun}},
  \bibinfo {author} {\bibfnamefont {B.}~\bibnamefont {Tregenna}},\ and\
  \bibinfo {author} {\bibfnamefont {P.~L.}\ \bibnamefont {Knight}},\ }\bibfield
   {title} {\bibinfo {title} {Quantum {{Computing Using Dissipation}} to
  {{Remain}} in a {{Decoherence-Free Subspace}}},\ }\href
  {https://doi.org/10.1103/PhysRevLett.85.1762} {\bibfield  {journal} {\bibinfo
   {journal} {Physical Review Letters}\ }\textbf {\bibinfo {volume} {85}},\
  \bibinfo {pages} {1762} (\bibinfo {year} {2000})}\BibitemShut {NoStop}%
\bibitem [{\citenamefont {S{\o}rensen}\ and\ \citenamefont
  {M{\o}lmer}(2003)}]{sorensen_measurement_2003}%
  \BibitemOpen
  \bibfield  {author} {\bibinfo {author} {\bibfnamefont {A.~S.}\ \bibnamefont
  {S{\o}rensen}}\ and\ \bibinfo {author} {\bibfnamefont {K.}~\bibnamefont
  {M{\o}lmer}},\ }\bibfield  {title} {\bibinfo {title} {Measurement {{Induced
  Entanglement}} and {{Quantum Computation}} with {{Atoms}} in {{Optical
  Cavities}}},\ }\href {https://doi.org/10.1103/PhysRevLett.91.097905}
  {\bibfield  {journal} {\bibinfo  {journal} {Physical Review Letters}\
  }\textbf {\bibinfo {volume} {91}},\ \bibinfo {pages} {097905} (\bibinfo
  {year} {2003})}\BibitemShut {NoStop}%
\bibitem [{\citenamefont {Zheng}(2004)}]{zheng_unconventional_2004}%
  \BibitemOpen
  \bibfield  {author} {\bibinfo {author} {\bibfnamefont {S.-B.}\ \bibnamefont
  {Zheng}},\ }\bibfield  {title} {\bibinfo {title} {Unconventional geometric
  quantum phase gates with a cavity {{QED}} system},\ }\href
  {https://doi.org/10.1103/PhysRevA.70.052320} {\bibfield  {journal} {\bibinfo
  {journal} {Physical Review A}\ }\textbf {\bibinfo {volume} {70}},\ \bibinfo
  {pages} {052320} (\bibinfo {year} {2004})}\BibitemShut {NoStop}%
\bibitem [{\citenamefont {Borregaard}\ \emph {et~al.}(2015)\citenamefont
  {Borregaard}, \citenamefont {K{\'o}m{\'a}r}, \citenamefont {Kessler},
  \citenamefont {S{\o}rensen},\ and\ \citenamefont
  {Lukin}}]{borregaard_heralded_2015}%
  \BibitemOpen
  \bibfield  {author} {\bibinfo {author} {\bibfnamefont {J.}~\bibnamefont
  {Borregaard}}, \bibinfo {author} {\bibfnamefont {P.}~\bibnamefont
  {K{\'o}m{\'a}r}}, \bibinfo {author} {\bibfnamefont {E.~M.}\ \bibnamefont
  {Kessler}}, \bibinfo {author} {\bibfnamefont {A.~S.}\ \bibnamefont
  {S{\o}rensen}},\ and\ \bibinfo {author} {\bibfnamefont {M.~D.}\ \bibnamefont
  {Lukin}},\ }\bibfield  {title} {\bibinfo {title} {Heralded {{Quantum Gates}}
  with {{Integrated Error Detection}} in {{Optical Cavities}}},\ }\href
  {https://doi.org/10.1103/PhysRevLett.114.110502} {\bibfield  {journal}
  {\bibinfo  {journal} {Physical Review Letters}\ }\textbf {\bibinfo {volume}
  {114}},\ \bibinfo {pages} {110502} (\bibinfo {year} {2015})}\BibitemShut
  {NoStop}%
\bibitem [{\citenamefont {Ramette}\ \emph {et~al.}(2021)\citenamefont
  {Ramette}, \citenamefont {Sinclair}, \citenamefont {Vendeiro}, \citenamefont
  {Rudelis}, \citenamefont {Cetina},\ and\ \citenamefont
  {Vuleti{\'c}}}]{ramette_any--any_2021}%
  \BibitemOpen
  \bibfield  {author} {\bibinfo {author} {\bibfnamefont {J.}~\bibnamefont
  {Ramette}}, \bibinfo {author} {\bibfnamefont {J.}~\bibnamefont {Sinclair}},
  \bibinfo {author} {\bibfnamefont {Z.}~\bibnamefont {Vendeiro}}, \bibinfo
  {author} {\bibfnamefont {A.}~\bibnamefont {Rudelis}}, \bibinfo {author}
  {\bibfnamefont {M.}~\bibnamefont {Cetina}},\ and\ \bibinfo {author}
  {\bibfnamefont {V.}~\bibnamefont {Vuleti{\'c}}},\ }\bibfield  {title}
  {\bibinfo {title} {Any-to-any connected cavity-mediated architecture for
  quantum computing with trapped ions or {{Rydberg}} arrays},\ }\href
  {http://arxiv.org/abs/2109.11551} {\bibfield  {journal} {\bibinfo  {journal}
  {arXiv:2109.11551 [physics, physics:quant-ph]}\ } (\bibinfo {year}
  {2021})}\BibitemShut {NoStop}%
\bibitem [{\citenamefont {Lewalle}\ \emph {et~al.}(2022)\citenamefont
  {Lewalle}, \citenamefont {Martin}, \citenamefont {Flurin}, \citenamefont
  {Zhang}, \citenamefont {Blumenthal}, \citenamefont {{Hacohen-Gourgy}},
  \citenamefont {Burgarth},\ and\ \citenamefont
  {Whaley}}]{lewalle_multi-qubit_2022}%
  \BibitemOpen
  \bibfield  {author} {\bibinfo {author} {\bibfnamefont {P.}~\bibnamefont
  {Lewalle}}, \bibinfo {author} {\bibfnamefont {L.~S.}\ \bibnamefont {Martin}},
  \bibinfo {author} {\bibfnamefont {E.}~\bibnamefont {Flurin}}, \bibinfo
  {author} {\bibfnamefont {S.}~\bibnamefont {Zhang}}, \bibinfo {author}
  {\bibfnamefont {E.}~\bibnamefont {Blumenthal}}, \bibinfo {author}
  {\bibfnamefont {S.}~\bibnamefont {{Hacohen-Gourgy}}}, \bibinfo {author}
  {\bibfnamefont {D.}~\bibnamefont {Burgarth}},\ and\ \bibinfo {author}
  {\bibfnamefont {K.~B.}\ \bibnamefont {Whaley}},\ }\bibfield  {title}
  {\bibinfo {title} {A {{Multi-Qubit Quantum Gate Using}} the {{Zeno
  Effect}}},\ }\href {http://arxiv.org/abs/2211.05988} {\bibfield  {journal}
  {\bibinfo  {journal} {arXiv:2211.05988}\ } (\bibinfo {year}
  {2022})}\BibitemShut {NoStop}%
\bibitem [{\citenamefont {Kastoryano}\ \emph {et~al.}(2011)\citenamefont
  {Kastoryano}, \citenamefont {Reiter},\ and\ \citenamefont
  {S{\o}rensen}}]{kastoryano_dissipative_2011}%
  \BibitemOpen
  \bibfield  {author} {\bibinfo {author} {\bibfnamefont {M.~J.}\ \bibnamefont
  {Kastoryano}}, \bibinfo {author} {\bibfnamefont {F.}~\bibnamefont {Reiter}},\
  and\ \bibinfo {author} {\bibfnamefont {A.~S.}\ \bibnamefont {S{\o}rensen}},\
  }\bibfield  {title} {\bibinfo {title} {Dissipative {{Preparation}} of
  {{Entanglement}} in {{Optical Cavities}}},\ }\href
  {https://doi.org/10.1103/PhysRevLett.106.090502} {\bibfield  {journal}
  {\bibinfo  {journal} {Physical Review Letters}\ }\textbf {\bibinfo {volume}
  {106}},\ \bibinfo {pages} {090502} (\bibinfo {year} {2011})}\BibitemShut
  {NoStop}%
\bibitem [{\citenamefont {Hein}\ \emph {et~al.}(2004)\citenamefont {Hein},
  \citenamefont {Eisert},\ and\ \citenamefont
  {Briegel}}]{hein_multi-party_2004}%
  \BibitemOpen
  \bibfield  {author} {\bibinfo {author} {\bibfnamefont {M.}~\bibnamefont
  {Hein}}, \bibinfo {author} {\bibfnamefont {J.}~\bibnamefont {Eisert}},\ and\
  \bibinfo {author} {\bibfnamefont {H.~J.}\ \bibnamefont {Briegel}},\
  }\bibfield  {title} {\bibinfo {title} {Multi-party entanglement in graph
  states},\ }\href {https://doi.org/10.1103/PhysRevA.69.062311} {\bibfield
  {journal} {\bibinfo  {journal} {Physical Review A}\ }\textbf {\bibinfo
  {volume} {69}},\ \bibinfo {pages} {062311} (\bibinfo {year}
  {2004})}\BibitemShut {NoStop}%
\bibitem [{\citenamefont {Leibfried}\ \emph {et~al.}(2004)\citenamefont
  {Leibfried}, \citenamefont {Barrett}, \citenamefont {Schaetz}, \citenamefont
  {Britton}, \citenamefont {Chiaverini}, \citenamefont {Itano}, \citenamefont
  {Jost}, \citenamefont {Langer},\ and\ \citenamefont
  {Wineland}}]{leibfried_toward_2004}%
  \BibitemOpen
  \bibfield  {author} {\bibinfo {author} {\bibfnamefont {D.}~\bibnamefont
  {Leibfried}}, \bibinfo {author} {\bibfnamefont {M.~D.}\ \bibnamefont
  {Barrett}}, \bibinfo {author} {\bibfnamefont {T.}~\bibnamefont {Schaetz}},
  \bibinfo {author} {\bibfnamefont {J.}~\bibnamefont {Britton}}, \bibinfo
  {author} {\bibfnamefont {J.}~\bibnamefont {Chiaverini}}, \bibinfo {author}
  {\bibfnamefont {W.~M.}\ \bibnamefont {Itano}}, \bibinfo {author}
  {\bibfnamefont {J.~D.}\ \bibnamefont {Jost}}, \bibinfo {author}
  {\bibfnamefont {C.}~\bibnamefont {Langer}},\ and\ \bibinfo {author}
  {\bibfnamefont {D.~J.}\ \bibnamefont {Wineland}},\ }\bibfield  {title}
  {\bibinfo {title} {Toward {{Heisenberg-Limited Spectroscopy}} with
  {{Multiparticle Entangled States}}},\ }\href
  {https://doi.org/10.1126/science.1097576} {\bibfield  {journal} {\bibinfo
  {journal} {Science}\ }\textbf {\bibinfo {volume} {304}},\ \bibinfo {pages}
  {1476} (\bibinfo {year} {2004})}\BibitemShut {NoStop}%
\bibitem [{\citenamefont {Zhao}\ \emph {et~al.}(2021)\citenamefont {Zhao},
  \citenamefont {Zhang}, \citenamefont {Chen}, \citenamefont {Wang},\ and\
  \citenamefont {Hu}}]{zhao_creation_2021}%
  \BibitemOpen
  \bibfield  {author} {\bibinfo {author} {\bibfnamefont {Y.}~\bibnamefont
  {Zhao}}, \bibinfo {author} {\bibfnamefont {R.}~\bibnamefont {Zhang}},
  \bibinfo {author} {\bibfnamefont {W.}~\bibnamefont {Chen}}, \bibinfo {author}
  {\bibfnamefont {X.-B.}\ \bibnamefont {Wang}},\ and\ \bibinfo {author}
  {\bibfnamefont {J.}~\bibnamefont {Hu}},\ }\bibfield  {title} {\bibinfo
  {title} {Creation of {{Greenberger-Horne-Zeilinger}} states with thousands of
  atoms by entanglement amplification},\ }\href
  {https://doi.org/10.1038/s41534-021-00364-8} {\bibfield  {journal} {\bibinfo
  {journal} {npj Quantum Information}\ }\textbf {\bibinfo {volume} {7}},\
  \bibinfo {pages} {24} (\bibinfo {year} {2021})}\BibitemShut {NoStop}%
\bibitem [{\citenamefont {Barkoutsos}\ \emph {et~al.}(2018)\citenamefont
  {Barkoutsos}, \citenamefont {Gonthier}, \citenamefont {Sokolov},
  \citenamefont {Moll}, \citenamefont {Salis}, \citenamefont {Fuhrer},
  \citenamefont {Ganzhorn}, \citenamefont {Egger}, \citenamefont {Troyer},
  \citenamefont {Mezzacapo}, \citenamefont {Filipp},\ and\ \citenamefont
  {Tavernelli}}]{barkoutsos_quantum_2018}%
  \BibitemOpen
  \bibfield  {author} {\bibinfo {author} {\bibfnamefont {P.~K.}\ \bibnamefont
  {Barkoutsos}}, \bibinfo {author} {\bibfnamefont {J.~F.}\ \bibnamefont
  {Gonthier}}, \bibinfo {author} {\bibfnamefont {I.}~\bibnamefont {Sokolov}},
  \bibinfo {author} {\bibfnamefont {N.}~\bibnamefont {Moll}}, \bibinfo {author}
  {\bibfnamefont {G.}~\bibnamefont {Salis}}, \bibinfo {author} {\bibfnamefont
  {A.}~\bibnamefont {Fuhrer}}, \bibinfo {author} {\bibfnamefont
  {M.}~\bibnamefont {Ganzhorn}}, \bibinfo {author} {\bibfnamefont {D.~J.}\
  \bibnamefont {Egger}}, \bibinfo {author} {\bibfnamefont {M.}~\bibnamefont
  {Troyer}}, \bibinfo {author} {\bibfnamefont {A.}~\bibnamefont {Mezzacapo}},
  \bibinfo {author} {\bibfnamefont {S.}~\bibnamefont {Filipp}},\ and\ \bibinfo
  {author} {\bibfnamefont {I.}~\bibnamefont {Tavernelli}},\ }\bibfield  {title}
  {\bibinfo {title} {Quantum algorithms for electronic structure calculations:
  {{Particle-hole Hamiltonian}} and optimized wave-function expansions},\
  }\href {https://doi.org/10.1103/PhysRevA.98.022322} {\bibfield  {journal}
  {\bibinfo  {journal} {Physical Review A}\ }\textbf {\bibinfo {volume} {98}},\
  \bibinfo {pages} {022322} (\bibinfo {year} {2018})}\BibitemShut {NoStop}%
\bibitem [{\citenamefont {Grimsley}\ \emph {et~al.}(2019)\citenamefont
  {Grimsley}, \citenamefont {Economou}, \citenamefont {Barnes},\ and\
  \citenamefont {Mayhall}}]{grimsley_adaptive_2019}%
  \BibitemOpen
  \bibfield  {author} {\bibinfo {author} {\bibfnamefont {H.~R.}\ \bibnamefont
  {Grimsley}}, \bibinfo {author} {\bibfnamefont {S.~E.}\ \bibnamefont
  {Economou}}, \bibinfo {author} {\bibfnamefont {E.}~\bibnamefont {Barnes}},\
  and\ \bibinfo {author} {\bibfnamefont {N.~J.}\ \bibnamefont {Mayhall}},\
  }\bibfield  {title} {\bibinfo {title} {An adaptive variational algorithm for
  exact molecular simulations on a quantum computer},\ }\href
  {https://doi.org/10.1038/s41467-019-10988-2} {\bibfield  {journal} {\bibinfo
  {journal} {Nature Communications}\ }\textbf {\bibinfo {volume} {10}},\
  \bibinfo {pages} {3007} (\bibinfo {year} {2019})}\BibitemShut {NoStop}%
\bibitem [{\citenamefont {{Paz-Silva}}\ \emph {et~al.}(2010)\citenamefont
  {{Paz-Silva}}, \citenamefont {Brennen},\ and\ \citenamefont
  {Twamley}}]{paz-silva_fault_2010}%
  \BibitemOpen
  \bibfield  {author} {\bibinfo {author} {\bibfnamefont {G.~A.}\ \bibnamefont
  {{Paz-Silva}}}, \bibinfo {author} {\bibfnamefont {G.~K.}\ \bibnamefont
  {Brennen}},\ and\ \bibinfo {author} {\bibfnamefont {J.}~\bibnamefont
  {Twamley}},\ }\bibfield  {title} {\bibinfo {title} {Fault {{Tolerance}} with
  {{Noisy}} and {{Slow Measurements}} and {{Preparation}}},\ }\href
  {https://doi.org/10.1103/PhysRevLett.105.100501} {\bibfield  {journal}
  {\bibinfo  {journal} {Physical Review Letters}\ }\textbf {\bibinfo {volume}
  {105}},\ \bibinfo {pages} {100501} (\bibinfo {year} {2010})}\BibitemShut
  {NoStop}%
\bibitem [{\citenamefont {Crow}\ \emph {et~al.}(2016)\citenamefont {Crow},
  \citenamefont {Joynt},\ and\ \citenamefont {Saffman}}]{crow_improved_2016}%
  \BibitemOpen
  \bibfield  {author} {\bibinfo {author} {\bibfnamefont {D.}~\bibnamefont
  {Crow}}, \bibinfo {author} {\bibfnamefont {R.}~\bibnamefont {Joynt}},\ and\
  \bibinfo {author} {\bibfnamefont {M.}~\bibnamefont {Saffman}},\ }\bibfield
  {title} {\bibinfo {title} {Improved {{Error Thresholds}} for
  {{Measurement-Free Error Correction}}},\ }\href
  {https://doi.org/10.1103/PhysRevLett.117.130503} {\bibfield  {journal}
  {\bibinfo  {journal} {Physical Review Letters}\ }\textbf {\bibinfo {volume}
  {117}},\ \bibinfo {pages} {130503} (\bibinfo {year} {2016})}\BibitemShut
  {NoStop}%
\bibitem [{\citenamefont {Ercan}\ \emph {et~al.}(2018)\citenamefont {Ercan},
  \citenamefont {Ghosh}, \citenamefont {Crow}, \citenamefont {Premakumar},
  \citenamefont {Joynt}, \citenamefont {Friesen},\ and\ \citenamefont
  {Coppersmith}}]{ercan_measurement-free_2018}%
  \BibitemOpen
  \bibfield  {author} {\bibinfo {author} {\bibfnamefont {H.~E.}\ \bibnamefont
  {Ercan}}, \bibinfo {author} {\bibfnamefont {J.}~\bibnamefont {Ghosh}},
  \bibinfo {author} {\bibfnamefont {D.}~\bibnamefont {Crow}}, \bibinfo {author}
  {\bibfnamefont {V.~N.}\ \bibnamefont {Premakumar}}, \bibinfo {author}
  {\bibfnamefont {R.}~\bibnamefont {Joynt}}, \bibinfo {author} {\bibfnamefont
  {M.}~\bibnamefont {Friesen}},\ and\ \bibinfo {author} {\bibfnamefont {S.~N.}\
  \bibnamefont {Coppersmith}},\ }\bibfield  {title} {\bibinfo {title}
  {Measurement-free implementations of small-scale surface codes for
  quantum-dot qubits},\ }\href {https://doi.org/10.1103/PhysRevA.97.012318}
  {\bibfield  {journal} {\bibinfo  {journal} {Physical Review A}\ }\textbf
  {\bibinfo {volume} {97}},\ \bibinfo {pages} {012318} (\bibinfo {year}
  {2018})}\BibitemShut {NoStop}%
\bibitem [{\citenamefont {Balauca}\ and\ \citenamefont
  {Arusoaie}(2022)}]{groen_efficient_2022}%
  \BibitemOpen
  \bibfield  {author} {\bibinfo {author} {\bibfnamefont {S.}~\bibnamefont
  {Balauca}}\ and\ \bibinfo {author} {\bibfnamefont {A.}~\bibnamefont
  {Arusoaie}},\ }\bibfield  {title} {\bibinfo {title} {Efficient
  {{Constructions}} for {{Simulating Multi Controlled Quantum Gates}}},\ }in\
  \href {https://doi.org/10.1007/978-3-031-08760-8\_16} {\emph {\bibinfo
  {booktitle} {Computational {{Science}} \textendash{} {{ICCS}} 2022}}},\ Vol.\
  \bibinfo {volume} {13353},\ \bibinfo {editor} {edited by\ \bibinfo {editor}
  {\bibfnamefont {D.}~\bibnamefont {Groen}}, \bibinfo {editor} {\bibfnamefont
  {C.}~\bibnamefont {{de Mulatier}}}, \bibinfo {editor} {\bibfnamefont
  {M.}~\bibnamefont {Paszynski}}, \bibinfo {editor} {\bibfnamefont {V.~V.}\
  \bibnamefont {Krzhizhanovskaya}}, \bibinfo {editor} {\bibfnamefont {J.~J.}\
  \bibnamefont {Dongarra}},\ and\ \bibinfo {editor} {\bibfnamefont {P.~M.~A.}\
  \bibnamefont {Sloot}}}\ (\bibinfo  {publisher} {{Springer International
  Publishing}},\ \bibinfo {address} {{Cham}},\ \bibinfo {year} {2022})\ pp.\
  \bibinfo {pages} {179--194}\BibitemShut {NoStop}%
\bibitem [{\citenamefont {Labuhn}\ \emph {et~al.}(2014)\citenamefont {Labuhn},
  \citenamefont {Ravets}, \citenamefont {Barredo}, \citenamefont {B{\'e}guin},
  \citenamefont {Nogrette}, \citenamefont {Lahaye},\ and\ \citenamefont
  {Browaeys}}]{labuhn_single-atom_2014}%
  \BibitemOpen
  \bibfield  {author} {\bibinfo {author} {\bibfnamefont {H.}~\bibnamefont
  {Labuhn}}, \bibinfo {author} {\bibfnamefont {S.}~\bibnamefont {Ravets}},
  \bibinfo {author} {\bibfnamefont {D.}~\bibnamefont {Barredo}}, \bibinfo
  {author} {\bibfnamefont {L.}~\bibnamefont {B{\'e}guin}}, \bibinfo {author}
  {\bibfnamefont {F.}~\bibnamefont {Nogrette}}, \bibinfo {author}
  {\bibfnamefont {T.}~\bibnamefont {Lahaye}},\ and\ \bibinfo {author}
  {\bibfnamefont {A.}~\bibnamefont {Browaeys}},\ }\bibfield  {title} {\bibinfo
  {title} {Single-atom addressing in microtraps for quantum-state engineering
  using {{Rydberg}} atoms},\ }\href
  {https://doi.org/10.1103/PhysRevA.90.023415} {\bibfield  {journal} {\bibinfo
  {journal} {Physical Review A}\ }\textbf {\bibinfo {volume} {90}},\ \bibinfo
  {pages} {023415} (\bibinfo {year} {2014})}\BibitemShut {NoStop}%
\bibitem [{\citenamefont {Wang}\ \emph {et~al.}(2016)\citenamefont {Wang},
  \citenamefont {Kumar}, \citenamefont {Wu},\ and\ \citenamefont
  {Weiss}}]{wang_single-qubit_2016}%
  \BibitemOpen
  \bibfield  {author} {\bibinfo {author} {\bibfnamefont {Y.}~\bibnamefont
  {Wang}}, \bibinfo {author} {\bibfnamefont {A.}~\bibnamefont {Kumar}},
  \bibinfo {author} {\bibfnamefont {T.-Y.}\ \bibnamefont {Wu}},\ and\ \bibinfo
  {author} {\bibfnamefont {D.~S.}\ \bibnamefont {Weiss}},\ }\bibfield  {title}
  {\bibinfo {title} {Single-qubit gates based on targeted phase shifts in a
  {{3D}} neutral atom array},\ }\href {https://doi.org/10.1126/science.aaf2581}
  {\bibfield  {journal} {\bibinfo  {journal} {Science}\ }\textbf {\bibinfo
  {volume} {352}},\ \bibinfo {pages} {1562} (\bibinfo {year}
  {2016})}\BibitemShut {NoStop}%
\bibitem [{\citenamefont {Pedersen}\ \emph {et~al.}(2007)\citenamefont
  {Pedersen}, \citenamefont {M{\o}ller},\ and\ \citenamefont
  {M{\o}lmer}}]{pedersen_fidelity_2007}%
  \BibitemOpen
  \bibfield  {author} {\bibinfo {author} {\bibfnamefont {L.~H.}\ \bibnamefont
  {Pedersen}}, \bibinfo {author} {\bibfnamefont {N.~M.}\ \bibnamefont
  {M{\o}ller}},\ and\ \bibinfo {author} {\bibfnamefont {K.}~\bibnamefont
  {M{\o}lmer}},\ }\bibfield  {title} {\bibinfo {title} {Fidelity of quantum
  operations},\ }\href {https://doi.org/10.1016/j.physleta.2007.02.069}
  {\bibfield  {journal} {\bibinfo  {journal} {Physics Letters A}\ }\textbf
  {\bibinfo {volume} {367}},\ \bibinfo {pages} {47} (\bibinfo {year}
  {2007})}\BibitemShut {NoStop}%
\bibitem [{\citenamefont {Hunger}\ \emph {et~al.}(2010)\citenamefont {Hunger},
  \citenamefont {Steinmetz}, \citenamefont {Colombe}, \citenamefont {Deutsch},
  \citenamefont {H{\"a}nsch},\ and\ \citenamefont
  {Reichel}}]{hunger_fiber_2010-1}%
  \BibitemOpen
  \bibfield  {author} {\bibinfo {author} {\bibfnamefont {D.}~\bibnamefont
  {Hunger}}, \bibinfo {author} {\bibfnamefont {T.}~\bibnamefont {Steinmetz}},
  \bibinfo {author} {\bibfnamefont {Y.}~\bibnamefont {Colombe}}, \bibinfo
  {author} {\bibfnamefont {C.}~\bibnamefont {Deutsch}}, \bibinfo {author}
  {\bibfnamefont {T.~W.}\ \bibnamefont {H{\"a}nsch}},\ and\ \bibinfo {author}
  {\bibfnamefont {J.}~\bibnamefont {Reichel}},\ }\bibfield  {title} {\bibinfo
  {title} {A fiber {{Fabry}}\textendash{{Perot}} cavity with high finesse},\
  }\href {https://doi.org/10.1088/1367-2630/12/6/065038} {\bibfield  {journal}
  {\bibinfo  {journal} {New Journal of Physics}\ }\textbf {\bibinfo {volume}
  {12}},\ \bibinfo {pages} {065038} (\bibinfo {year} {2010})}\BibitemShut
  {NoStop}%
\bibitem [{\citenamefont {Uphoff}\ \emph {et~al.}(2015)\citenamefont {Uphoff},
  \citenamefont {Brekenfeld}, \citenamefont {Rempe},\ and\ \citenamefont
  {Ritter}}]{uphoff_frequency_2015}%
  \BibitemOpen
  \bibfield  {author} {\bibinfo {author} {\bibfnamefont {M.}~\bibnamefont
  {Uphoff}}, \bibinfo {author} {\bibfnamefont {M.}~\bibnamefont {Brekenfeld}},
  \bibinfo {author} {\bibfnamefont {G.}~\bibnamefont {Rempe}},\ and\ \bibinfo
  {author} {\bibfnamefont {S.}~\bibnamefont {Ritter}},\ }\bibfield  {title}
  {\bibinfo {title} {Frequency splitting of polarization eigenmodes in
  microscopic {{Fabry}}\textendash{{Perot}} cavities},\ }\href
  {https://doi.org/10.1088/1367-2630/17/1/013053} {\bibfield  {journal}
  {\bibinfo  {journal} {New Journal of Physics}\ }\textbf {\bibinfo {volume}
  {17}},\ \bibinfo {pages} {013053} (\bibinfo {year} {2015})}\BibitemShut
  {NoStop}%
\bibitem [{\citenamefont {Barontini}\ \emph {et~al.}(2015)\citenamefont
  {Barontini}, \citenamefont {Hohmann}, \citenamefont {Haas}, \citenamefont
  {Est{\`e}ve},\ and\ \citenamefont {Reichel}}]{barontini_deterministic_2015}%
  \BibitemOpen
  \bibfield  {author} {\bibinfo {author} {\bibfnamefont {G.}~\bibnamefont
  {Barontini}}, \bibinfo {author} {\bibfnamefont {L.}~\bibnamefont {Hohmann}},
  \bibinfo {author} {\bibfnamefont {F.}~\bibnamefont {Haas}}, \bibinfo {author}
  {\bibfnamefont {J.}~\bibnamefont {Est{\`e}ve}},\ and\ \bibinfo {author}
  {\bibfnamefont {J.}~\bibnamefont {Reichel}},\ }\bibfield  {title} {\bibinfo
  {title} {Deterministic generation of multiparticle entanglement by quantum
  {{Zeno}} dynamics},\ }\href {https://doi.org/10.1126/science.aaa0754}
  {\bibfield  {journal} {\bibinfo  {journal} {Science}\ }\textbf {\bibinfo
  {volume} {349}},\ \bibinfo {pages} {1317} (\bibinfo {year}
  {2015})}\BibitemShut {NoStop}%
\bibitem [{\citenamefont {Jaksch}\ \emph {et~al.}(2000)\citenamefont {Jaksch},
  \citenamefont {Cirac}, \citenamefont {Zoller}, \citenamefont {Rolston},
  \citenamefont {C{\^o}t{\'e}},\ and\ \citenamefont
  {Lukin}}]{jaksch_fast_2000}%
  \BibitemOpen
  \bibfield  {author} {\bibinfo {author} {\bibfnamefont {D.}~\bibnamefont
  {Jaksch}}, \bibinfo {author} {\bibfnamefont {J.~I.}\ \bibnamefont {Cirac}},
  \bibinfo {author} {\bibfnamefont {P.}~\bibnamefont {Zoller}}, \bibinfo
  {author} {\bibfnamefont {S.~L.}\ \bibnamefont {Rolston}}, \bibinfo {author}
  {\bibfnamefont {R.}~\bibnamefont {C{\^o}t{\'e}}},\ and\ \bibinfo {author}
  {\bibfnamefont {M.~D.}\ \bibnamefont {Lukin}},\ }\bibfield  {title} {\bibinfo
  {title} {Fast {{Quantum Gates}} for {{Neutral Atoms}}},\ }\href
  {https://doi.org/10.1103/PhysRevLett.85.2208} {\bibfield  {journal} {\bibinfo
   {journal} {Physical Review Letters}\ }\textbf {\bibinfo {volume} {85}},\
  \bibinfo {pages} {2208} (\bibinfo {year} {2000})}\BibitemShut {NoStop}%
\bibitem [{\citenamefont {Morgado}\ and\ \citenamefont
  {Whitlock}(2021)}]{morgado_quantum_2021}%
  \BibitemOpen
  \bibfield  {author} {\bibinfo {author} {\bibfnamefont {M.}~\bibnamefont
  {Morgado}}\ and\ \bibinfo {author} {\bibfnamefont {S.}~\bibnamefont
  {Whitlock}},\ }\bibfield  {title} {\bibinfo {title} {Quantum simulation and
  computing with {{Rydberg-interacting}} qubits},\ }\href
  {https://doi.org/10.1116/5.0036562} {\bibfield  {journal} {\bibinfo
  {journal} {AVS Quantum Science}\ }\textbf {\bibinfo {volume} {3}},\ \bibinfo
  {pages} {023501} (\bibinfo {year} {2021})}\BibitemShut {NoStop}%
\bibitem [{\citenamefont {Ramette}\ \emph {et~al.}(2023)\citenamefont
  {Ramette}, \citenamefont {Sinclair}, \citenamefont {Breuckmann},\ and\
  \citenamefont {Vuleti{\'c}}}]{ramette_fault-tolerant_2023}%
  \BibitemOpen
  \bibfield  {author} {\bibinfo {author} {\bibfnamefont {J.}~\bibnamefont
  {Ramette}}, \bibinfo {author} {\bibfnamefont {J.}~\bibnamefont {Sinclair}},
  \bibinfo {author} {\bibfnamefont {N.~P.}\ \bibnamefont {Breuckmann}},\ and\
  \bibinfo {author} {\bibfnamefont {V.}~\bibnamefont {Vuleti{\'c}}},\ }\href
  {http://arxiv.org/abs/2302.01296} {\bibinfo {title} {Fault-{{Tolerant
  Connection}} of {{Error-Corrected Qubits}} with {{Noisy Links}}}} (\bibinfo
  {year} {2023})\BibitemShut {NoStop}%
\bibitem [{\citenamefont {Pritchard}\ \emph {et~al.}(2014)\citenamefont
  {Pritchard}, \citenamefont {Isaacs}, \citenamefont {Beck}, \citenamefont
  {McDermott},\ and\ \citenamefont {Saffman}}]{pritchard_hybrid_2014}%
  \BibitemOpen
  \bibfield  {author} {\bibinfo {author} {\bibfnamefont {J.~D.}\ \bibnamefont
  {Pritchard}}, \bibinfo {author} {\bibfnamefont {J.~A.}\ \bibnamefont
  {Isaacs}}, \bibinfo {author} {\bibfnamefont {M.~A.}\ \bibnamefont {Beck}},
  \bibinfo {author} {\bibfnamefont {R.}~\bibnamefont {McDermott}},\ and\
  \bibinfo {author} {\bibfnamefont {M.}~\bibnamefont {Saffman}},\ }\bibfield
  {title} {\bibinfo {title} {Hybrid atom-photon quantum gate in a
  superconducting microwave resonator},\ }\href
  {https://doi.org/10.1103/PhysRevA.89.010301} {\bibfield  {journal} {\bibinfo
  {journal} {Physical Review A}\ }\textbf {\bibinfo {volume} {89}},\ \bibinfo
  {pages} {010301} (\bibinfo {year} {2014})}\BibitemShut {NoStop}%
\bibitem [{\citenamefont {Lei}\ \emph {et~al.}(2020)\citenamefont {Lei},
  \citenamefont {Krayzman}, \citenamefont {Ganjam}, \citenamefont {Frunzio},\
  and\ \citenamefont {Schoelkopf}}]{lei_high_2020}%
  \BibitemOpen
  \bibfield  {author} {\bibinfo {author} {\bibfnamefont {C.~U.}\ \bibnamefont
  {Lei}}, \bibinfo {author} {\bibfnamefont {L.}~\bibnamefont {Krayzman}},
  \bibinfo {author} {\bibfnamefont {S.}~\bibnamefont {Ganjam}}, \bibinfo
  {author} {\bibfnamefont {L.}~\bibnamefont {Frunzio}},\ and\ \bibinfo {author}
  {\bibfnamefont {R.~J.}\ \bibnamefont {Schoelkopf}},\ }\bibfield  {title}
  {\bibinfo {title} {High coherence superconducting microwave cavities with
  indium bump bonding},\ }\href {https://doi.org/10.1063/5.0003907} {\bibfield
  {journal} {\bibinfo  {journal} {Applied Physics Letters}\ }\textbf {\bibinfo
  {volume} {116}},\ \bibinfo {pages} {154002} (\bibinfo {year}
  {2020})}\BibitemShut {NoStop}%
\bibitem [{\citenamefont {Andr{\'e}}\ \emph {et~al.}(2006)\citenamefont
  {Andr{\'e}}, \citenamefont {DeMille}, \citenamefont {Doyle}, \citenamefont
  {Lukin}, \citenamefont {Maxwell}, \citenamefont {Rabl}, \citenamefont
  {Schoelkopf},\ and\ \citenamefont {Zoller}}]{andre_coherent_2006}%
  \BibitemOpen
  \bibfield  {author} {\bibinfo {author} {\bibfnamefont {A.}~\bibnamefont
  {Andr{\'e}}}, \bibinfo {author} {\bibfnamefont {D.}~\bibnamefont {DeMille}},
  \bibinfo {author} {\bibfnamefont {J.~M.}\ \bibnamefont {Doyle}}, \bibinfo
  {author} {\bibfnamefont {M.~D.}\ \bibnamefont {Lukin}}, \bibinfo {author}
  {\bibfnamefont {S.~E.}\ \bibnamefont {Maxwell}}, \bibinfo {author}
  {\bibfnamefont {P.}~\bibnamefont {Rabl}}, \bibinfo {author} {\bibfnamefont
  {R.~J.}\ \bibnamefont {Schoelkopf}},\ and\ \bibinfo {author} {\bibfnamefont
  {P.}~\bibnamefont {Zoller}},\ }\bibfield  {title} {\bibinfo {title} {A
  coherent all-electrical interface between polar molecules and mesoscopic
  superconducting resonators},\ }\href {https://doi.org/10.1038/nphys386}
  {\bibfield  {journal} {\bibinfo  {journal} {Nature Physics}\ }\textbf
  {\bibinfo {volume} {2}},\ \bibinfo {pages} {636} (\bibinfo {year}
  {2006})}\BibitemShut {NoStop}%
\bibitem [{\citenamefont {Rabl}\ \emph {et~al.}(2006)\citenamefont {Rabl},
  \citenamefont {DeMille}, \citenamefont {Doyle}, \citenamefont {Lukin},
  \citenamefont {Schoelkopf},\ and\ \citenamefont {Zoller}}]{rabl_hybrid_2006}%
  \BibitemOpen
  \bibfield  {author} {\bibinfo {author} {\bibfnamefont {P.}~\bibnamefont
  {Rabl}}, \bibinfo {author} {\bibfnamefont {D.}~\bibnamefont {DeMille}},
  \bibinfo {author} {\bibfnamefont {J.~M.}\ \bibnamefont {Doyle}}, \bibinfo
  {author} {\bibfnamefont {M.~D.}\ \bibnamefont {Lukin}}, \bibinfo {author}
  {\bibfnamefont {R.~J.}\ \bibnamefont {Schoelkopf}},\ and\ \bibinfo {author}
  {\bibfnamefont {P.}~\bibnamefont {Zoller}},\ }\bibfield  {title} {\bibinfo
  {title} {Hybrid {{Quantum Processors}}: {{Molecular Ensembles}} as {{Quantum
  Memory}} for {{Solid State Circuits}}},\ }\href
  {https://doi.org/10.1103/PhysRevLett.97.033003} {\bibfield  {journal}
  {\bibinfo  {journal} {Physical Review Letters}\ }\textbf {\bibinfo {volume}
  {97}},\ \bibinfo {pages} {033003} (\bibinfo {year} {2006})}\BibitemShut
  {NoStop}%
\bibitem [{\citenamefont {Sawant}\ \emph {et~al.}(2020)\citenamefont {Sawant},
  \citenamefont {Blackmore}, \citenamefont {Gregory}, \citenamefont
  {{Mur-Petit}}, \citenamefont {Jaksch}, \citenamefont {Aldegunde},
  \citenamefont {Hutson}, \citenamefont {Tarbutt},\ and\ \citenamefont
  {Cornish}}]{sawant_ultracold_2020}%
  \BibitemOpen
  \bibfield  {author} {\bibinfo {author} {\bibfnamefont {R.}~\bibnamefont
  {Sawant}}, \bibinfo {author} {\bibfnamefont {J.~A.}\ \bibnamefont
  {Blackmore}}, \bibinfo {author} {\bibfnamefont {P.~D.}\ \bibnamefont
  {Gregory}}, \bibinfo {author} {\bibfnamefont {J.}~\bibnamefont
  {{Mur-Petit}}}, \bibinfo {author} {\bibfnamefont {D.}~\bibnamefont {Jaksch}},
  \bibinfo {author} {\bibfnamefont {J.}~\bibnamefont {Aldegunde}}, \bibinfo
  {author} {\bibfnamefont {J.~M.}\ \bibnamefont {Hutson}}, \bibinfo {author}
  {\bibfnamefont {M.~R.}\ \bibnamefont {Tarbutt}},\ and\ \bibinfo {author}
  {\bibfnamefont {S.~L.}\ \bibnamefont {Cornish}},\ }\bibfield  {title}
  {\bibinfo {title} {Ultracold polar molecules as qudits},\ }\href
  {https://doi.org/10.1088/1367-2630/ab60f4} {\bibfield  {journal} {\bibinfo
  {journal} {New Journal of Physics}\ }\textbf {\bibinfo {volume} {22}},\
  \bibinfo {pages} {013027} (\bibinfo {year} {2020})}\BibitemShut {NoStop}%
\bibitem [{\citenamefont {Childs}\ \emph {et~al.}(1981)\citenamefont {Childs},
  \citenamefont {Cok}, \citenamefont {Goodman},\ and\ \citenamefont
  {Goodman}}]{childs_hyperfine_1981}%
  \BibitemOpen
  \bibfield  {author} {\bibinfo {author} {\bibfnamefont {W.~J.}\ \bibnamefont
  {Childs}}, \bibinfo {author} {\bibfnamefont {D.~R.}\ \bibnamefont {Cok}},
  \bibinfo {author} {\bibfnamefont {G.~L.}\ \bibnamefont {Goodman}},\ and\
  \bibinfo {author} {\bibfnamefont {L.~S.}\ \bibnamefont {Goodman}},\
  }\bibfield  {title} {\bibinfo {title} {Hyperfine and spin\textendash
  rotational structure of {{CaBr X}} {\textsuperscript{2}} {{$\Sigma$}} (v = 0)
  by molecular-beam laser-rf double resonance},\ }\href
  {https://doi.org/10.1063/1.442057} {\bibfield  {journal} {\bibinfo  {journal}
  {The Journal of Chemical Physics}\ }\textbf {\bibinfo {volume} {75}},\
  \bibinfo {pages} {501} (\bibinfo {year} {1981})}\BibitemShut {NoStop}%
\bibitem [{\citenamefont {Buhmann}\ \emph {et~al.}(2008)\citenamefont
  {Buhmann}, \citenamefont {Tarbutt}, \citenamefont {Scheel},\ and\
  \citenamefont {Hinds}}]{buhmann_surface-induced_2008}%
  \BibitemOpen
  \bibfield  {author} {\bibinfo {author} {\bibfnamefont {S.~Y.}\ \bibnamefont
  {Buhmann}}, \bibinfo {author} {\bibfnamefont {M.~R.}\ \bibnamefont
  {Tarbutt}}, \bibinfo {author} {\bibfnamefont {S.}~\bibnamefont {Scheel}},\
  and\ \bibinfo {author} {\bibfnamefont {E.~A.}\ \bibnamefont {Hinds}},\
  }\bibfield  {title} {\bibinfo {title} {Surface-induced heating of cold polar
  molecules},\ }\href {https://doi.org/10.1103/PhysRevA.78.052901} {\bibfield
  {journal} {\bibinfo  {journal} {Physical Review A}\ }\textbf {\bibinfo
  {volume} {78}},\ \bibinfo {pages} {052901} (\bibinfo {year}
  {2008})}\BibitemShut {NoStop}%
\bibitem [{\citenamefont {Manucharyan}\ \emph {et~al.}(2009)\citenamefont
  {Manucharyan}, \citenamefont {Koch}, \citenamefont {Glazman},\ and\
  \citenamefont {Devoret}}]{Manucharyan_2009}%
  \BibitemOpen
  \bibfield  {author} {\bibinfo {author} {\bibfnamefont {V.~E.}\ \bibnamefont
  {Manucharyan}}, \bibinfo {author} {\bibfnamefont {J.}~\bibnamefont {Koch}},
  \bibinfo {author} {\bibfnamefont {L.~I.}\ \bibnamefont {Glazman}},\ and\
  \bibinfo {author} {\bibfnamefont {M.~H.}\ \bibnamefont {Devoret}},\
  }\bibfield  {title} {\bibinfo {title} {Fluxonium: Single cooper-pair circuit
  free of charge offsets},\ }\href {https://doi.org/10.1126/science.1175552}
  {\bibfield  {journal} {\bibinfo  {journal} {Science}\ }\textbf {\bibinfo
  {volume} {\bf 326}},\ \bibinfo {pages} {113} (\bibinfo {year}
  {2009})}\BibitemShut {NoStop}%
\bibitem [{\citenamefont {Abdelhafez}\ \emph {et~al.}(2020)\citenamefont
  {Abdelhafez}, \citenamefont {Baker}, \citenamefont {Gyenis}, \citenamefont
  {Mundada}, \citenamefont {Houck}, \citenamefont {Schuster},\ and\
  \citenamefont {Koch}}]{Abdelhafez_2020}%
  \BibitemOpen
  \bibfield  {author} {\bibinfo {author} {\bibfnamefont {M.}~\bibnamefont
  {Abdelhafez}}, \bibinfo {author} {\bibfnamefont {B.}~\bibnamefont {Baker}},
  \bibinfo {author} {\bibfnamefont {A.}~\bibnamefont {Gyenis}}, \bibinfo
  {author} {\bibfnamefont {P.}~\bibnamefont {Mundada}}, \bibinfo {author}
  {\bibfnamefont {A.~A.}\ \bibnamefont {Houck}}, \bibinfo {author}
  {\bibfnamefont {D.}~\bibnamefont {Schuster}},\ and\ \bibinfo {author}
  {\bibfnamefont {J.}~\bibnamefont {Koch}},\ }\bibfield  {title} {\bibinfo
  {title} {Universal gates for protected superconducting qubits using optimal
  control},\ }\bibfield  {journal} {\bibinfo  {journal} {Physical Review A}\
  }\textbf {\bibinfo {volume} {101}},\ \href
  {https://doi.org/10.1103/physreva.101.022321} {10.1103/physreva.101.022321}
  (\bibinfo {year} {2020})\BibitemShut {NoStop}%
\bibitem [{\citenamefont {Somoroff}\ \emph {et~al.}(2023)\citenamefont
  {Somoroff}, \citenamefont {Ficheux}, \citenamefont {Mencia}, \citenamefont
  {Xiong}, \citenamefont {Kuzmin},\ and\ \citenamefont
  {Manucharyan}}]{PhysRevLett.130.267001}%
  \BibitemOpen
  \bibfield  {author} {\bibinfo {author} {\bibfnamefont {A.}~\bibnamefont
  {Somoroff}}, \bibinfo {author} {\bibfnamefont {Q.}~\bibnamefont {Ficheux}},
  \bibinfo {author} {\bibfnamefont {R.~A.}\ \bibnamefont {Mencia}}, \bibinfo
  {author} {\bibfnamefont {H.}~\bibnamefont {Xiong}}, \bibinfo {author}
  {\bibfnamefont {R.}~\bibnamefont {Kuzmin}},\ and\ \bibinfo {author}
  {\bibfnamefont {V.~E.}\ \bibnamefont {Manucharyan}},\ }\bibfield  {title}
  {\bibinfo {title} {Millisecond coherence in a superconducting qubit},\ }\href
  {https://doi.org/10.1103/PhysRevLett.130.267001} {\bibfield  {journal}
  {\bibinfo  {journal} {Phys. Rev. Lett.}\ }\textbf {\bibinfo {volume} {130}},\
  \bibinfo {pages} {267001} (\bibinfo {year} {2023})}\BibitemShut {NoStop}%
\bibitem [{\citenamefont {Groszkowski}\ and\ \citenamefont
  {Koch}(2021)}]{Groszkowski_2021}%
  \BibitemOpen
  \bibfield  {author} {\bibinfo {author} {\bibfnamefont {P.}~\bibnamefont
  {Groszkowski}}\ and\ \bibinfo {author} {\bibfnamefont {J.}~\bibnamefont
  {Koch}},\ }\bibfield  {title} {\bibinfo {title} {Scqubits: a python package
  for superconducting qubits},\ }\href
  {https://doi.org/10.22331/q-2021-11-17-583} {\bibfield  {journal} {\bibinfo
  {journal} {Quantum}\ }\textbf {\bibinfo {volume} {5}},\ \bibinfo {pages}
  {583} (\bibinfo {year} {2021})}\BibitemShut {NoStop}%
\bibitem [{\citenamefont {Chitta}\ \emph {et~al.}(2022)\citenamefont {Chitta},
  \citenamefont {Zhao}, \citenamefont {Huang}, \citenamefont {Mondragon-Shem},\
  and\ \citenamefont {Koch}}]{chitta2022computeraided}%
  \BibitemOpen
  \bibfield  {author} {\bibinfo {author} {\bibfnamefont {S.~P.}\ \bibnamefont
  {Chitta}}, \bibinfo {author} {\bibfnamefont {T.}~\bibnamefont {Zhao}},
  \bibinfo {author} {\bibfnamefont {Z.}~\bibnamefont {Huang}}, \bibinfo
  {author} {\bibfnamefont {I.}~\bibnamefont {Mondragon-Shem}},\ and\ \bibinfo
  {author} {\bibfnamefont {J.}~\bibnamefont {Koch}},\ }\href@noop {} {\bibinfo
  {title} {Computer-aided quantization and numerical analysis of
  superconducting circuits}} (\bibinfo {year} {2022}),\ \Eprint
  {https://arxiv.org/abs/2206.08320} {arXiv:2206.08320 [quant-ph]} \BibitemShut
  {NoStop}%
\bibitem [{\citenamefont {Nguyen}\ \emph {et~al.}(2019)\citenamefont {Nguyen},
  \citenamefont {Lin}, \citenamefont {Somoroff}, \citenamefont {Mencia},
  \citenamefont {Grabon},\ and\ \citenamefont
  {Manucharyan}}]{PhysRevX.9.041041}%
  \BibitemOpen
  \bibfield  {author} {\bibinfo {author} {\bibfnamefont {L.~B.}\ \bibnamefont
  {Nguyen}}, \bibinfo {author} {\bibfnamefont {Y.-H.}\ \bibnamefont {Lin}},
  \bibinfo {author} {\bibfnamefont {A.}~\bibnamefont {Somoroff}}, \bibinfo
  {author} {\bibfnamefont {R.}~\bibnamefont {Mencia}}, \bibinfo {author}
  {\bibfnamefont {N.}~\bibnamefont {Grabon}},\ and\ \bibinfo {author}
  {\bibfnamefont {V.~E.}\ \bibnamefont {Manucharyan}},\ }\bibfield  {title}
  {\bibinfo {title} {High-coherence fluxonium qubit},\ }\href
  {https://doi.org/10.1103/PhysRevX.9.041041} {\bibfield  {journal} {\bibinfo
  {journal} {Phys. Rev. X}\ }\textbf {\bibinfo {volume} {9}},\ \bibinfo {pages}
  {041041} (\bibinfo {year} {2019})}\BibitemShut {NoStop}%
\bibitem [{\citenamefont {Moskalenko}\ \emph {et~al.}(2022)\citenamefont
  {Moskalenko}, \citenamefont {Simakov}, \citenamefont {Abramov}, \citenamefont
  {Grigorev}, \citenamefont {Moskalev}, \citenamefont {Pishchimova},
  \citenamefont {Smirnov}, \citenamefont {Zikiy}, \citenamefont {Rodionov},\
  and\ \citenamefont {Besedin}}]{Moskalenko2022}%
  \BibitemOpen
  \bibfield  {author} {\bibinfo {author} {\bibfnamefont {I.~N.}\ \bibnamefont
  {Moskalenko}}, \bibinfo {author} {\bibfnamefont {I.~A.}\ \bibnamefont
  {Simakov}}, \bibinfo {author} {\bibfnamefont {N.~N.}\ \bibnamefont
  {Abramov}}, \bibinfo {author} {\bibfnamefont {A.~A.}\ \bibnamefont
  {Grigorev}}, \bibinfo {author} {\bibfnamefont {D.~O.}\ \bibnamefont
  {Moskalev}}, \bibinfo {author} {\bibfnamefont {A.~A.}\ \bibnamefont
  {Pishchimova}}, \bibinfo {author} {\bibfnamefont {N.~S.}\ \bibnamefont
  {Smirnov}}, \bibinfo {author} {\bibfnamefont {E.~V.}\ \bibnamefont {Zikiy}},
  \bibinfo {author} {\bibfnamefont {I.~A.}\ \bibnamefont {Rodionov}},\ and\
  \bibinfo {author} {\bibfnamefont {I.~S.}\ \bibnamefont {Besedin}},\
  }\bibfield  {title} {\bibinfo {title} {High fidelity two-qubit gates on
  fluxoniums using a tunable coupler},\ }\href
  {https://doi.org/10.1038/s41534-022-00644-x} {\bibfield  {journal} {\bibinfo
  {journal} {npj Quantum Information}\ }\textbf {\bibinfo {volume} {8}},\
  \bibinfo {pages} {130} (\bibinfo {year} {2022})}\BibitemShut {NoStop}%
\bibitem [{\citenamefont {Cheng}\ \emph {et~al.}(2022)\citenamefont {Cheng},
  \citenamefont {Zhang}, \citenamefont {Zhou},\ and\ \citenamefont
  {Zhou}}]{dynamical_dec}%
  \BibitemOpen
  \bibfield  {author} {\bibinfo {author} {\bibfnamefont {J.-M.}\ \bibnamefont
  {Cheng}}, \bibinfo {author} {\bibfnamefont {Y.}~\bibnamefont {Zhang}},
  \bibinfo {author} {\bibfnamefont {X.-F.}\ \bibnamefont {Zhou}},\ and\
  \bibinfo {author} {\bibfnamefont {Z.-W.}\ \bibnamefont {Zhou}},\ }\bibfield
  {title} {\bibinfo {title} {Enhancing quantum coherence of a fluxonium qubit
  by employing flux modulation with tunable-complex-amplitude},\ }\href
  {https://doi.org/10.1088/1367-2630/acacbd} {\bibfield  {journal} {\bibinfo
  {journal} {New Journal of Physics}\ }\textbf {\bibinfo {volume} {24}}
  (\bibinfo {year} {2022})}\BibitemShut {NoStop}%
\bibitem [{\citenamefont {DiVincenzo}\ and\ \citenamefont
  {Shor}(1996)}]{PhysRevLett.77.3260}%
  \BibitemOpen
  \bibfield  {author} {\bibinfo {author} {\bibfnamefont {D.~P.}\ \bibnamefont
  {DiVincenzo}}\ and\ \bibinfo {author} {\bibfnamefont {P.~W.}\ \bibnamefont
  {Shor}},\ }\bibfield  {title} {\bibinfo {title} {Fault-tolerant error
  correction with efficient quantum codes},\ }\href
  {https://doi.org/10.1103/PhysRevLett.77.3260} {\bibfield  {journal} {\bibinfo
   {journal} {Phys. Rev. Lett.}\ }\textbf {\bibinfo {volume} {77}},\ \bibinfo
  {pages} {3260} (\bibinfo {year} {1996})}\BibitemShut {NoStop}%
\bibitem [{\citenamefont {DiVincenzo}\ and\ \citenamefont
  {Aliferis}(2007)}]{divincenzo_effective_2007}%
  \BibitemOpen
  \bibfield  {author} {\bibinfo {author} {\bibfnamefont {D.~P.}\ \bibnamefont
  {DiVincenzo}}\ and\ \bibinfo {author} {\bibfnamefont {P.}~\bibnamefont
  {Aliferis}},\ }\bibfield  {title} {\bibinfo {title} {Effective
  {{Fault-Tolerant Quantum Computation}} with {{Slow Measurements}}},\ }\href
  {https://doi.org/10.1103/PhysRevLett.98.020501} {\bibfield  {journal}
  {\bibinfo  {journal} {Physical Review Letters}\ }\textbf {\bibinfo {volume}
  {98}},\ \bibinfo {pages} {020501} (\bibinfo {year} {2007})}\BibitemShut
  {NoStop}%
\bibitem [{\citenamefont {Chao}\ and\ \citenamefont
  {Reichardt}(2018)}]{Chao_2018}%
  \BibitemOpen
  \bibfield  {author} {\bibinfo {author} {\bibfnamefont {R.}~\bibnamefont
  {Chao}}\ and\ \bibinfo {author} {\bibfnamefont {B.~W.}\ \bibnamefont
  {Reichardt}},\ }\bibfield  {title} {\bibinfo {title} {Quantum error
  correction with only two extra qubits},\ }\href
  {https://doi.org/10.1103/PhysRevLett.121.050502} {\bibfield  {journal}
  {\bibinfo  {journal} {Phys. Rev. Lett.}\ }\textbf {\bibinfo {volume} {121}},\
  \bibinfo {pages} {050502} (\bibinfo {year} {2018})}\BibitemShut {NoStop}%
\bibitem [{\citenamefont {Glaser}\ \emph {et~al.}(2015)\citenamefont {Glaser},
  \citenamefont {Boscain}, \citenamefont {Calarco}, \citenamefont {Koch},
  \citenamefont {K{\"o}ckenberger}, \citenamefont {Kosloff}, \citenamefont
  {Kuprov}, \citenamefont {Luy}, \citenamefont {Schirmer}, \citenamefont
  {{Schulte-Herbr{\"u}ggen}}, \citenamefont {Sugny},\ and\ \citenamefont
  {Wilhelm}}]{glaser_training_2015}%
  \BibitemOpen
  \bibfield  {author} {\bibinfo {author} {\bibfnamefont {S.~J.}\ \bibnamefont
  {Glaser}}, \bibinfo {author} {\bibfnamefont {U.}~\bibnamefont {Boscain}},
  \bibinfo {author} {\bibfnamefont {T.}~\bibnamefont {Calarco}}, \bibinfo
  {author} {\bibfnamefont {C.~P.}\ \bibnamefont {Koch}}, \bibinfo {author}
  {\bibfnamefont {W.}~\bibnamefont {K{\"o}ckenberger}}, \bibinfo {author}
  {\bibfnamefont {R.}~\bibnamefont {Kosloff}}, \bibinfo {author} {\bibfnamefont
  {I.}~\bibnamefont {Kuprov}}, \bibinfo {author} {\bibfnamefont
  {B.}~\bibnamefont {Luy}}, \bibinfo {author} {\bibfnamefont {S.}~\bibnamefont
  {Schirmer}}, \bibinfo {author} {\bibfnamefont {T.}~\bibnamefont
  {{Schulte-Herbr{\"u}ggen}}}, \bibinfo {author} {\bibfnamefont
  {D.}~\bibnamefont {Sugny}},\ and\ \bibinfo {author} {\bibfnamefont {F.~K.}\
  \bibnamefont {Wilhelm}},\ }\bibfield  {title} {\bibinfo {title} {Training
  {{Schr\"odinger}}'s cat: Quantum optimal control: {{Strategic}} report on
  current status, visions and goals for research in {{Europe}}},\ }\href
  {https://doi.org/10.1140/epjd/e2015-60464-1} {\bibfield  {journal} {\bibinfo
  {journal} {The European Physical Journal D}\ }\textbf {\bibinfo {volume}
  {69}},\ \bibinfo {pages} {279} (\bibinfo {year} {2015})}\BibitemShut
  {NoStop}%
\bibitem [{\citenamefont {Wilhelm}\ \emph {et~al.}(2020)\citenamefont
  {Wilhelm}, \citenamefont {Kirchhoff}, \citenamefont {Machnes}, \citenamefont
  {Wittler},\ and\ \citenamefont {Sugny}}]{wilhelm_introduction_2020}%
  \BibitemOpen
  \bibfield  {author} {\bibinfo {author} {\bibfnamefont {F.~K.}\ \bibnamefont
  {Wilhelm}}, \bibinfo {author} {\bibfnamefont {S.}~\bibnamefont {Kirchhoff}},
  \bibinfo {author} {\bibfnamefont {S.}~\bibnamefont {Machnes}}, \bibinfo
  {author} {\bibfnamefont {N.}~\bibnamefont {Wittler}},\ and\ \bibinfo {author}
  {\bibfnamefont {D.}~\bibnamefont {Sugny}},\ }\bibfield  {title} {\bibinfo
  {title} {An introduction into optimal control for quantum technologies},\
  }\bibfield  {journal} {\bibinfo  {journal} {arXiv:2003.10132}\ }\href
  {https://doi.org/10.48550/arXiv.2003.10132} {10.48550/arXiv.2003.10132}
  (\bibinfo {year} {2020})\BibitemShut {NoStop}%
\end{thebibliography}%

\end{document}